\documentclass[aps,twocolumn,prd,superscriptaddress,floatfix,showpacs,showkeys]{revtex4-2}

\usepackage[latin1]{inputenc}
\usepackage[T1]{fontenc}
\usepackage{lmodern}
\usepackage{newtxtext}

\usepackage{amsmath}
\usepackage{amssymb}
\usepackage{amsthm}
\usepackage{amscd}
\usepackage{mathtools}
\usepackage{empheq}
\usepackage{mathrsfs}
\usepackage{dsfont}
\usepackage{scalerel}

\usepackage{graphicx}
\usepackage{grffile}
\usepackage{xcolor}
\usepackage{subcaption}
\usepackage{sidecap}

\usepackage{tikz}
\usetikzlibrary{arrows.meta, positioning, shapes.geometric, decorations.pathmorphing}

\usepackage{etoolbox}
\usepackage[mathlines]{lineno}
\usepackage[ruled]{algorithm2e}

\usepackage[colorlinks=true, pdfstartview=FitV, linkcolor=blue, citecolor=blue, urlcolor=blue]{hyperref}

\makeatletter
\renewcommand{\p@subsubsection}{\thesection\,}

\renewcommand{\subsubsection}{\@startsection{subsubsection}{3}{\z@}%
  {2ex \@plus 1ex \@minus .1ex}%
  {1ex \@plus .1ex}%
  {\normalfont\small\bfseries}}
\makeatother

\makeatletter
\def\appendix@toc@setup{%
  \let\oldl@section\l@section
  \def\l@section##1##2{%
    \begingroup
    \def\numberline####1{\makebox[6.5em][l]{Appendix ####1.\quad}}%
    \oldl@section{##1}{##2}%
    \endgroup
  }%
}
\apptocmd{\appendix}{\addtocontents{toc}{\protect\appendix@toc@setup}}{}{}
\makeatother

\newtheorem{thm}{Theorem}[section]
\newtheorem{prop}[thm]{Proposition}

\newtheorem{rem}{Remark}[section]

\def\bprop{\begin{prop}}
\def\eprop{\end{prop}}
\def\bt{\begin{thm}}
\def\et{\end{thm}}
\def\br{\begin{rem}}
\def\er{\end{rem}}

\usepackage{mathrsfs}
\usepackage{bm}
\newcommand{\Rstyle}{\mathscr{R}}
\newcommand{\Cstyle}{\mathscr{C}}
\newcommand{\vq}{\mathbf{v}}
\def\be{\begin{equation}}
\def\ee{\end{equation}}
\def\bes{\begin{equation*}}
\def\ees{\end{equation*}}
\def\bea{\begin{equation} \begin{aligned}}
\def\eea{\end{aligned} \end{equation}}
\def\beas{\begin{equation*} \begin{aligned}}
\def\eeas{\end{aligned} \end{equation*}}

\def\bi{\begin{itemize}}
\def\ei{\end{itemize}}

\def\d{\, \mathrm{d}}

\newcommand{\eps}{\varepsilon}

\newcommand{\pd}[2]{\frac{\partial #1}{\partial #2}}

\definecolor{rred}{rgb}{0.7,0,0.1}
\definecolor{ccyan}{rgb}{0,.5,1}
\definecolor{greenrb}{rgb}{0.2,0.6,0.2}

\definecolor{ppink}{rgb}{1,0.1,0.6}

\begin{document}

\title{Ruelle--Pollicott Theory for Metastable Systems: A Unified Framework for Tipping Transitions}

\author{Micka{\"e}l D. Chekroun}
\email{mchekroun@atmos.ucla.edu}
\affiliation{Department of Atmospheric and Oceanic Sciences, University of California, Los Angeles, CA 90095-1565, USA}
\affiliation{Department of Earth and Planetary Sciences, Weizmann Institute of Science, Rehovot 76100, Israel}

\author{Valerio Lucarini}
\affiliation{School of Computing and Mathematical Sciences, University of Leicester, Leicester, LE17RH, UK}

\date{\today}

\begin{abstract}

Tipping points---abrupt and potentially irreversible reorganizations of the statistical state of a system---pose a fundamental challenge across climate science, ecology, and other complex systems. Classical early-warning theory rests on critical slowing down: as a stability threshold is approached, recovery slows, autocorrelations and variance increase, and spectral power shifts toward low frequencies. This paradigm is powerful near simple equilibrium bifurcations, but it does not provide a general theory for stochastic systems with non-equilibrium attractors, multistability, or metastable transitions. We develop such a theory from the Ruelle--Pollicott (RP) spectrum of Kolmogorov generators of hypoelliptic It\^o diffusions.

Resolving the sensitivity of invariant statistics on RP spectral blocks---including spectral projectors and nilpotent Jordan structure---reveals that each contribution factorizes into a small spectral denominator and a residue determined jointly by the observable and the direction of perturbation. Small denominators permit large responses; residues produce them. A closing RP gap is therefore not sufficient for an early warning: vanishing residues can silence a critical block, while growing residues can generate the full classical warning signature with no gap closure at all. An early-warning signal is thus a property of a triple---critical RP block, observable, and perturbation direction---rather than of the dynamics alone. Variance, autocorrelation, susceptibility, and spectral power are different contractions of this RP structure and need not select the same blocks.

We demonstrate the distinction on a stochastic non-normal system whose RP spectrum is exactly frozen. Along a slow parameter ramp, variance and low-frequency power grow dramatically and lag-one autocorrelation approaches unity, producing a record more alarming than a genuine approach to gap closure although no resonance moves. The RP decomposition attributes the effect entirely to residue growth and provides two diagnostics: a residue-selected Kolmogorov mode that removes the false rate-based warning, and a dimensionless index \(\mathcal N(f)\) comparing the observed relaxation rate with that implied by the variance budget. We prove that \(\mathcal N(f)\leq1\) for reversible dynamics, so that \(\mathcal N(f)>1\) certifies residue-driven amplification from a single ramped record.

For metastable systems, we show that local critical slowing down and transition risk are distinct spectral objects and place bifurcation-induced (B-) and noise-induced (N-) tipping on a common RP footing. A single killed local problem generates two complementary reductions: its Doob \(Q\)-process isolates the local critical block governing in-well recovery, whereas its escape clocks combined with committor-based destination probabilities yield a reduced generator governing transitions between metastable states. In a one-dimensional metastable fold, the two mechanisms exhibit distinct scalings, \((\epsilon_c-\epsilon)^{1/2}\) for local recovery and \((\epsilon_c-\epsilon)^{3/2}\) for the quasipotential barrier, demonstrating that loss of recovery and escape risk are spectrally and operationally separable. Finally, the Doob drift emerges as the long-horizon limit of an optimal Girsanov importance-sampling control, providing a direct computational route to both diagnostics.

\end{abstract}

\maketitle

{\small
\tableofcontents
}

\paragraph*{\bf Significance Statement:}
{\it \small 
This work advances tipping-point theory by replacing the usual gap-centered view of early-warning signals with a fully Ruelle--Pollicott spectral framework. Its central originality is to show that an early warning is not determined by a closing spectral gap alone: it depends jointly on the critical spectral block, the measured observable, and the perturbation direction through spectral residues. We demonstrate that classical warning indicators can become strongly amplified even when the entire RP spectrum is frozen, and introduce a dimensionless diagnostic that certifies such residue-driven amplification. For metastable systems, we further show that loss of local recovery and increasing transition risk are governed by distinct spectral objects derived from the same killed process, thereby placing bifurcation-induced and noise-induced tipping on a common mathematical footing. By connecting these constructions to Doob transforms, committors, and optimal Girsanov sampling, the framework also provides a practical route from operator theory to data analysis and rare-event computation.}

\section{Introduction}
\label{sec:intro}

\subsection{Tipping points and the critical-slowing-down paradigm}
\label{subsec:intro_tipping}
Many natural and engineered systems can undergo abrupt, and often irreversible, reorganizations of their statistical state under a (possibly) slow change of external conditions. In the climate sciences such \emph{tipping points} have been identified in the Atlantic meridional overturning circulation, in polar ice sheets, in monsoon systems and in vegetation cover \cite{lenton2008tipping,lenton2012early}, and analogous transitions occur in ecology, in neuroscience and in finance \cite{scheffer2009early,scheffer2012anticipating}.
 The literature on the subject is extensive, on the theoretical as well as on the empirical side \cite{kuehn2011mathematical,Chek_al14_RP,Feudel2018,Ghil2020,LucariniChekroun2023,Boers2025,Lenton2025,Hastings2026}.

 Abrupt transitions are commonly sorted by what causes them \cite{ashwin2012tipping}. In \emph{bifurcation-induced} (B-)tipping the state the system occupies ceases to exist, or ceases to be stable, because a control parameter has been pushed past a bifurcation; the transition would occur without noise, which only advances it. In \emph{noise-induced} (N-)tipping the state remains stable, but a sufficiently large random fluctuation carries the system out of its basin anyway; the transition is then a rare event whose likelihood depends on how stable the state is, not on whether it is about to disappear. Further classes are tied to nonautonomous forcing \cite{Ashwin2026}: in rate-induced (R-)tipping the transition is caused by how \emph{fast} the forcing changes rather than by where it ends up \cite{alkhayuon2018rate}; in phase-induced (P-)tipping it occurs preferentially in a particular phase of a periodically modulated system \cite{Alkhayuon2021}; in shock-induced (S-)tipping a single finite perturbation sends the system from one basin of attraction to a competing one \cite{Halekotte2020}. This paper is concerned with B- and N-tipping, and with a common spectral language in which the two can be described and told apart.

The prospect of anticipating tipping events from observational or model time series has motivated an extensive literature on \emph{early-warning signals} (EWS). The dominant conceptual framework for EWS is \emph{critical slowing down} (CSD). If the observed state fluctuates around a stable equilibrium whose linearization has a leading eigenvalue approaching zero, then recovery from perturbations becomes slower, lag-one autocorrelation increases, variance may grow, and spectral power shifts toward low frequencies \cite{scheffer2009early,dakos2008slowing,scheffer2012anticipating}. In the climate context this idea underlies degenerate fingerprinting \cite{Held2004} and its many descendants, and it has been used to argue that the Greenland ice sheet \cite{boers2021critical} and the AMOC \cite{boers2021observation,ditlevsen2023warning,vanWesten2024} are approaching thresholds. Its mathematical backbone is the normal-form theory of local bifurcations perturbed by small noise \cite{kuehn2011mathematical,berglund2006noise}.

This picture is powerful precisely because it is simple. It is also, for the same reason, fragile. It presupposes (i) that the reference state is a fixed point, so that a single linear rate exists; (ii) that whatever slows down is visible in the particular observable that happens to be measured; and (iii) that the loss of stability, rather than a noise-induced escape over a still-finite barrier, is what actually terminates the metastable state. Each of these assumptions can fail, and each failure has been documented. Regime shifts can occur with no detectable warning \cite{hastings2010regime}; non-catastrophic transitions can produce warnings that are not followed by tipping \cite{kefi2013early}; and nonstationary correlated noise generates spurious CSD \cite{boettner2022critical}.  When tipping is associated with high-dimensional boundary crises, the possibility of long chaotic transients of arbitrary duration  can make it extremely hard to define any useful EWS \cite{Borner2026}. Most strikingly, stochastic \emph{non-normal} systems generate the entire canonical EWS battery --- rising variance, rising autocorrelation, dimensional collapse --- through transient amplification alone, while remaining uniformly far from any bifurcation \cite{troude2026pseudo,farrell1996generalized,trefethen2005spectra}. We take this last failure as the sharpest available test of any proposed early-warning theory, and return to it in Sec.~\ref{sec:nonnormal}. Conversely, boundary crises of chaotic attractors can be approached without the decay of correlations flagging anything at all \cite{Tantet2018}. In high-dimensional systems the question of \emph{which} observable carries the signal becomes acute, and recent work shows that observables aligned with the relevant edge state are dramatically more informative than generic ones \cite{Lohmann2025}.

Three questions therefore remain open, and they are the questions this paper is written to answer. What exactly is the spectral object that slows down, when the reference state is not a fixed point? Through which observables is that object visible, and through which is it invisible? And how is the loss of local recovery to be separated, operationally, from the growth of escape risk?  In other words: is there a single spectral language in which B- and N-tipping can both be described, and told apart?

\subsection{Ruelle--Pollicott resonances as the right spectral object}

Ruelle-Pollicott (RP) resonances, which were initially introduced for deterministic chaotic systems \cite{pollicott1986meromorphic,ruelle1986locating,keller1999stability,baladi2000positive,dyatlov2019mathematical,gaspard2005chaos},  have since been shown to extend seamlessly to stochastic systems, where they provide
a powerful tool for analyzing temporal correlations, spectral densities, and bifurcations of complex systems subject to inherent stochasticity \cite{Chekroun_al_RP2,Tantet_al_Hopf,Tantet_al_ENSO,chekroun2025kolmogorov}, such as those encountered in climate modeling \cite{Chek_al14_RP,Tantet_al_ENSO,Lucarini_Chekroun_PRL24}, turbulent flows \cite{Kondrashov_al2018_QG}, and biological systems \cite{LucariniChekroun2023}. These resonances reveal the fundamental time scales and oscillatory modes of the system, acting as characteristic frequencies and decay rates that govern the relaxation of any correlations, even in the presence of significant noise. Theoretically compelling, RP resonances bridge the gap between observable behavior and the underlying governing equations: they can be extracted from time series data through methods like Markov chain modeling \cite{Pande2010,Lucarini2025PTRSA} and are mathematically defined as the eigenvalues of the generator of the stochastic process, the Kolmogorov operator \cite{crommelin2006b,Chek_al14_RP,Chekroun_al_RP2}.

 Crucially, RP resonances characterize the system's relaxation towards its statistical equilibrium and their corresponding eigenfunctions, often referred to as Kolmogorov modes, provide a natural basis for understanding and quantifying the response of complex systems to perturbations \cite{Chek_al14_RP,Santos2022,LucariniChekroun2023,Lucarini_Chekroun_PRL24,Zagli2024,chekroun2025kolmogorov,Lucarini2025PTRSA,Lucarini2026CSF}.

Two features of the RP spectrum make it the natural carrier of an early-warning theory. First, it is defined for the \emph{invariant dynamics} rather than for a distinguished trajectory: the reference state may be a noisy equilibrium, a limit cycle, a chaotic invariant set, or a metastable region, and in every case the resonances are the eigenvalues of one and the same object, the Kolmogorov generator \cite{Chekroun_al_RP2,chekroun2025kolmogorov}. The classical CSD eigenvalue is recovered as the special case of a noisy fixed point. Second, the RP spectrum comes equipped with spectral \emph{projectors} and, when resonances collide, with \emph{nilpotent} Jordan parts \cite{kato1995perturbation}. These are exactly the objects needed to say through which observable a slow mode is visible, and they are systematically discarded when one summarizes the spectrum by its gap alone. Transfer-operator and Markov-state estimation of RP spectra from data is by now standard \cite{dellnitz1999approximation,froyland2009almost,schutte2001transfer,schutte2013metastability,Chekroun_al_RP2}, and RP-based early-warning indicators have already been explored empirically \cite{tantet2015early,tantet2018crisis,Tantet2018}.

\subsection{What this paper does}
\begin{figure*}[t]
\centering
\includegraphics[width=\textwidth]{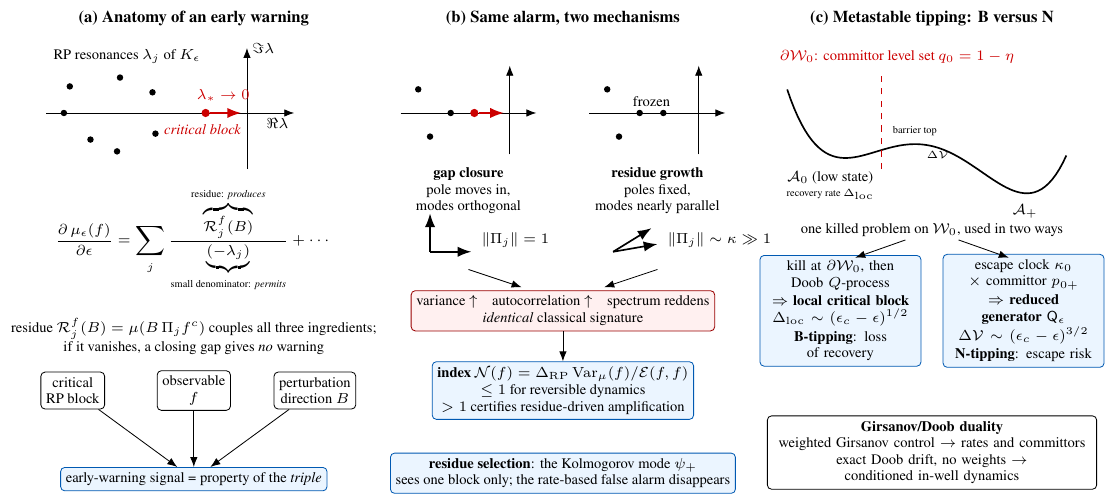}
\caption{\label{fig:schematic} Map of the paper for the physics-oriented reader. (a) \emph{Anatomy of an early warning} (Secs.~\ref{sec:L2mu_setting}--\ref{sec:RP_EWS_tipping}). The sensitivity of any invariant statistic $\mu_\epsilon(f)$ is a sum over the Ruelle--Pollicott (RP) resonances $\lambda_j$ of the generator, each entering through a small denominator, the distance of $\lambda_j$ to the origin, times a residue $\mathcal R_j^f(B)$ that pairs the block with the measured observable $f$ and with the direction $B$ in which the control parameter perturbs the dynamics. A closing gap \emph{permits} a large response; only a nonzero residue \emph{produces} it. An early-warning signal is therefore a property of the triple (block, observable, perturbation), not of the dynamics alone. (b) \emph{Same alarm, two mechanisms} (Sec.~\ref{sec:nonnormal}). Rising variance, rising autocorrelation and a reddening spectrum are produced identically by a pole moving toward the imaginary axis (genuine critical slowing down) and by non-normal growth of the residues at frozen poles, when the RP eigenmodes become nearly parallel and the spectral projectors ill-conditioned. Two RP-based tools separate the mechanisms: the index $\mathcal N(f)$, which compares the observed relaxation rate with the one implied by the variance budget and cannot exceed one for reversible dynamics, and residue selection, which measures the Kolmogorov mode of the critical block and is blind to the spurious alarm. (c) \emph{Metastable tipping} (Secs.~\ref{Sec_EWS_Metastable}--\ref{sec:1D_fold_example}). Near a metastable state $\mathcal A_0$, let $q_0(x)$ be the committor, the probability that a trajectory started at $x$ returns to a small core around $\mathcal A_0$ before reaching a competing state; the metastable neighborhood $\mathcal W_0=\{q_0>1-\eta\}$, with $0<\eta\ll1$, is the region of high commitment to $\mathcal A_0$, and its boundary $\partial\mathcal W_0$ is a smooth, noise-adapted interface that lies uphill of $\mathcal A_0$ but below the barrier top. One kills the process on $\partial\mathcal W_0$. Passing to the Doob $Q$-process removes escape and isolates the local critical block, whose gap $\Delta_{\rm loc}$ measures in-well recovery: its closure is B-tipping. Retaining the escape clock and weighting it by the committor yields instead the reduced generator $\mathsf Q_\epsilon$ on metastable labels, whose transition rates are controlled by the quasipotential barrier $\Delta\mathcal V$: its decrease raises the N-tipping risk. In a one-dimensional fold the two carry different exponents, $(\epsilon_c-\epsilon)^{1/2}$ and $(\epsilon_c-\epsilon)^{3/2}$. The same Girsanov/Doob control serves both purposes: with its weights it estimates rates and committors, without them it simulates the conditioned in-well dynamics.}
\end{figure*}

We develop, in an $L^2_\mu$ spectral setting for hypoelliptic It\^o diffusions, a systematic theory of the sensitivity of invariant statistics in terms of RP blocks, and we use it to
reformulate the early-warning problem ; none of the results depends on the reference state being a fixed point. Four results structure the paper.  Figure~\ref{fig:schematic} collects them, together with their physical reading, in one page; the reader more interested in the consequences for tipping than in the spectral machinery may use it as a map of what follows.

\emph{(i) A residue-resolved sensitivity formula.} In Sec.~\ref{sec:L2mu_setting} we express the Poisson corrector $u_f$ of an observable $f$ through the RP blocks of the generator, including spectral projectors and nilpotent components, and obtain the first-order sensitivity of $\mu_\epsilon(f)$ under a perturbation $B$ of the generator as a sum of block contributions weighted by $(-\lambda_j)^{-(k+1)}$ and by \emph{RP response residues} $\Rstyle_{j,k}^{f}(B)=\mu(BN_j^k\Pi_jf^c)$. Small denominators \emph{permit} large sensitivity; nonzero residues \emph{produce} it. Because nonsemisimple blocks are retained, the attainable singular order is $|\lambda_j|^{-m_j}$ rather than $|\lambda_j|^{-1}$, and it is reached only if the top nilpotent residue survives --- the stochastic analogue of the amplification associated with exceptional points of non-Hermitian operators \cite{kato1995perturbation,heiss2012physics,trefethen2005spectra}. This is also the spectral mechanism through which the non-normal ``pseudo-bifurcations'' of \cite{troude2026pseudo} become legible: transient amplification without gap closure is amplification carried by projectors and nilpotents, not by denominators.

\emph{(ii) A triple criterion for early warnings.} In Sec.~\ref{sec:RP_EWS_tipping} we argue that an EWS is not a property of the dynamics alone, but of a triple: a critical RP block, a measured observable, and a perturbation direction. Variance, autocorrelation time, integrated autocorrelation, parameter susceptibility, and spectral power are shown to be \emph{different contractions} of the same RP expansion, which may select different blocks and weight them by different residues. In particular, spectral reddening is not the general frequency-domain mechanism of warning; it is the zero-frequency, real-resonance, residue-visible corner of a broader RP-resolved theory, the general case involving pairs of blocks through two resolvents.

\emph{(iii) A counterexample that decides the matter.} Section~\ref{sec:nonnormal} exhibits a two-dimensional linear diffusion, taken from \cite{troude2026pseudo}, whose RP resonances are \emph{exactly} independent of a non-normality parameter $\kappa$ while its spectral projectors become ill-conditioned like $\kappa$. Every classical early-warning indicator then grows like $\kappa^2$ with the spectrum frozen to machine precision, and along a slow ramp the variance rises by three orders of magnitude and the lag-one autocorrelation reaches $0.9997$ --- a more alarming record than a genuine approach to a fold produces. The RP decomposition attributes all of it to the numerators, and supplies two repairs: a residue-selected observable, the Kolmogorov mode itself, whose normalized autocorrelation is exactly $\kappa$-independent; and a dimensionless index
\be
\mathcal N(f)=\frac{\Delta_{\rm RP}\,{\rm Var}_\mu(f)}{\mathcal E(f,f)} .
\label{eq:intro_index}
\ee
Here $\Delta_{\rm RP}$ is the leading RP gap, i.e.~the rate at which the slowest mode actually decays; ${\rm Var}_\mu(f)$ is the stationary variance of the measured observable, i.e.~how much fluctuation it carries; and $\mathcal E(f,f)$ is the Dirichlet form \cite{bakry2014analysis}, i.e.~the rate at which the noise supplies that variance, equal to minus the initial slope of the autocovariance of $f$. The ratio ${\rm Var}_\mu(f)/\mathcal E(f,f)$ is thus the relaxation time one would infer from a variance budget alone --- how long the observable would need to dissipate what the noise puts in --- and multiplying by $\Delta_{\rm RP}$ compares it with the relaxation time the spectrum actually exhibits. If a single mode carries the observable the two agree and $\mathcal N=1$. We prove that $\mathcal N(f)\le1$ for every observable whenever the generator is self-adjoint in $L^2_\mu$, so that a measured value above one means the system relaxes more slowly than its own variance budget permits, which can only happen through cancellation between signed residues. The index is therefore a certificate that amplification is residue-driven rather than gap-driven. This is, to our knowledge, the first quantitative discriminator between critical slowing down and non-normal transient amplification that is formulated at the level of the generator rather than of a fitted indicator.

\emph{(iv) Local recovery versus escape risk in metastable systems.}
Section~\ref{Sec_EWS_Metastable}  contains the main application: a unified spectral treatment of B- and N-tipping. In a metastable regime, the dominant nonzero RP block of the full generator is typically an \emph{interwell transition block}, which measures residence times, not recovery inside a well. To isolate local recovery we kill the process at a committor-defined, noise-regularized boundary and pass to the associated Doob $Q$-process \cite{Doob1957conditional,ColletMartinezSanMartin2013,ChampagnatVillemonais2016,meleard2012quasi}. This removes the survival decay carried by the principal Dirichlet eigenvalue and yields a conservative in-well dynamics whose leading nonzero RP block we call the \emph{local critical block}. The complementary reduction keeps the killed escape clocks and weights them by committor-based destination probabilities, producing a continuous-time Markov generator $\mathsf Q_\epsilon$ on metastable labels whose entries carry the Freidlin--Wentzell/Eyring--Kramers asymptotics \cite{Freidlin1984,kramers1940,hanggi1990reaction,bovier2004metastability,bovier2005metastability,BouchetReygner2016,BouchetReygner2022,LePeutrec2024}. The two reductions use the same killed problem in complementary ways, and they answer different questions.

 \emph{(v) B-tipping and N-tipping as two spectral objects.} The distinction just described maps precisely onto the tipping taxonomy recalled in Sec.~\ref{subsec:intro_tipping} \cite{ashwin2012tipping,alkhayuon2018rate,Alkhayuon2021}. The two mechanisms are usually described in dynamical terms; the point of what follows is that each has a clean spectral counterpart. Bifurcation-induced (B-)tipping is the closure of the conditioned local RP gap; noise-induced (N-)tipping is a property of the transition blocks, i.e.~of escape clocks, committors and quasipotential barriers. Both mechanisms produce long memory and enhanced low-frequency power, and a global lag-one autocorrelation estimate mixes them. The RP framework makes the separation operational, and Sec.~\ref{sec:1D_fold_example} makes it quantitative on a scalar metastable fold, where the local conditioned gap closes like $(\epsilon_c-\epsilon)^{1/2}$ while the quasipotential barrier collapses like $(\epsilon_c-\epsilon)^{3/2}$. The same example shows that variance, correlation time and zero-frequency power carry three different exponents of the same closing block, and that the committor weighting is not decorative: it is exactly the classical factor two between the mean first-passage time to the barrier top and the Kramers transition rate \cite{kramers1940,matkowski1981}.

A final, more practical thread runs through Sec.~\ref{sec:Girsanov_Doob}. The Doob drift of the $Q$-process is identified with the long-horizon limit of the logarithmic-gradient control that solves the optimal Girsanov importance-sampling problem for Koopman expectations \cite{sikorski2024learning,hartmann2012efficient,hartmann2017variational}. This yields a dual computational prescription: keep the Girsanov weights and one estimates escape probabilities, exit distributions and transition rates; discard them and use the exact Doob drift and one samples the conditioned in-well dynamics, hence the local critical block. Confusing these two uses is, we argue, one concrete route by which transition risk gets misreported as local critical slowing down. Rare-event samplers such as adaptive multilevel splitting \cite{cerou2007adaptive,bouchet2019rare,ragone2018computation} occupy the first branch; Fleming--Viot and Doob simulations occupy the second.  Both of the latter are ways of simulating a process \emph{conditioned not to have escaped}, which is what the local theory needs. A \emph{Fleming--Viot} scheme does this with an ensemble and no extra modelling: one runs many copies of the original dynamics and, whenever a copy leaves the region, immediately restarts it at the current position of another randomly chosen surviving copy, so that the population size is held fixed and its empirical distribution converges to the quasi-stationary distribution \cite{burdzy2000fleming,villemonais2014general,delmoral2004feynman}. A \emph{Doob} simulation does the same job with a single trajectory, by adding to the drift the extra term that repels the process from the boundary; it requires knowing the principal Dirichlet eigenfunction, but returns unweighted samples of the conditioned dynamics.

\subsection{Relation to the companion study and outline}
This paper is the theoretical member of a pair. Its companion \cite{Chekroun_Lucarini26_EBM} analyzes the approach to a warm-to-cold transition in a stochastic Ghil--Sellers energy balance model through reduced RP resonances and Green's functions, extreme-value statistics, and full-field data-adaptive harmonic modes. That study is deliberately empirical, and it raises --- without resolving --- exactly the questions addressed here: near tipping, \emph{several} decay rates compress toward the origin rather than one, response amplification is conditional on nonvanishing residues, extremes probing the escape direction towards the  competing state organize differently from bulk relaxation, and full-field phase coherence degrades while reduced modes harmonize. The present framework supplies the spectral vocabulary in which those observations are statements about distinct RP blocks, distinct contractions, and distinct residues, rather than about a single scalar indicator.

The paper is organized as follows. Section~\ref{sec:L2mu_setting} sets up the $L^2_\mu$ RP framework, derives the residue-resolved sensitivity formula and its frequency-resolved two-resolvent counterpart. Section~\ref{sec:RP_EWS_tipping} formulates the triple criterion and contrasts it with classical CSD. Section~\ref{sec:nonnormal} tests the whole construction on a non-normal counterexample with a frozen spectrum, and introduces the non-normality index. Section~\ref{Sec_EWS_Metastable} develops the metastable theory: committor-defined neighborhoods, killed dynamics and the $Q$-process, local critical blocks, transition blocks and the reduced generator $\mathsf Q_\epsilon$, and the metastable fold scenario. Section~\ref{sec:Girsanov_Doob} develops the dual Girsanov--Doob view and the unified account of B- and N-tipping. Section~\ref{sec:1D_fold_example} works out a scalar metastable fold in closed form. Section~\ref{sec:discussion} discusses scope, limitations and open problems.

\section{Ruelle-Pollicott Sensitivity Formulas}
\label{sec:L2mu_setting}

\subsection{Unperturbed Stochastic Dynamics in the $L^2_\mu$-Spectral Setting}
\label{subsec:unperturbed_L2mu_setting}
We consider an unperturbed It\^o diffusion of the form
\be
\d X_t
= F(X_t)\d t+\sigma D(X_t)\d W_t, \qquad
X_t\in\mathbb R^d,
\label{Eq_unperturbed_SDE}
\ee
where $W_t$ is an $\mathbb R^q$-valued Brownian motion whose components are mutually independent and $D$ is a matrix valued function  from $\mathbb{R}^d$ to $\textrm{Mat}_{\mathbb{R}}(d\times q)$. The drift part is provided by a vector field $F$ on $\mathbb{R}^d$.
We denote by $a(x)=D(x)D(x)^T$
the unperturbed diffusion tensor.  Here $\sigma>0$ is a scalar noise amplitude that has been factored out of $D$, so that $D$ carries only the (normalized) noise geometry and the full diffusion tensor of Eq.~\eqref{Eq_unperturbed_SDE} is $\sigma^2a$. This convention is kept throughout: $\sigma$ is the small parameter of the metastable/Freidlin--Wentzell asymptotics of Secs.~\ref{Sec_EWS_Metastable}--\ref{sec:1D_fold_example}, whereas $\epsilon$ always denotes the control parameter driving the system toward tipping. The Markov semigroup acting on observables is
$P_t\varphi(x)
= \mathbb E\big[\varphi(X_t^x)\big],$
where $X_t^x$ denotes the stochastic process solving Eq.~\eqref{Eq_unperturbed_SDE}. The  Kolmogorov generator is
\be
K\varphi
= F\cdot \nabla\varphi
+
\frac{\sigma^2}{2}
a:\nabla^2\varphi,
\label{eq:unperturbed_generator}
\ee
where
\be
a:\nabla^2\varphi
= \sum_{i,j=1}^d a_{ij}\partial_{ij}^2\varphi .
\label{eq:colon_hessian}
\ee
We assume that the unperturbed diffusion is non-explosive and satisfies a Lyapunov dissipativity condition ensuring the existence of an invariant probability measure \cite[e.g.][]{bellet2006ergodic,douc2009subgeometric,Chekroun_al_RP2}. We assume furthermore that the diffusion is hypoelliptic, in the sense of H\"ormander \cite{hormander1967hypoelliptic,norris1986simplified}  and that the Markov process is irreducible \cite[e.g.][]{eckmann2003spectral,Hairer_Majda,Chekroun_al_RP2}.

Under these assumptions, the invariant probability measure is unique and ergodic; we denote it by $\mu$. Moreover, $\mu$ is absolutely continuous with respect to Lebesgue measure,
\be
\d\mu(x)=\rho \d x,
\label{eq:mu_density}
\ee
with a smooth density $\rho$.

 The adjoint picture is obtained by following probability densities rather than observables. If $X_0$ has density $\nu_0$, the density $\nu_t$ of $X_t$ solves the Fokker--Planck equation \cite{pavliotis2014stochastic}
\begin{equation}
\partial_t \nu= -\nabla\cdot (F \nu)+ \frac{\sigma^2}{2}
\nabla^2: (a \nu)\equiv \mathcal{L}\nu,
\label{eq:Fokker_Planck}
\end{equation}
where $\nabla^2:(a\nu)=\sum_{i,j}\partial^2_{ij}(a_{ij}\nu)$, and the invariant density is characterized by $\mathcal{L}\rho=0$. The Fokker--Planck operator $\mathcal{L}$ is the formal adjoint of $K$ in the unweighted space $L^2(\mathbb R^d,\d x)$.

All spectral objects below are defined in the Hilbert space
\be
\mathcal H=L^2_\mu(\mathbb R^d),
\qquad
\langle f,g\rangle_\mu
= \int_{\mathbb R^d} f\overline g \d\mu .
\label{eq:L2mu}
\ee
For real-valued observables, the conjugation is of course immaterial; it is included only because the RP eigenfunctions and eigenvalues may be complex.

Thus $P_t$ is viewed as a strongly continuous Markov semigroup on $L^2_\mu$, and $K$ denotes its generator in $L^2_\mu$ \cite{Engel_Nagel,Chekroun_al_RP2}. The constants form the invariant eigenspace associated with the eigenvalue $0$ (which corresponds to the invariant measure in the adjoint picture), and we write
\be
\Pi_0 f
= \mu(f)\mathds{1},
\qquad
f^c
= (I-\Pi_0)f
= f-\mu(f)\mathds{1}.
\label{eq:centering_L2mu}
\ee
The centered space is $L^2_{\mu,0}
= \left\{
f\in L^2_\mu:
\mu(f)=0
\right\}.$

We assume that $0$ is a simple eigenvalue of $K$ and that the nonzero spectral values of $K$ in the relevant strip
\be
\textrm{Re}\, z>\omega_{\rm ess}(P)
\label{eq:relevant_strip_L2}
\ee
 are isolated, where $\omega_{\rm ess}(P)$ denotes the essential growth bound of the semigroup $(P_t)_{t\ge0}$ \cite{Engel_Nagel}. These spectral values are the
 RP resonances
\be
\lambda_1,\ldots,\lambda_N,
\qquad
\textrm{Re}\lambda_j<0,
\label{eq:RP_resonances_L2mu}
\ee
 where $N$ may be finite or infinite. The same resonances govern the adjoint picture. The adjoint of $K$ in $L^2_\mu$ is $K^*g=\rho^{-1}\mathcal{L}(\rho g)$, which is similar to the Fokker--Planck operator $\mathcal L$ of Eq.~\eqref{eq:Fokker_Planck} through the multiplication map $g\mapsto\rho g$; hence $\mathcal L$, acting on densities, has eigenvalue $0$ with eigenfunction $\rho$ and, for its remaining isolated spectrum, the complex conjugates $\overline{\lambda_j}$ of the RP resonances of $K$.

For each $\lambda_j$, let
\be
\Pi_j
= \frac{1}{2\pi i}
\int_{\Gamma_j}
(zI-K)^{-1}\d z
\label{eq:Pi_j_L2}
\ee
be the spectral projection in $L^2_\mu$, where $\Gamma_j$ is a small positively oriented contour enclosing $\lambda_j$ and no other point of the spectrum.
We introduce the nilpotent operator $N_j$ defined as
\be
N_j=(K-\lambda_j I)\Pi_j .
\label{eq:nilpotent_L2}
\ee
If $m_j$ is the pole order of the resolvent at $\lambda_j$, equivalently the size of the largest Jordan block associated with $\lambda_j$, then
\be
N_j^{m_j}=0,
\qquad
N_j^{m_j-1}\neq0
\quad
\text{when }m_j>1.
\label{eq:nilpotent_order_L2}
\ee
Thus, on the generalized eigenspace $\Pi_jL^2_\mu$,
\be
K\Pi_j=(\lambda_j I+N_j)\Pi_j.
\label{eq:block_generator_L2}
\ee
With this notation, the RP contribution to the semigroup is not, in general, merely
$\sum_j e^{\lambda_jt}\Pi_j$. Rather, the finite RP-block expansion reads \cite[Theorem 1]{Chekroun_al_RP2}:
\be
P_t
= \Pi_0
+
\sum_{j=1}^N
e^{\lambda_j t}
\sum_{k=0}^{m_j-1}
\frac{t^k}{k!}N_j^k\Pi_j
+
R_N(t),
\label{eq:RP_semigroup_L2}
\ee
where $R_N(t)$ denotes the residual spectral contribution. We assume that the residual part is sufficiently stable so that, for some $M_N>0$ and $\epsilon_N>0$,
\be
\left|R_N(t)\right|_{L^2_\mu\to L^2_\mu}
\le
M_Ne^{-\epsilon_Nt},
\qquad
t\ge0.
\label{eq:RN_bound_L2}
\ee
This condition is satisfied when the retained RP window is separated from the remaining spectrum by a spectral gap in $L^2_\mu$  \cite[Theorem 1]{Chekroun_al_RP2}.

For $f\in L^2_\mu$, define the Poisson corrector $u_f$ by
\be
K u_f=f^c,
\qquad
\mu(u_f)=0.
\label{eq:Poisson_L2}
\ee
This object plays a central role in what follows. It is not an additional dynamical variable, but a diagnostic observable associated with $f$. It records how the centered anomaly $f^c$ relaxes under the unperturbed dynamics. Indeed, whenever the integral converges in $L^2_\mu$,
\be
u_f
= -\int_0^\infty P_t f^c\d t.
\label{eq:Poisson_integral_L2}
\ee
Thus $u_f(x)$ is, up to sign, the total expected future anomaly of $f$ accumulated along trajectories initialized at $x$ before they return to stationarity.

The centering condition is essential. Since
\be
\Pi_0 f^c=0,
\label{Eq_centering}
\ee
the non-decaying invariant component has been removed from the time integral. Without this subtraction, the integral in Eq.~\eqref{eq:Poisson_integral_L2} would  diverge, since it would accumulate the stationary mean of $f$  indefinitely, and would not describe relaxation back to equilibrium.

The reason the Poisson corrector is useful is that it turns questions about invariant statistics into questions about dynamical relaxation. The invariant mean $\mu(f)$ is a static number, but its sensitivity to changes in the dynamics depends on how perturbations are propagated and accumulated before the system equilibrates. The corrector $u_f$ is precisely this accumulated relaxation observable. Its RP expansion will therefore reveal which resonant decay modes contribute to the sensitivity of $\mu(f)$ once a perturbation of the generator is introduced.

Inserting the RP-block expansion Eq.~\eqref{eq:RP_semigroup_L2} into Eq.~\eqref{eq:Poisson_integral_L2} gives the block-resolvent representation of $u_f$. Indeed, for each RP block and each $0\le k\le m_j-1$,
\be
-\int_0^\infty
e^{\lambda_jt}
\frac{t^k}{k!}\d t
= \frac{(-1)^k}{\lambda_j^{k+1}},
\qquad
\textrm{Re}(\lambda_j)<0.
\label{eq:time_integral_RP_block}
\ee
Therefore
\be
u_f
= \sum_{j=1}^N
\sum_{k=0}^{m_j-1}
\frac{(-1)^k}{\lambda_j^{k+1}}
N_j^k\Pi_j f^c
+
\mathcal R_N^{(0)}f^c,
\label{eq:Poisson_RP_L2}
\ee
where the residual reduced resolvent is
\be
\mathcal R_N^{(0)}f^c
= K^{-1}
\left(
I-\Pi_0-\sum_{j=1}^N\Pi_j
\right)f^c.
\label{eq:poisson_remainder_L2}
\ee
Equivalently,
\be
\mathcal R_N^{(0)}f^c
= -\int_0^\infty R_N(t)f^c\d t,
\label{eq:poisson_remainder_integral_L2}
\ee
and the bound \eqref{eq:RN_bound_L2} gives
\be
\left|\mathcal R_N^{(0)}f^c\right|_{L^2_\mu}
\le
\frac{M_N}{\epsilon_N}
\left|f^c\right|_{L^2_\mu}.
\label{eq:poisson_remainder_bound_L2}
\ee
The same formula is obtained directly by inverting $K$ on each generalized eigenspace. Due to Eq.~\eqref{eq:block_generator_L2}
and  since $N_j^{m_j}=0$, one has
\be
K^{-1}\Pi_j
= (\lambda_j I+N_j)^{-1}\Pi_j
= \sum_{k=0}^{m_j-1}
\frac{(-1)^k}{\lambda_j^{k+1}}
N_j^k\Pi_j .
\label{eq:block_inverse_L2}
\ee
This is the basic $L^2_\mu$ RP-resolvent representation of the Poisson corrector.

It is also the first place where algebraic multiplicity matters. If $m_j=1$, the $j$th RP block contributes only
$\frac{1}{\lambda_j}\Pi_j f^c.$
If $m_j>1$, the same resonance contributes the higher nilpotent levels
\beas
&\frac{1}{\lambda_j}\Pi_j f^c
- \frac{1}{\lambda_j^2}N_j\Pi_j f^c
+ \frac{1}{\lambda_j^3}N_j^2\Pi_j f^c
- \cdot \cdot  \cdot \\
&\qquad\qquad
+ \frac{(-1)^{m_j-1}}{\lambda_j^{m_j}}
N_j^{m_j-1}\Pi_j f^c .
\eeas
Thus, near a critical RP resonance $\lambda_j\simeq0$, the relevant sensitivity scale is not necessarily $|\lambda_j|^{-1}$; in a nonsemisimple block it may be as singular as $|\lambda_j|^{-m_j}$, provided the corresponding nilpotent residue is seen by the observable and by the perturbation.  We stress that it is the modulus $|\lambda_j|$, and not the decay rate $-\Re\lambda_j$, that sets this scale. A complex resonance approaching the imaginary axis at nonzero frequency slows the decay of correlations, which is governed by $\Re\lambda_j$, without making the static response singular; the two situations are distinguished in Sec.~\ref{sec:RP_EWS_tipping}.

\subsection{RP Sensitivity and Residue Selection}
\label{subsec:summary_RP_sensitivity_theorem}
 The differentiation of invariant statistics with respect to a perturbation of the generator is carried out in Appendix~\ref{sec:differentiating_statistics}, where the response formulas of \cite{Santos2022,Lucarini2017b,Chekroun_al_RP2} are recast in the RP-block notation of this paper; Appendix~\ref{subsec:direct_covariance_perturbation} treats direct perturbations of the noise covariance. Resolved on the RP blocks, that computation yields the following sensitivity theorem, formulated here  in the $L^2_\mu$ setting of Section~\ref{subsec:unperturbed_L2mu_setting}, under the hypoellipticity and dissipativity assumptions ensuring that the invariant measure $\mu$ is unique and smooth, given by Eq.~\eqref{eq:mu_density}.
It uses the RP block notation introduced in Eqns.~\eqref{eq:Pi_j_L2}--\eqref{eq:nilpotent_order_L2}, the  representation of the Poisson corrector (Eq.~\eqref{eq:Poisson_RP_L2}), and the perturbation functional $\mathscr{L}_B$  defined in Eq.~\eqref{Eq_summary_ellB} below.

\bt[RP sensitivity formula and residue selection]
\label{thm:RP_sensitivity_residue_selection}
Let $K_\eps$ be a differentiable perturbation of the Kolmogorov generator $K$ on a common core, with
\be
K_\eps
= K+\eps B+o(\eps),
\label{eq:summary_Keps_expansion}
\ee
and let $\mu_\eps$ be the corresponding invariant probability measure, with $\mu_0=\mu$. Assume that the stationary identity can be differentiated at $\eps=0$ against the Poisson corrector $u_f$ solving
\be
Ku_f=f^c,
\qquad
\mu(u_f)=0,
\label{eq:summary_poisson}
\ee
and assume that $u_f\in D(B)$.

Then
\be
\left.\pd{}{\eps}\mu_\eps(f)\right|_{\eps=0}
= -\mu(Bu_f).
\label{eq:summary_poisson_response}
\ee
Moreover, using the RP-block expansion of $u_f$, one obtains
\bea
\left.\pd{}{\eps}\mu_\eps(f)\right|_{\eps=0}
&=
\sum_{j=1}^N
\sum_{k=0}^{m_j-1}
\frac{1}{(-\lambda_j)^{k+1}}
\Rstyle_{j,k}^{f}(B)\\
&\qquad \quad- \mathscr{L}_B\left(
\mathcal R_N^{(0)}f^c
\right),
\label{eq:summary_RP_sensitivity}
\eea
where $\mathscr{L}_B$ denotes the perturbation functional
\be
\mathscr{L}_B(g)
= \mu(Bg)
= \int_{\mathbb R^d}(Bg)(x)\rho(x)\d x,
\label{Eq_summary_ellB}
\ee
and for  $0\le k\le m_j-1$, $\mathscr{R}_{j,k}^{f}(B)$ denotes the {\bf $k$th RP response residue} of the observable $f$ under the perturbation $B$, associated with the RP block $\lambda_j$:
\be
\mathscr{R}_{j,k}^{f}(B)
= \mathscr{L}_B\left(N_j^k\Pi_j f^c\right)
= \mu\left(
BN_j^k\Pi_j f^c
\right),
\label{eq:summary_residue}
\ee
Thus the contribution of the $j$th RP block is determined not only by the denominator $(-\lambda_j)^{-(k+1)}$, but by the RP residues $\Rstyle_{j,k}^{f}(B)$.

If at least one of the residues associated with the $j$th block is nonzero, we define the {\bf active nilpotent level} as
\be
\kappa_j^f(B)
= \max
\left\{
0\le k\le m_j-1:
\Rstyle_{j,k}^{f}(B)\neq0
\right\}.
\label{eq:summary_dominant_level}
\ee
Then, near a critical RP resonance $\lambda_j\simeq0$, the leading singular contribution of this block is
\be
\frac{\Rstyle_{j,\kappa_j^f(B)}^{f}(B)}{(-\lambda_j)^{\kappa_j^f(B)+1}} .
\label{eq:summary_dominant_singularity}
\ee
In particular, the maximal possible singular order associated with a pole of order $m_j$ is $|\lambda_j|^{-m_j}$, and it occurs only if
\be
\Rstyle_{j,m_j-1}^{f}(B)
= \mu\left(
BN_j^{m_j-1}\Pi_j f^c
\right)
\neq0.
\label{eq:summary_maximal_residue_condition}
\ee
If all residues $\Rstyle_{j,k}^{f}(B)$ vanish, then the $j$th RP block does not contribute to the first-order response of the statistic $\mu(f)$.

Finally, if the $j$th RP block is semisimple, then $m_j=1$, $N_j=0$, and
\be
\Rstyle_{j,0}^{f}(B)
= \mu\left(
B\Pi_j f^c
\right).
\label{eq:summary_semisimple_residue}
\ee
The corresponding contribution reduces to
$\mu\left(B\Pi_j f^c\right)/(-\lambda_j).$
\et

\begin{proof}
Equation \eqref{eq:summary_poisson_response} is obtained by differentiating the stationary identity
\be
\int K_\eps\varphi\d\mu_\eps=0
\label{eq:summary_stationary_identity}
\ee
at $\eps=0$ and taking $\varphi=u_f$. This gives
\be
\dot\mu(Ku_f)+\mu(Bu_f)=0.
\label{eq:summary_linearized_stationarity}
\ee
Since $Ku_f=f^c$ and $\dot\mu(\mathds{1})=0$, we have
\be
\dot\mu(Ku_f)
= \dot\mu(f^c)
= \dot\mu(f),
\label{eq:summary_mudot_fc}
\ee
which proves Eq.~\eqref{eq:summary_poisson_response}. Substituting the RP representation Eq.~\eqref{eq:Poisson_RP_L2} of $u_f$ into Eq.~\eqref{eq:summary_poisson_response} gives Eq.~\eqref{eq:summary_RP_sensitivity}. The residue-selection statements follow directly from the finite nilpotent expansion on each generalized eigenspace.
\end{proof}

\br
 Note that the sign carried by the resolvent expansion of $u_f$ in Eq.~\eqref{eq:Poisson_RP_L2} and the minus sign of Eq.~\eqref{eq:summary_poisson_response} combine into the single positive-power weight
$-(-1)^k\lambda_j^{-(k+1)}=(-\lambda_j)^{-(k+1)}$, which is real and positive for a real negative resonance. Writing the RP weights as $(-\lambda_j)^{-(k+1)}$ rather than $(-1)^k\lambda_j^{-(k+1)}$ is therefore not cosmetic: it makes the sign of each block contribution unambiguous and shows that a slowly decaying block ($-\lambda_j\downarrow0$) enters with a large \emph{positive} weight, the sign of the contribution being carried entirely by the residue.
\er

The collection $\{\mathscr{R}_{j,k}^{f}(B) : 0\le k\leq m_j-1 \}$
will be called the \emph{RP residue ladder} of the $j$th block. The terminology emphasizes that these coefficients are not merely projections of $f$ onto the generalized eigenspace $\Pi_jL^2_\mu$. They also include the contraction against the perturbation functional $\mathscr L_B=\mu\circ B$. Thus $\Pi_j f^c$ determines whether the observable has a component in the $j$th RP block, whereas $\mathscr{R}_{j,k}^{f}(B)$ determines whether that component is actually activated by the perturbation.

 The $j$th RP block is  said to be response-visible for the pair $(f,B)$ if at least one element of its residue ladder is nonzero:
\be
\mathscr{R}_{j,k}^{f}(B)\neq0
\quad
\text{for some}
\quad
0\le k\le m_j-1.
\label{eq:response_visible_block}
\ee
If this condition fails, the block does not contribute to the first-order response of $\mu(f)$, even if $\Pi_j f^c\neq0$.

The theorem separates three mechanisms that are often collapsed in global perturbation bounds:
\begin{itemize}
\item The small denominator $(-\lambda_j)^{-(k+1)}$ measures the dynamical persistence of the RP block;
\item The projection $\Pi_j f^c$ measures whether the observable has a component in that generalized eigenspace;
\item The functional $\mathscr{L}_B=\mu\circ B$ measures whether the perturbation couples that component to the invariant statistic.
\end{itemize}

Thus a small RP gap is not, by itself, an early-warning or sensitivity criterion. It becomes one only through a nonzero residue. In the context of tipping, this is the mathematical origin of residue-conditioned early-warning signals.

 A vanishing residue may result from an \emph{unlucky} combination of observable and dynamics. The mechanism is most transparent in systems with symmetries, where an observable of the wrong parity is orthogonal to the critical eigenspace. In applications, where records are finite and noisy, a residue need not vanish exactly to be harmful: a residue that is merely small already suppresses the practical efficacy of the warning.

\subsection{Frequency-Resolved Sensitivity}
\label{sec:frequency_resolved}
The same RP block structure controls correlation spectra and power spectral densities. For $z$ in the resolvent set of $K$, we define the resolvent as  $R(z;K)
= (zI-K)^{-1}.$
Since $0$ is an eigenvalue associated with constant eigenfunctions, it is convenient to work on centered observables and to use the reduced resolvent
\be
R_c(z;K)
= R(z;K)(I-\Pi_0).
\label{eq:reduced_resolvent_def}
\ee
For $g^c\in L^2_{\mu,0}$, $R_c(z;K)g^c = R(z;K)g^c.$ On the RP block associated with $\lambda_j$, one has
\be
R(z;K)\Pi_j
= \sum_{k=0}^{m_j-1}
\frac{1}{(z-\lambda_j)^{k+1}}
N_j^k\Pi_j .
\label{eq:resolvent_RP_block}
\ee
Thus a resonance whose resolvent pole has order $m_j$ produces an order-$m_j$ pole in the frequency-resolvent.

For centered observables $f^c$ and $g^c$, define the one-sided stationary correlation transform by
\be
\widehat C_{f,g}(z)
= \int_0^\infty
e^{-zt}
\mu\left(
f^c P_tg^c
\right)
\d t,
\qquad
\textrm{Re}(z)>0.
\label{eq:one_sided_corr_transform}
\ee
Analytic continuation, or the boundary value $z=\eta+i\omega$ with $\eta\downarrow0$, gives the frequency-resolved object. From the definition
$R(z;K)=(zI-K)^{-1},$
we get
\be
\widehat C_{f,g}(z)
= \mu\left(
f^c R(z;K)g^c
\right).
\label{eq:corr_resolvent}
\ee
For $g=f$, the power spectral density is obtained from the corresponding symmetrized or real-part convention; these normalization choices play no role in the RP denominators below.

We now perturb the generator as in Eq.~\eqref{eq:summary_Keps_expansion}, i.e.~$
K_\eps = K+\eps B+o(\eps),$ and write $R_\eps(z) = (zI-K_\eps)^{-1}.$

 As in the response theory of correlation functions \cite{Lucarini2017b}, the full derivative of the correlation transform has two conceptually distinct parts: the invariant measure $\mu_\eps$ changes, and the generator, hence the propagator, changes. To display this separation, define
$h_{f,g,z} = f^cR(z;K)g^c.$
The derivative of the invariant measure contribution is
\be
\dot\mu(h_{f,g,z})
= -\mathscr{L}_B(u_{h_{f,g,z}}),
\label{eq:measure_contribution_frequency}
\ee
where $u_{h_{f,g,z}}$ is the Poisson corrector associated with the observable $h_{f,g,z}$, and $\mathscr{L}_B$ is the perturbation functional defined in Eq.~\eqref{Eq_summary_ellB}.

The resolvent identity gives
\be
\left.
\pd{}{\eps}
R_\eps(z)
\right|_{\eps=0}
= R(z;K)BR(z;K).
\label{eq:resolvent_derivative}
\ee

The derivative of the resolvent contribution is
\be
\mathcal G_{f,g}^{B}(z)
= \mu\left(
f^c R(z;K)B R(z;K)g^c
\right).
\label{eq:generator_contribution_frequency}
\ee
Therefore, we have
\be
\left.
\pd{}{\eps}
\widehat C_{f,g}^{\eps}(z)
\right|_{\eps=0}
= -\mathscr{L}_B(u_{h_{f,g,z}})
+
\mathcal G_{f,g}^{B}(z),
\label{eq:freq_sensitivity_decomposition}
\ee
up to the harmless changes in centering induced by $\mu_\eps(f)$ and $\mu_\eps(g)$, which vanish in \eqref{eq:freq_sensitivity_decomposition} because the resolvent is applied to centered observables.

The first term in \eqref{eq:freq_sensitivity_decomposition} is already covered by the invariant-statistic formula \eqref{eq:RP_sensitivity_residue_form}, applied to the observable $h_{f,g,z}$. Namely,
\be
-\mathscr{L}_B(u_{h_{f,g,z}})
=
\sum_{j=1}^N
\sum_{k=0}^{m_j-1}
\frac{\Rstyle_{j,k}^{h_{f,g,z}}(B)}{(-\lambda_j)^{k+1}}
- \mathscr{L}_B\left(
\mathcal R_N^{(0)} h_{f,g,z}^{c}
\right).
\label{eq:measure_contribution_RP_frequency}
\ee

Thus the genuinely new frequency-resolved structure is contained in the second term $\mathcal G_{f,g}^{B}(z)$, which has two resolvents and therefore couples pairs of RP blocks.

Using \eqref{eq:resolvent_RP_block}, the RP contribution to $\mathcal G_{f,g}^{B}(z)$ is
\begin{widetext}
\bea
\mathcal G_{f,g}^{B}(z)
&=
\sum_{j,\ell=1}^N
\sum_{k=0}^{m_j-1}
\sum_{r=0}^{m_\ell-1}
\frac{
\mathfrak s_{j,k;\ell,r}^{f,g}(B)
}{
(z-\lambda_j)^{k+1}
(z-\lambda_\ell)^{r+1}
}
+
\mathcal G_{f,g,N}^{B}(z),
\label{eq:freq_sensitivity_blocks}
\eea
\end{widetext}
where $\mathcal G_{f,g,N}^{B}(z)$ collects all terms involving at least one residual resolvent, and
\be
\mathfrak s_{j,k;\ell,r}^{f,g}(B)
= \mu\left(
f^c
N_j^k\Pi_j
B
N_\ell^r\Pi_\ell g^c
\right).
\label{eq:frequency_residue_def}
\ee
The scalars $\mathfrak s_{j,k;\ell,r}^{f,g}(B)$ are the frequency-resolved RP residues. They are the direct analogue, for correlation spectra, of the RP response residues $\Rstyle_{j,k}^{f}(B)$ in Eq.~\eqref{eq:RP_residues_general_B}. The difference is that frequency sensitivity contains two resolvents, one on each side of the perturbation $B$.

Equation \eqref{eq:freq_sensitivity_blocks} shows that a spectral statistic can be sensitive near frequencies satisfying
\be
z\simeq \lambda_j
\qquad
\text{or}
\qquad
z\simeq \lambda_\ell,
\label{eq:frequency_near_resonances}
\ee
and, after taking boundary values $z=\eta+i\omega$, near
\be
\omega\simeq \Im\lambda_j
\qquad
\text{or}
\qquad
\omega\simeq \Im\lambda_\ell .
\label{eq:frequency_near_imaginary_parts}
\ee
However, the size of the sensitivity is not determined by the denominators alone. It is determined by the frequency residues
\be
\mathfrak s_{j,k;\ell,r}^{f,g}(B)
= \mu\left(
f^c
N_j^k\Pi_j
B
N_\ell^r\Pi_\ell g^c
\right).
\label{eq:frequency_residue_repeat}
\ee
If these residues vanish, the corresponding pair of RP blocks does not contribute to the generator-resolvent sensitivity, even if one or both resonances are close to the imaginary axis.

In the semisimple case, $m_j=m_\ell=1$ and $N_j=N_\ell=0$, the pairwise contribution reduces to
\be
\frac{
\mu\left(
f^c\Pi_jB\Pi_\ell g^c
\right)
}{
(z-\lambda_j)(z-\lambda_\ell)
}.
\label{eq:frequency_semisimple_pair}
\ee
For nonsemisimple RP blocks, higher nilpotent levels produce higher-order resonant denominators:
\be
(z-\lambda_j)^{-(k+1)}
(z-\lambda_\ell)^{-(r+1)}.
\label{eq:frequency_higher_order_denominators}
\ee
Thus algebraic multiplicity can amplify frequency-resolved sensitivity in the same way that it amplifies invariant-statistic sensitivity, but now through pairwise residues involving $f$, $g$, and $B$.

The central conclusion is therefore sharper than the statement that small RP gaps imply rough spectral statistics \cite{Chek_al14_RP}. Small denominators permit large frequency-resolved sensitivity; nonzero RP residues produce it. The frequency-resolved RP susceptibility includes the distance to the relevant resonances, the nilpotent structure of their generalized eigenspaces, the conditioning of the corresponding spectral projectors, and the coupling coefficients
$\mathfrak s_{j,k;\ell,r}^{f,g}(B).$
In this sense, the frequency-domain theory is the two-resolvent analogue of the invariant-statistic sensitivity formula.

\subsection{Two kinds of small denominator: Resonance collisions and the spectral susceptibility}
\label{subsec:pairwise_interference}

Equation~\eqref{eq:freq_sensitivity_blocks} deserves more emphasis than a formal remark, because it says something that has no counterpart in the invariant-statistic theory of Sec.~\ref{subsec:summary_RP_sensitivity_theorem} and, to our knowledge, no counterpart in the early-warning literature. Frequency-resolved sensitivity is not a sum of independent single-block Lorentzians: it is a \emph{bilinear} object in the RP blocks, and it is therefore controlled by \emph{two} distinct families of small denominators.
 It is, in this respect, akin to the second-order susceptibility of nonlinear response theory, which is likewise bilinear in the resolvent \cite{Lucarini2008}.

To see this, consider the semisimple case and  decompose Eq.~\eqref{eq:freq_sensitivity_blocks} into partial fractions. For $j\neq\ell$,
\be
\frac{1}{(z-\lambda_j)(z-\lambda_\ell)}
=
\frac{1}{\lambda_j-\lambda_\ell}
\left[
\frac{1}{z-\lambda_j}
-
\frac{1}{z-\lambda_\ell}
\right],
\label{eq:pairwise_partial_fractions}
\ee
so that the generator-resolvent sensitivity takes the pole-resolved form
\begin{widetext}
\be
\mathcal G_{f,g}^{B}(z)
=
\sum_{j=1}^N
\frac{1}{z-\lambda_j}
\left[
\underbrace{\vphantom{\sum_{\ell\neq j}}\ \mathfrak s_{j,j}^{f,g}(B)\ \frac{1}{z-\lambda_j}}_{\text{self-term: double pole}}
+
\underbrace{\sum_{\ell\neq j}
\frac{
\mathfrak s_{j,\ell}^{f,g}(B)+\mathfrak s_{\ell,j}^{f,g}(B)
}{
\lambda_j-\lambda_\ell
}}_{\text{cross-terms: inter-block denominators}}
\right]
+
\mathcal G_{f,g,N}^{B}(z),
\label{eq:pairwise_pole_resolved}
\ee
\end{widetext}
where we abbreviate $\mathfrak s_{j,\ell}^{f,g}(B)=\mathfrak s_{j,0;\ell,0}^{f,g}(B)=\mu(f^c\Pi_jB\Pi_\ell g^c)$. Note that only the \emph{symmetrized} pair residue $\mathfrak s_{j,\ell}+\mathfrak s_{\ell,j}$ survives at the pole $\lambda_j$, weighted by the antisymmetric denominator $(\lambda_j-\lambda_\ell)^{-1}$; the antisymmetric part of the pair residue cancels identically. Two structural facts follow.

\emph{(i) The spectral susceptibility is one order more singular than the spectrum.} The correlation transform $\widehat C_{f,g}$ has a simple pole at a semisimple resonance $\lambda_j$, by Eq.~\eqref{eq:resolvent_RP_block}. Its derivative with respect to the control parameter has a \emph{double} pole there, produced by the self-term $j=\ell$ in \eqref{eq:pairwise_pole_resolved}. The double pole comes from the generator-resolvent term alone: the invariant-measure term \eqref{eq:measure_contribution_frequency} depends on $z$ only through the observable $h_{f,g,z}=f^cR(z;K)g^c$, and therefore contributes simple poles only. Near a real critical resonance $\lambda_*(\epsilon)\uparrow0$ this means
\be
\widehat C_{f,f}(0)=O\!\left(\frac{1}{-\lambda_*}\right),
\qquad
\pd{}{\epsilon}\widehat C_{f,f}(0)=O\!\left(\frac{1}{\lambda_*^2}\right),
\label{eq:susceptibility_one_order_more_singular}
\ee
provided $\mathfrak s_{*,*}^{f,f}(B_\epsilon)\neq0$. The practical reading is that the \emph{rate of change} of low-frequency power is a sharper indicator than the low-frequency power itself, and that it is governed by its own residue, which may vanish when the correlation residue does not. This is a testable statement about the derivative of a spectral EWS, not about its level.

\emph{(ii) Resonance collisions amplify spectral sensitivity even far from criticality.} The cross-terms in Eq.\eqref{eq:pairwise_pole_resolved} carry the denominators $\lambda_j-\lambda_\ell$, which measure the distance \emph{between} RP blocks rather than the distance of any block to the imaginary axis. Two resonances may therefore be deep in the left half-plane, so that every classical CSD diagnostic is quiet, and still produce a large frequency-resolved susceptibility because they are close to each other. This is the RP mechanism behind mode clustering: a spectral early-warning signal can be created by a near-collision of two fast modes, with no slow mode anywhere.

The two amplification routes are moreover distinguishable in shape. A closing gap moves a pole toward the imaginary axis and therefore \emph{narrows} the spectral feature while raising it; a resonance collision leaves the pole locations, and hence the widths, essentially unchanged and raises the \emph{amplitude} of the susceptibility only. Width and height, which classical reddening ties together, decouple.

These two mechanisms are not independent, and their meeting point is the nilpotent structure that Sec.~\ref{subsec:unperturbed_L2mu_setting} was careful to retain. As $\lambda_\ell\to\lambda_j$ the fraction  decomposition \eqref{eq:pairwise_partial_fractions} degenerates, the two simple poles merge, and the limit is precisely a second-order pole with a nilpotent numerator, i.e.~a nonsemisimple block with $m_j=2$. The inter-block denominator $(\lambda_j-\lambda_\ell)^{-1}$ and the nilpotent weight $(-\lambda_j)^{-2}$ are thus two faces of the same object, and the residue-selection rules of Theorem~\ref{thm:RP_sensitivity_residue_selection} govern both. In operator-theoretic language, the collision point is an exceptional point of the Kolmogorov generator \cite{kato1995perturbation,heiss2012physics}, and the divergence of $\|\Pi_j\|$ there is the RP expression of the transient, non-normal amplification that generates early-warning-like statistics without any loss of stability \cite{trefethen2005spectra,farrell1996generalized,troude2026pseudo}.

Finally, this is not  only an academic  construct. The companion study \cite{Chekroun_Lucarini26_EBM} reports that, on approach to tipping in a stochastic energy balance model, it is not one but \emph{several} reduced decay rates that compress toward the origin, with the associated Kolmogorov modes becoming geometrically harmonized along a common slow direction. Compression of a cluster is exactly the regime $|\lambda_j-\lambda_\ell|\ll|\lambda_j|$ in which the cross-terms of \eqref{eq:pairwise_pole_resolved} dominate the self-terms. The theory therefore predicts that, in such a regime, the frequency-resolved susceptibility is carried by inter-block interference rather than by any single Lorentzian, and that fitting a single slow mode to the spectrum will systematically misestimate the warning. Testing for this pairwise interference in spectral EWS --- for instance through the sign structure of $\partial_\epsilon S_f(\omega)$ across the cluster --- appears to be an open and accessible question.

\section{Ruelle-Pollicott Early-Warning Theory Near Tipping Points}
\label{sec:RP_EWS_tipping}

\subsection{The Triple Criterion}
We now translate the RP sensitivity formula into an early-warning theory for tipping points. The classical critical-slowing-down narrative is most transparent near a stable steady state, where the approach to a threshold is diagnosed through the loss of a local relaxation rate. That picture is useful, but it is not the natural language for more general invariant dynamics. A system may fluctuate around a noisy equilibrium, cycle around an oscillatory attractor, wander on a chaotic invariant set, or switch between metastable regions. In all these cases, the relevant question is not only how a local linear relaxation rate behaves, but how correlations, responses, and spectra are organized by the resonances of the evolution operator.

This is precisely the role of RP resonances. They are the spectral objects that encode decay of correlations and response for invariant dynamics beyond the steady-state setting. For deterministic chaotic systems, this statement is conceptually natural but technically delicate: the appropriate RP spectrum is typically defined on anisotropic spaces adapted to stable and unstable directions, not on a naive Hilbert space of square-integrable functions. In the hypoelliptic stochastic setting considered here, noise regularizes the invariant measure and allows us to work directly in the smoother framework developed above, namely with a smooth invariant density $\rho_\epsilon$ and spectral objects of the Kolmogorov generator $K_\epsilon$ acting on $L^2_{\mu_\epsilon}$ \cite{Chekroun_al_RP2}.

Thus the early-warning object in this paper is not a local steady-state eigenvalue, but a \emph{critical Kolmogorov block}. Let $\epsilon$ denote the control parameter and let $K_\epsilon$ be the generator with invariant measure $\d\mu_\epsilon=\rho_\epsilon\d x$. Moving the control parameter changes the generator through $B_\epsilon=\partial_\epsilon K_\epsilon$. For each $\epsilon$ before tipping, let $(\lambda_*(\epsilon),\Pi_*^\epsilon,N_*^\epsilon)$ denote the dominant nonzero RP block of $K_\epsilon$ in the relevant $L^2_{\mu_\epsilon}$ spectral window; equivalently, $\lambda_*(\epsilon)$ is the RP resonance controlling the slowest observable relaxation after constants have been removed. Here
\bes
N_*^\epsilon=(K_\epsilon-\lambda_*(\epsilon)I)\Pi_*^\epsilon, \textrm{ and  } (N_*^\epsilon)^{m_*}=0,
\ees
where $m_*$ is the pole order of the block.

We call this block \emph{critical} when the corresponding RP gap closes as the control parameter approaches the tipping threshold. For the static and local critical-slowing-down problem considered first, this means $\lambda_*(\epsilon)\to0$ as $\epsilon\to\epsilon_c$, typically through a real negative resonance. In that case the critical Kolmogorov block is the stochastic/RP counterpart of a soft relaxation channel. More generally, for frequency-resolved warnings, the relevant critical behavior may instead be $\Re\lambda_*(\epsilon)\uparrow0$ with $\Im\lambda_*(\epsilon)\neq0$, producing an oscillatory spectral warning rather than a singular static response.

The RP sensitivity formula now gives the physical content of an early-warning signal. For a measured observable $f$, write $f_\epsilon^c=f-\mu_\epsilon(f)\mathds{1}$. The critical Kolmogorov block is visible in the response of the invariant statistic $\mu_\epsilon(f)$ only through the RP response residues
\be
\Rstyle_{*,k}^{f}(\epsilon) = \mu_\epsilon\left( B_\epsilon (N_*^\epsilon)^k \Pi_*^\epsilon f_\epsilon^c \right), \qquad 0\le k\le m_*-1.
\label{eq:EWS_residues_general}
\ee
These residues are the actual early-warning amplitudes. The critical RP block supplies the small denominator; the observable and the perturbation direction supply the numerator.

If the largest nonzero residue occurs at the nilpotent level
\be
\kappa_*^f(\epsilon) = \max \left\{ 0\le k\le m_*-1: \Rstyle_{*,k}^{f}(\epsilon)\neq0 \right\},
\label{eq:EWS_dominant_residue_level}
\ee
then, with the sign convention $K_\epsilon u=f_\epsilon^c$, the critical block contributes
\be
\pd{}{\epsilon}\mu_\epsilon(f) \sim \frac{ \Rstyle_{*,\kappa_*^f}^{f}(\epsilon) }{ \left(-\lambda_*(\epsilon)\right)^{\kappa_*^f+1} },
\label{eq:EWS_dominant_sensitivity}
\ee
up to less singular RP blocks and residual spectral contributions. This is the residue-conditioned RP early-warning criterion.

The message is simple but important: a critical RP resonance is not enough. It permits a large response, but it does not guarantee one. The warning becomes visible only if at least one residue in \eqref{eq:EWS_residues_general} is nonzero. Thus an early-warning signal is not a property of the dynamics alone. It is a property of a triple:
\be
\boxed{
\text{critical RP block}+ \text{observable}+ \text{perturbation direction}.
}
\label{eq:EWS_triple}
\ee
 In a one-degree-of-freedom caricature the three ingredients are indistinguishable, since there is a single block, a single observable and a single direction; they separate as soon as the dynamics has more than one degree of freedom. This is, in one sentence, both the merit and the limitation of the classical picture.

The block supplies $\lambda_*(\epsilon)$ and the nilpotent ladder $(N_*^\epsilon)^k\Pi_*^\epsilon$. The observable supplies the projected component $\Pi_*^\epsilon f_\epsilon^c$. The perturbation supplies the functional $g\mapsto\mu_\epsilon(B_\epsilon g)$. Their contraction is the RP response residue.

In the semisimple case, $m_*=1$ and $N_*^\epsilon=0$. The residue ladder collapses to one number,
$\Rstyle_*^{f}(\epsilon)=\mu_\epsilon(B_\epsilon\Pi_*^\epsilon f_\epsilon^c)$, and the critical-block contribution reduces to
\[
\pd{}{\epsilon}\mu_\epsilon(f) \sim \Rstyle_*^{f}(\epsilon)/\left(-\lambda_*(\epsilon)\right) .
\]
Thus a semisimple critical Kolmogorov block is an effective EWS for $f$ only when $\mu_\epsilon(B_\epsilon\Pi_*^\epsilon f_\epsilon^c)\neq0$. If this residue is small or vanishes, the measured statistic may show little warning even though the system possesses a critical spectral mechanism.

If the block is nonsemisimple, the possible amplification is stronger. A pole of order $m_*$ can in principle produce a sensitivity as large as $|\lambda_*(\epsilon)|^{-m_*}$, but only if the highest nilpotent residue
$\mu_\epsilon(B_\epsilon(N_*^\epsilon)^{m_*-1}\Pi_*^\epsilon f_\epsilon^c)$ is nonzero. Otherwise the dominant order is lower, selected by the highest nilpotent level actually seen by the observable--perturbation pair.

This viewpoint reframes classical early-warning theory. The warning signal is not tied to a universal scalar indicator, nor to a local steady-state diagnostic. It is tied to a critical Kolmogorov block approaching spectral degeneracy, but only after projection onto the measured observable and contraction against the perturbation direction. In this sense, the RP formulation replaces the question ``is there a slow mode?'' by the sharper question ``which critical RP block is seen by this observable under this perturbation?''

The same distinction applies across EWS metrics. For invariant means, the singular response is strongest when $\lambda_*(\epsilon)$ approaches the origin. For correlation times, the decay is controlled by $\Re\lambda_*(\epsilon)$. For frequency-resolved quantities, a resonance approaching the imaginary axis at nonzero frequency produces a warning near $\omega=\Im\lambda_*(\epsilon)$ rather than necessarily a singularity in the static mean. Thus the RP early-warning principle is not simply ``slowing down implies warning.'' It is: identify the critical RP block, identify what the observable sees, and identify how the control parameter couples to it.

\subsection{The Paradigm Shift from Critical Slowing Down}
\label{subsec:EWS_vs_CSD}
The RP formulation changes the status of critical slowing down. A slow spectral value is not, by itself, an early-warning signal. It becomes a warning only after it is selected by an observable and activated by a perturbation direction, as summarized by the triple criterion in Eq.~\eqref{eq:EWS_triple}. Thus the relevant object is not merely a slow rate, but a critical RP block of the generator $K_\epsilon$ acting on observables in $L^2_{\mu_\epsilon}$, together with the residues through which this block is seen.

This immediately changes how one interprets classical EWS diagnostics. Variance, autocorrelation time, parameter susceptibility, and spectral power are not interchangeable symptoms of the same object. They are different contractions of the RP expansion, and they can select different blocks.

For a centered observable $f_\epsilon^c=f-\mu_\epsilon(f)\mathds{1}$, the stationary autocorrelation is $C_f^\epsilon(t)=\mu_\epsilon(f_\epsilon^cP_t^\epsilon f_\epsilon^c)$. The RP expansion gives
\bea
C_f^\epsilon(t)
&=
\sum_{j=1}^N
e^{\lambda_j(\epsilon)t}
\sum_{k=0}^{m_j-1}
\frac{t^k}{k!}
\mathcal A_{j,k}^{f}(\epsilon)\\
&\quad+
\text{residual contribution},
\label{eq:correlation_RP_EWS}
\eea
where the correlation residues are
\be
\mathcal A_{j,k}^{f}(\epsilon) = \mu_\epsilon\left( f_\epsilon^c (N_j^\epsilon)^k \Pi_j^\epsilon f_\epsilon^c \right).
\label{eq:correlation_residue_EWS}
\ee
Therefore a slow RP resonance affects the observed autocorrelation only if the corresponding correlation residue is nonzero. A small RP gap that is invisible to $f$ is not an autocorrelation-based warning for $f$.

The same separation appears more sharply when variance and autocorrelation time are compared. The variance is simply the equal-time object ${\rm Var}_{\mu_\epsilon}(f)=C_f^\epsilon(0)=\mu_\epsilon(|f_\epsilon^c|^2)$. It measures the spread of the invariant measure through the observable $f$, but it is not itself a resolvent. By contrast, the integrated autocorrelation
$\mathcal I_f(\epsilon)=\int_0^\infty C_f^\epsilon(t)\d t$ is a reduced-resolvent quantity:
\be
\mathcal I_f(\epsilon) = -\mu_\epsilon\left( f_\epsilon^cK_\epsilon^{-1}f_\epsilon^c \right).
\label{eq:integrated_autocorrelation_EWS}
\ee
Using the RP blocks, this becomes
\bes
\mathcal I_f(\epsilon)
=\sum_{j=1}^N
\sum_{k=0}^{m_j-1}
\frac{\mathcal A_{j,k}^{f}(\epsilon)}{\left(-\lambda_j(\epsilon)\right)^{k+1}}
+
\text{residual contribution}.
\ees
Thus autocorrelation time is governed by the RP denominators weighted by correlation residues, not by the spectrum alone.

Parameter susceptibility is a different contraction again. With $B_\epsilon=\partial_\epsilon K_\epsilon$, the response of the invariant mean is controlled by the RP response residues
$\Rstyle_{j,k}^{f}(\epsilon)=\mu_\epsilon(B_\epsilon(N_j^\epsilon)^k\Pi_j^\epsilon f_\epsilon^c)$. With the convention $K_\epsilon u=f_\epsilon^c$, one has
\beas
\pd{}{\epsilon}\mu_\epsilon(f)
&=
\sum_{j=1}^N
\sum_{k=0}^{m_j-1}
\frac{\Rstyle_{j,k}^{f}(\epsilon)}{\left(-\lambda_j(\epsilon)\right)^{k+1}}\\
&\quad+
\text{residual contribution}.
\eeas
The coefficients entering these three diagnostics are different. The variance contains no RP denominator and no response residue; it is the equal-time quantity $\mu_\epsilon(|f_\epsilon^c|^2)$. The autocorrelation and the integrated autocorrelation are controlled by the correlation residues
$\mathcal A_{j,k}^{f}(\epsilon)=\mu_\epsilon(f_\epsilon^c(N_j^\epsilon)^k\Pi_j^\epsilon f_\epsilon^c)$, with the integrated autocorrelation adding the RP weights $\left(-\lambda_j(\epsilon)\right)^{-(k+1)}$. By contrast, the parameter sensitivity of the invariant mean is controlled by the response residues
$\Rstyle_{j,k}^{f}(\epsilon)=\mu_\epsilon(B_\epsilon(N_j^\epsilon)^k\Pi_j^\epsilon f_\epsilon^c)$.
 The response residues are themselves correlation-type quantities. Writing $\mu_\epsilon(B_\epsilon g)=\int g\,(B_\epsilon^{\dagger}\rho_\epsilon)\d x$, with $B_\epsilon^{\dagger}$ the formal adjoint of $B_\epsilon$ in $L^2(\d x)$, one has $\Rstyle_{j,k}^{f}(\epsilon)=\mu_\epsilon\big(g_{B_\epsilon}(N_j^\epsilon)^k\Pi_j^\epsilon f_\epsilon^c\big)$ with $g_{B_\epsilon}=\rho_\epsilon^{-1}B_\epsilon^{\dagger}\rho_\epsilon$, and the susceptibility itself reads $\partial_\epsilon\mu_\epsilon(f)=\int_0^\infty\mu_\epsilon\big(g_{B_\epsilon}P_t^\epsilon f_\epsilon^c\big)\d t$: it is the time-integrated cross-correlation between $f$ and the conjugate observable $g_{B_\epsilon}$ determined by the perturbation and by the invariant density \cite{Santos2022,LucariniChekroun2023}. Response residues thus differ from correlation residues only in that the left factor $f_\epsilon^c$ is replaced by $g_{B_\epsilon}$.

Thus the same RP block may be visible to one diagnostic and invisible to another. An observable can have large correlation residues $\mathcal A_{j,k}^{f}(\epsilon)$, and hence a growing autocorrelation time, while having small response residues $\Rstyle_{j,k}^{f}(\epsilon)$, and hence weak sensitivity to the control parameter. Conversely, the invariant mean can be highly parameter-sensitive because $\Rstyle_{j,k}^{f}(\epsilon)$ is large, even when the raw variance $\mu_\epsilon(|f_\epsilon^c|^2)$ does not provide a reliable warning signal.

The practical implication is that an EWS metric must declare what it is designed to detect: (i) variance probes the geometry and spread of $\mu_\epsilon$ through $f$; (ii) autocorrelation time probes observable-weighted RP relaxation rates; (iii) mean susceptibility probes RP response residues involving $B_\epsilon$; and (iv) spectral indicators probe frequency-resolved RP residues. There is therefore no universal scalar early-warning signal independent of the critical RP block, the observable, and the perturbation direction.

 Of these four, (ii)--(iv) are of one kind and (i) is of another. Autocorrelation, susceptibility and spectral indicators are all resolvent quantities: each carries RP denominators weighted by residues of correlation type, and they differ only in which observable is paired with $f$ and at which frequency the resolvent is evaluated. The variance is an equal-time quantity with no denominator at all. This is why, in Sec.~\ref{sec:nonnormal}, the variance is the one indicator that residue selection cannot rescue.

\section{Early Warnings Without a Critical Block: A Worked Non-Normal Counterexample}
\label{sec:nonnormal}

The residue-conditioned viewpoint predicts something uncomfortable for the classical theory: because a diagnostic is a product of a denominator and a residue, it can be made to grow by inflating the residue alone, with the spectrum held \emph{exactly} fixed. If this is possible, then the canonical early-warning battery is not merely observable-dependent, it is not diagnostic of criticality at all. This section shows that the possibility is realized, in a model that is analytically transparent, and that the RP framework both explains the failure and repairs it.

 The key notion is non-normality. An operator $K$ is non-normal when $KK^*\neq K^*K$; its eigenvectors are then not orthogonal, and the eigenbasis can be arbitrarily ill-conditioned \cite{trefethen2005spectra}. In the language of Sec.~\ref{subsec:unperturbed_L2mu_setting}, retain the isolated point spectrum, assume for simplicity that every block is semisimple, and neglect the residual part; the resolvent then reads $(zI-K)^{-1}=\Pi_0/z+\sum_j\Pi_j/(z-\lambda_j)$, so that the residues of the resolvent at its poles are precisely the spectral projectors. For a normal operator the $\Pi_j$ are orthogonal projections, of norm one in $L^2_\mu$; for a non-normal operator they are oblique, and $\|\Pi_j\|$ can be arbitrarily large. Large projector norms are the spectral signature of non-normality, and they act in the time domain: even when every $\Re\lambda_j<0$, the expansion $P_t=\Pi_0+\sum_j e^{\lambda_jt}\Pi_j$ can display large transient growth at intermediate times, through cancellation between large terms of opposite sign, a phenomenon invisible to the eigenvalues alone \cite{trefethen2005spectra,farrell1996generalized}. In the vocabulary of Sec.~\ref{subsec:summary_RP_sensitivity_theorem}, large projectors mean large residues.

 The example below is the stochastic non-normal system recently used by Troude \emph{et al.} \cite{troude2026pseudo} to argue that widely used early-warning indicators are not uniquely diagnostic of tipping points. Their empirical claim and the present theory meet exactly here, and we take the opportunity to make the mechanism explicit.

\begin{table}[t]
\caption{\label{tab:nonnormal} Exact RP decomposition of the non-normal model \eqref{eq:nonnormal_SDE} for $\alpha=3/2$, $\delta=1/2$, observable $f=r$. The resonances are $\lambda_\pm=-1/2,-5/2$ at every $\kappa$ (constant to machine precision). Everything that grows is a residue. The last column is the integrated correlation time measured through the residue-selected Kolmogorov mode $\psi_+$, for which $\mathcal A_-^{\psi_+}$ vanishes identically (numerically below $10^{-16}$) and the normalized autocorrelation is exactly $e^{\lambda_+t}$. With this normalization $\Delta_{\rm RP}=\mathcal E(r,r)=1/2$, so the index $\mathcal N(r)$ of Eq.~\eqref{eq:nonnormality_index} coincides numerically with ${\rm Var}(r)$.}
\begin{ruledtabular}
\begin{tabular}{rrrrrrr}
$\kappa$ & ${\rm cond}\,\Pi$ & $\mathcal A_+^{r}$ & $\mathcal A_-^{r}$ & ${\rm Var}(r)$ & $\tau_{\rm int}(r)$ & $\tau_{\rm int}(\psi_+)$\\
\hline
1   & 1   & $0.500$ & $\phantom{-}0.100$ & $0.600$ & $1.733$ & $2.000$\\
3   & 3   & $1.833$ & $-0.167$ & $1.667$ & $2.160$ & $2.000$\\
10  & 10  & $17.00$ & $-3.200$ & $13.80$ & $2.371$ & $2.000$\\
30  & 30  & $150.3$ & $-29.87$ & $120.5$ & $2.397$ & $2.000$\\
100 & 100 & $1667$  & $-333.2$ & $1334$  & $2.400$ & $2.000$\\
\end{tabular}
\end{ruledtabular}
\end{table}

\begin{figure*}[t]
\centering
\includegraphics[width=\textwidth]{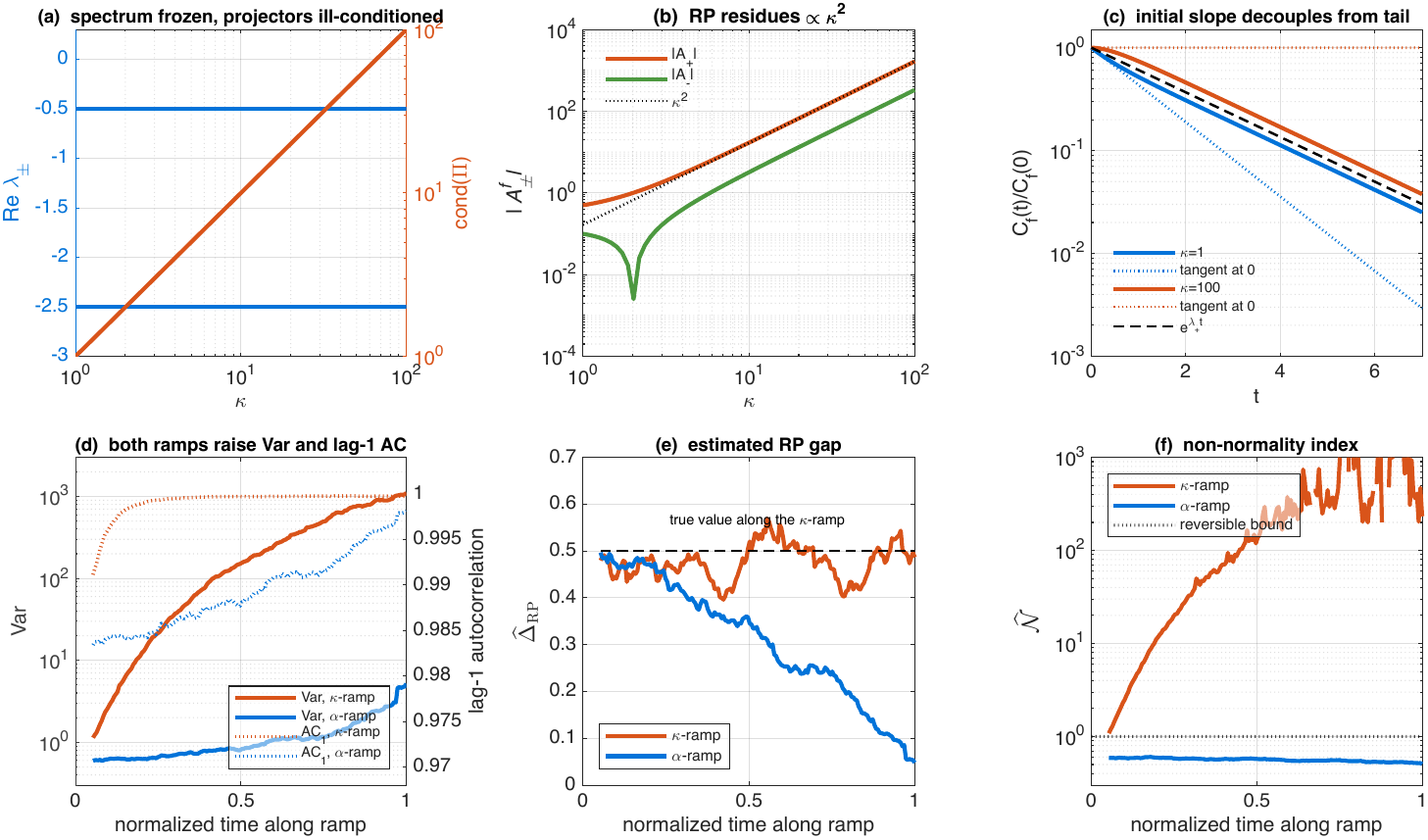}
\caption{\label{fig:nonnormal} Early warnings without a critical block, for the non-normal model \eqref{eq:nonnormal_SDE}.  (a) The two RP resonances are exactly independent of $\kappa$ (left axis, blue) while the conditioning of the spectral projectors grows linearly (right axis, orange). (b) The RP correlation residues of the mean-field observable grow like $\kappa^2$; the dip near $\kappa\simeq2$ is the sign change of $\mathcal A_-^{r}$. (c) Normalized autocorrelation of $r$: the initial log-slope, which equals $-\mathcal E(f,f)/{\rm Var}(f)$, decouples from the asymptotic slope $\lambda_+$ as $\kappa$ grows; the ratio of the two is the index $\mathcal N$. Panels (d--f) show rolling-window diagnostics along two slow ramps of equal length, described in Sec.~\ref{subsec:nonnormal_ramped}: the \emph{$\kappa$-ramp} (orange) increases the non-normality from $\kappa=1$ to $100$ at fixed spectrum, and the \emph{$\alpha$-ramp} (blue) closes the gap from $0.5$ to $0.01$ at $\kappa=1$. (d) The two classical indicators, running variance (solid, left axis) and running lag-one autocorrelation (dotted, right axis), rise along \emph{both} ramps, and rise \emph{more} along the $\kappa$-ramp, where the spectrum never changes: variance grows by a factor $9.8\times10^{2}$ against $8.7$, and the lag-one autocorrelation reaches $0.9997$ against $0.9981$. Ranking the two records by either indicator therefore inverts the correct answer. (e) The RP gap estimated from a VAR(1) fit stays at its true value $0.5$ along the $\kappa$-ramp (dashed line) and falls by an order of magnitude along the $\alpha$-ramp. (f) The estimated index crosses the reversible bound $\mathcal N=1$ within the first fifth of the $\kappa$-ramp and remains below it throughout the $\alpha$-ramp. Panels (e) and (f) are the two diagnostics that separate the mechanisms; panel (d) is the pair that does not.}
\end{figure*}

\subsection{The model and its frozen spectrum}
\label{subsec:nonnormal_model}
Consider the two-dimensional Ornstein--Uhlenbeck process
\be
\d X_t = A_\kappa X_t\d t + \sigma\d W_t,
\qquad
A_\kappa =
\begin{pmatrix}
-\alpha & \kappa^{-1}\\[2pt]
\kappa & -\alpha
\end{pmatrix},
\label{eq:nonnormal_SDE}
\ee
with $\alpha>1$ and $\kappa\ge1$, written in the reduced coordinates $X=(n,r)$ of \cite{troude2026pseudo}, where $r$ plays the role of a mean-field observable. We set $\delta=\sigma^2/2$. The Kolmogorov generator is $K_\kappa\varphi=A_\kappa x\cdot\nabla\varphi+\delta\Delta\varphi$, and its RP resonances at the first Hermite level are the eigenvalues of $A_\kappa$ \cite{metafunes2002},
\be
\lambda_\pm=-\alpha\pm1 ,
\label{eq:nonnormal_resonances}
\ee
which are \emph{independent of $\kappa$}. The whole RP spectrum  is known for such Ornstein-Uhlenbeck processes and is given by $\{n_+\lambda_++n_-\lambda_-\}$ \cite{metafunes2002,Tantet_al_Hopf}. As such, it is therefore frozen: the gap $\Delta_{\rm RP}=\alpha-1$ does not move, and no resonance approaches the imaginary axis at any $\kappa$.

What does move is the geometry of the eigenspaces. The right and left eigenvectors of $A_\kappa$ are
\be
v_\pm=\begin{pmatrix}1\\ \pm\kappa\end{pmatrix},
\qquad
\ell_\pm=\begin{pmatrix}1\\ \pm\kappa^{-1}\end{pmatrix},
\qquad
\langle\ell_\pm,v_\pm\rangle=2 ,
\label{eq:nonnormal_eigenvectors}
\ee
so that the RP eigenfunctions are $\psi_\pm(x)=\langle\ell_\pm,x\rangle$ and the spectral projectors, restricted to linear observables, have norm
\be
\left\|\Pi_\pm\right\|=\frac{1+\kappa^2}{2\kappa}\ \sim\ \frac{\kappa}{2},
\qquad
\kappa\to\infty .
\label{eq:nonnormal_projector_norm}
\ee
The angle between $\ell_+$ and $\ell_-$ closes like $\kappa^{-2}$: the two RP eigenspaces become nearly parallel while their eigenvalues stay a fixed distance apart. This is precisely the situation the block formalism of Sec.~\ref{subsec:unperturbed_L2mu_setting} was built to describe, and precisely the one that a gap-based diagnostic cannot see.

\subsection{All the growth is in the residues}
\label{subsec:nonnormal_residues}
Let $f(x)=\langle c,x\rangle$ and let $\Sigma_\kappa$ be the stationary covariance. The RP correlation residues of Eq.~\eqref{eq:correlation_residue_EWS} are, at this level,
\be
\mathcal A_\pm^{f}(\kappa)=c^{T}P_\pm\Sigma_\kappa c,
\qquad
P_\pm=\frac{v_\pm\ell_\pm^{T}}{2},
\label{eq:nonnormal_residues}
\ee
and the autocorrelation is exactly the two-block sum
$C_f(t)=\mathcal A_+^{f}e^{\lambda_+t}+\mathcal A_-^{f}e^{\lambda_-t}$.
Table~\ref{tab:nonnormal} and Fig.~\ref{fig:nonnormal}(a,b) report the outcome for $f=r$ and $\alpha=3/2$, $\delta=1/2$; panels (d--f) of the same figure anticipate the ramped experiments of Sec.~\ref{subsec:nonnormal_ramped}. The resonances are constant to machine precision over $\kappa\in[1,100]$, while
\bea
& \left|\mathcal A_\pm^{r}(\kappa)\right| \propto \kappa^{2},
\qquad
{\rm Var}(r)\propto\kappa^2,\\
& \mathcal I_r= \int_0^\infty C_r^{\kappa}(t)\d t\propto\kappa^2,
\label{eq:nonnormal_scalings}
\eea
with fitted exponents $1.996$, $1.993$ and $1.995$. Over this range the variance rises by a factor $2.2\times10^{3}$ and the zero-frequency power by the same factor. Every ingredient of the classical battery fires, and the RP sensitivity formula \eqref{eq:summary_RP_sensitivity} attributes all of it to the numerators.

Two further features of Table~\ref{tab:nonnormal} deserve comment. First, the fast residue $\mathcal A_-^{r}$ changes sign near $\kappa\simeq2$: there is a value of the non-normality at which the mean-field observable becomes exactly blind to the fast block, an isolated residue zero of the kind anticipated in Sec.~\ref{subsec:summary_RP_sensitivity_theorem}. Second, beyond that point $\mathcal A_-^{r}<0$, so the two residues have opposite signs and partially cancel. This is what allows the \emph{normalized} correlation time
$\tau_{\rm int}=\mathcal I_f/{\rm Var}(f)$
to exceed $1/|\lambda_+|$, which a positive-residue expansion could never do. It also caps it: $\tau_{\rm int}$ rises from $1.73$ to $2.40$ and then \emph{saturates}, because with frozen $\lambda_\pm$ the attainable residue ratio is bounded. Genuine gap closure has no such ceiling.

\subsection{Residue selection removes the false positive}
\label{subsec:nonnormal_selection}
The theory does not merely diagnose the failure; it says what to measure instead. Take as observable the Kolmogorov mode itself, $f=\psi_+=\langle\ell_+,x\rangle$. Since $\ell_+\perp v_-$, one has $c^{T}P_-=0$ identically, hence
\be
\mathcal A_-^{\psi_+}\equiv0,
\qquad
\frac{C_{\psi_+}(t)}{C_{\psi_+}(0)}=e^{\lambda_+t}
\quad\text{for every }\kappa .
\label{eq:nonnormal_blind_observable}
\ee
The residue-selected observable sees a single block, and its normalized autocorrelation, its correlation time $\tau_{\rm int}=1/(\alpha-1)$ and its spectral width are \emph{exactly} $\kappa$-independent (Table~\ref{tab:nonnormal}, lower block).  The spurious warning disappears from every rate-based diagnostic computed through $\psi_+$.

It does not disappear from the variance, which still grows like $\kappa^2$. That is not a defect of the construction but the prediction of Sec.~\ref{subsec:EWS_vs_CSD}: the variance is the one classical indicator that contains no RP denominator at all, being the equal-time quantity $\mu_\epsilon(|f^c_\epsilon|^2)$. Having no denominator, it cannot be residue-selected, and it cannot distinguish a slow block from a large one. The worked example thus confirms the taxonomy quantitatively: among the classical EWS, variance is structurally the least diagnostic, and rate-based indicators become diagnostic only after the observable has been chosen to isolate a block.

\subsection{A non-normality index, and a bound that certifies it}
\label{subsec:nonnormal_index}
Choosing the Kolmogorov mode requires knowing it. We now give a test that does not, and that follows from the residue ladder alone.

For a centered observable $f$ define the Dirichlet form
\bea
\mathcal E(f,f)
&=
-\left.\frac{\d}{\d t}C_f(t)\right|_{t=0^+}
=
-\mu\!\left(f^cKf^c\right)\\
&=
\frac{\sigma^2}{2}\mu\!\left(\nabla f\cdot a\nabla f\right),
\label{eq:dirichlet_form}
\eea
the last equality holding for a nondegenerate diffusion with smooth invariant density \cite{bakry2014analysis}. The key structural fact is that $\mathcal E$ depends only on the \emph{symmetric part} of $K$ in $L^2_\mu$: the circulatory, non-normal part of the drift is annihilated in $\mu(f^cKf^c)$. The Dirichlet form is therefore blind to non-normality, whereas the variance and the RP gap are not. Their mismatch is what we exploit. Define the \emph{RP non-normality index}
\be
\boxed{\ \
\mathcal N(f)
=
\frac{\Delta_{\rm RP}\ {\rm Var}_\mu(f)}{\mathcal E(f,f)}
\ \ }
\label{eq:nonnormality_index}
\ee
or, resolved on the RP blocks,
\be
\mathcal N(f)
=
\Delta_{\rm RP}\,
\frac{\sum_j\mathcal A_j^{f}}{\sum_j(-\lambda_j)\,\mathcal A_j^{f}} \ .
\label{eq:nonnormality_index_blocks}
\ee
The second form exhibits $\mathcal N$ as a ratio of two \emph{different contractions of the same residue ladder}, weighted by $1$ and by $(-\lambda_j)$ respectively. It is exactly the kind of object Sec.~\ref{subsec:EWS_vs_CSD} argues one should compare.

\bprop
\label{prop:nonnormality_bound}
Let $K$ be self-adjoint in $L^2_\mu$, with nonzero spectrum contained in $\{\Re z\le-\Delta_{\rm RP}\}$, $\Delta_{\rm RP}>0$. Then for every centered  observable $f$,
\be
\mathcal N(f)\le1 ,
\label{eq:nonnormality_bound}
\ee
with equality if and only if $f$ lies in the eigenspace of the leading nonzero eigenvalue. Consequently $\mathcal N(f)>1$ for some observable $f$ is a certificate that the generator is not self-adjoint in $L^2_\mu$.
\eprop
\begin{proof}
By the spectral theorem the correlation residues are $\mathcal A_j^{f}=\mu(f^c\Pi_jf^c)=\|\Pi_jf^c\|_\mu^2\ge0$, so ${\rm Var}_\mu(f)=\sum_j\mathcal A_j^{f}$ and
$\mathcal E(f,f)=\sum_j(-\lambda_j)\mathcal A_j^{f}\ge\Delta_{\rm RP}\sum_j\mathcal A_j^{f}$,
since $-\lambda_j\ge\Delta_{\rm RP}$ for every $j$. Equality forces $\mathcal A_j^{f}=0$ whenever $-\lambda_j>\Delta_{\rm RP}$.
\end{proof}

Inequality \eqref{eq:nonnormality_bound} is the Poincar\'e inequality with its optimal constant \cite{bakry2014analysis}, read backwards. Its diagnostic force comes from what breaks it. In the reversible case the residues are squared norms, hence positive, and no cancellation is available; the observed relaxation can never be slower than the variance-to-dissipation budget allows. When $K$ is not self-adjoint the residues are signed, cancellation is available, and $\mathcal N$ is unbounded. Thus
\be
\mathcal N(f)>1
\ \Longrightarrow\
\begin{minipage}{0.52\columnwidth}
\centering
amplification is carried by\\ residues, not by the gap.
\end{minipage}
\label{eq:index_reading}
\ee
 For the model \eqref{eq:nonnormal_SDE} the index can be evaluated in closed form, so nothing here rests on sampling. We restrict attention to the \emph{linear} observables $f(x)=\langle c,x\rangle$, which is the natural class: they span the first Hermite level, where the two RP blocks $\lambda_\pm$ live, and the spurious warning is carried by those blocks. Higher Hermite levels have resonances $n_+\lambda_++n_-\lambda_-$ that are strictly more negative and play no role. For such observables $\mathcal E(f,f)=\delta|c|^2$ and ${\rm Var}_\mu(f)=c^{T}\Sigma_\kappa c$, with $\Sigma_\kappa$ the stationary covariance, so that
\be
\mathcal N(c)=\frac{(\alpha-1)\,c^{T}\Sigma_\kappa c}{\delta\,|c|^{2}} ,
\label{eq:index_linear_observables}
\ee
and the extremes over the whole class are the extreme eigenvalues of a $2\times2$ matrix,
\bea
\max_{c\neq0}\mathcal N&=\frac{\alpha-1}{\delta}\,\lambda_{\max}(\Sigma_\kappa),\\
\min_{c\neq0}\mathcal N&=\frac{\alpha-1}{\delta}\,\lambda_{\min}(\Sigma_\kappa).
\label{eq:index_extremes}
\eea
Solving the Lyapunov equation $A_\kappa\Sigma_\kappa+\Sigma_\kappa A_\kappa^{T}=-2\delta I$ by hand gives
\bea
\Sigma_\kappa&=\frac{\delta}{2\alpha(\alpha^2-1)}
\begin{pmatrix}
\beta+\kappa^{-2} & \alpha(\kappa+\kappa^{-1})\\[2pt]
\alpha(\kappa+\kappa^{-1}) & \beta+\kappa^{2}
\end{pmatrix},\\
&\qquad\qquad\qquad\beta=2\alpha^{2}-1,
\label{eq:sigma_kappa_closed}
\eea
whence, for the mean-field observable $f=r$,
\be
\mathcal N(r)=\frac{\kappa^{2}+2\alpha^{2}-1}{2\alpha(\alpha+1)}
\ \sim\ \frac{\kappa^{2}}{2\alpha(\alpha+1)} .
\label{eq:index_closed_form}
\ee
The quadratic growth is therefore exact, not fitted. At $\alpha=3/2$ and $\kappa=100$ it gives $\mathcal N(r)=1333.8$, and $\max_c\mathcal N=1334.1$, so the mean-field observable is within $0.03\%$ of the worst case.

For the reversible control family the corresponding statement is sharper still, and it is an identity rather than a numerical check. At $\kappa=1$ the matrix $A_1$ is symmetric, $\Sigma_1=\delta\,{\rm diag}\big((\alpha-1)^{-1},(\alpha+1)^{-1}\big)$ in its eigenbasis, and therefore
\be
\max_{c\neq0}\mathcal N=\frac{\alpha-1}{\delta}\cdot\frac{\delta}{\alpha-1}=1,
\qquad
\min_{c\neq0}\mathcal N=\frac{\alpha-1}{\alpha+1},
\label{eq:index_reversible_exact}
\ee
for \emph{every} $\alpha>1$. The bound \eqref{eq:nonnormality_bound} is thus attained exactly, and attained by the leading eigenvector, as Proposition~\ref{prop:nonnormality_bound} predicts; evaluating \eqref{eq:index_extremes} numerically along the control family reproduces $\max_c\mathcal N=1$ to twelve digits. We stress that the bound itself is a theorem valid for all observables in the form domain, so these formulas verify where it is saturated, not whether it holds.

\subsection{The test on ramped records}
\label{subsec:nonnormal_ramped}

\emph{Goal.} Everything above is exact, and exactness is precisely what a practitioner does not have. The question of this subsection is therefore narrow and operational: given one finite, singly-sampled trajectory of a system whose control parameter drifts slowly, can the two mechanisms be told apart in practice? We ask this of the three quantities that enter the theory --- the leading RP gap, the index $\mathcal N$, and the conditioning of the spectral projectors --- and we compare them against the two indicators an early-warning study would normally report, the running variance and the running lag-one autocorrelation.

\emph{The two experiments.} We integrate Eq.~\eqref{eq:nonnormal_SDE} twice, over records of equal length and with the control parameter moved slowly and monotonically in each.
\bi
\item The \emph{$\kappa$-ramp} increases the non-normality from $\kappa=1$ to $\kappa=100$ at fixed $\alpha=3/2$. By Eq.~\eqref{eq:nonnormal_resonances} the RP spectrum is unchanged throughout: there is no bifurcation, no loss of stability, and no slow mode at any point of this record.
\item The \emph{$\alpha$-ramp} decreases $\alpha$ from $3/2$ to $1.01$ at fixed $\kappa=1$, so the matrix stays symmetric and the leading resonance $\lambda_+=-(\alpha-1)$ moves from $-0.5$ to $-0.01$. This is a genuine, if elementary, approach to a gap closure, and it is the behaviour that critical-slowing-down theory is built to detect.
\ei
A word on terminology. The model \eqref{eq:nonnormal_SDE} is linear and its fixed point is stable throughout, so \emph{neither} ramp contains a bifurcation: the $\alpha$-ramp is the elementary caricature of B-tipping --- a relaxation rate sent to zero --- rather than an actual loss of stability, and the effect produced by the $\kappa$-ramp is what \cite{troude2026pseudo} call a pseudo-bifurcation. We therefore label the experiments by the parameter that is moved, and reserve the tipping vocabulary for Secs.~\ref{Sec_EWS_Metastable}--\ref{sec:1D_fold_example}, where genuinely metastable dynamics is at stake. The measured observable is the mean-field coordinate $f=r$ in both cases. Diagnostics are computed on rolling windows of length $T_{\rm w}=2000$ stepped by $100$ time units; with $|\lambda_+|=0.5$ at the start of each ramp, a window spans $O(10^3)$ correlation times, so window-length bias is not the limiting factor.

\emph{Estimation.} The sampled process is recorded at $\Delta t=0.02$. For a linear diffusion the sampled chain is \emph{exactly} a first-order vector autoregression, $X_{k+1}=\Phi X_k+\xi_k$ with $\Phi=e^{A\Delta t}$ and $\xi_k$ i.i.d.~Gaussian, so the natural estimator is ordinary least squares on each window followed by a matrix logarithm,
\be
\widehat\Phi=\Big(\textstyle\sum_k X_{k+1}X_k^{T}\Big)\Big(\sum_k X_kX_k^{T}\Big)^{-1},
\quad
\widehat A=\frac{\log\widehat\Phi}{\Delta t},
\label{eq:var1_estimator}
\ee
from which $\widehat\Delta_{\rm RP}=-\max_j\Re\,\widehat\lambda_j(\widehat A)$ and ${\rm cond}(\widehat V)$ follow \cite{lutkepohl2005,crommelin2011diffusion}. Equation~\eqref{eq:var1_estimator} is the multivariate version of the discrete-to-generator relation \eqref{eq:discrete_to_generator_transition_resonance} already used for transition blocks, and the same branch caveat applies: the principal logarithm is the correct one provided $\Delta t$ is short enough that no eigenvalue of $\widehat\Phi$ approaches the negative real axis.

The sampling interval therefore has a genuine role, and it is bounded on both sides. It must be short enough to resolve the \emph{fastest} mode, since information about $\lambda_-$ is lost once $e^{\lambda_-\Delta t}$ is indistinguishable from zero, and long enough that $\Phi$ is not numerically close to the identity. Here $|\lambda_-|\Delta t=0.05$ and $|\lambda_+|\Delta t=0.01$, comfortably inside that window. Its adequacy is checked directly rather than assumed. On stationary runs at fixed parameters, averaged over eight replicates of length $2\times10^{4}$, the estimator returns $\widehat\Delta_{\rm RP}$ within $1.2\%$ of the true value $\alpha-1=0.5$ at every $\kappa\in\{1,3,10,30,100\}$, with individual replicates falling in $[0.487,0.513]$; the eigenvector conditioning is recovered to better than $1\%$ over the same range. The estimator thus resolves a gap of $0.5$ reliably, which is the accuracy the ramped experiments require. The Dirichlet form is obtained from the same fit through the symmetrized Lyapunov identity \eqref{eq:dirichlet_lyapunov_estimator} below, and the index from Eq.~\eqref{eq:nonnormality_index}.

\emph{What the two indicator families report.} The classical pair does not separate the experiments, and points the wrong way. Along the $\kappa$-ramp the running variance rises by a factor $9.8\times10^{2}$ and the lag-one autocorrelation climbs from $0.991$ to $0.9997$; along the $\alpha$-ramp the same quantities rise by a factor $8.7$ and from $0.983$ to $0.9981$ respectively (Fig.~\ref{fig:nonnormal}(d)). Both records therefore display the textbook signature. Worse, the record in which nothing whatever happens to the spectrum displays it two orders of magnitude more strongly, so that ranking the two by variance, or by autocorrelation, inverts the correct answer.

The RP diagnostics separate them cleanly (Fig.~\ref{fig:nonnormal}(e,f)). The estimated gap moves from $0.480$ to $0.487$ along the $\kappa$-ramp, i.e.~it stays flat at the true value $\alpha-1=0.5$, and from $0.497$ to $0.048$ along the $\alpha$-ramp, tracking the true closure $0.5\to0.01$. The index rises from $\widehat{\mathcal N}\simeq1.1$ to $\widehat{\mathcal N}\simeq2.3\times10^{2}$ along the $\kappa$-ramp, crossing the reversible bound within the first fifth of the record, and stays at $\widehat{\mathcal N}\simeq0.5$ throughout the $\alpha$-ramp without ever approaching it. The eigenvector conditioning recovered from the same fit rises from $2.1$ to $89$ along the $\kappa$-ramp, against a true $1\to100$, and remains at $1.0$ along the $\alpha$-ramp. Read together, the gap says whether anything is happening to the spectrum and the index says whether the amplification is instead being manufactured by the residues.

\emph{A caveat on the scalar route.} If only a single scalar channel is recorded, $\Delta_{\rm RP}$ can still be read off the autocovariance tail and $\mathcal E$ from the quadratic variation, and the same conclusions follow with more scatter. The quadratic-variation estimate does, however, require a sampling interval fine enough to resolve the non-normal transient, $\Delta t\ll2\delta/(\kappa^2{\rm Var}(n))$; when this fails the Dirichlet form is overestimated and $\widehat{\mathcal N}$ is biased \emph{downward}. The estimator is therefore conservative --- coarse sampling can hide non-normality but cannot manufacture it, so a detection $\widehat{\mathcal N}>1$ remains trustworthy. The bias is avoided altogether by estimating $\mathcal E$ from the fitted generator through the symmetrized Lyapunov identity
\be
\mathcal E(f,f)
=
-\tfrac12\,c^{T}\!\left(\widehat A\widehat\Sigma+\widehat\Sigma\widehat A^{T}\right)c ,
\label{eq:dirichlet_lyapunov_estimator}
\ee
which is exact for a linear diffusion and much better conditioned than the one-sided form $-c^{T}\widehat A\widehat\Sigma c$; this is the route used in Fig.~\ref{fig:nonnormal}(e,f). A systematic study of how the two routes degrade with sampling interval and record length is left for future work.

\subsection{What the counterexample establishes}
\label{subsec:nonnormal_lesson}
The example is deliberately the simplest possible: linear, two-dimensional, Gaussian, with a spectrum that is constant by construction. That is what makes it decisive. It exhibits, in closed form, a system in which variance, autocorrelation, integrated autocorrelation and low-frequency power all diverge while no RP resonance moves at all. No amount of care in estimating those four quantities could distinguish it from an approach to a fold, because the difference is not in them.

Three conclusions follow, and they are the three claims of this paper in miniature:
\begin{enumerate}
\item Small denominators permit large diagnostics but do not produce them, and conversely large diagnostics do not imply small denominators;
\item An early-warning signal is a property of the triple \eqref{eq:EWS_triple} and not of the dynamics alone, so that a residue-selected observable can be constructed which is exactly blind to a signal that dominates a generic one;
\item The classical indicators are inequivalent contractions of one spectral object, ordered here by how much of the RP structure they retain: variance retains none and fails outright, rate-based indicators retain the denominator and succeed once the observable is chosen, and the index \eqref{eq:nonnormality_index} retains enough of the residue ladder to certify the mechanism.
\end{enumerate}

We note finally that this is the non-metastable face of the distinction developed in Secs.~\ref{Sec_EWS_Metastable}--\ref{sec:1D_fold_example}. There, a warning could be an escape-risk signal rather than a loss of local recovery; here, it can be a residue-amplification signal rather than either. In both cases the corrective question is the same, and it is the question the RP formulation is designed to make answerable: not whether the correlations are long, but which block produced them.

\section{Rethinking  Early-Warnings for Metastable Systems}
\label{Sec_EWS_Metastable}

\subsection{Metastability and Phase Space Partitioning}
\label{Sec_metastability_partition}
The RP formulation also clarifies a second ambiguity of classical early-warning signals. In a small-noise system with several coexisting attracting sets, slow correlation decay may have two different origins: local critical slowing down inside the neighborhood of a single attracting set, or slow noise-induced exchange of probability mass between metastable regions.  This is the setting in which B- and N-tipping coexist and may interfere. Both mechanisms can increase lag-one autocorrelation, enhance low-frequency power, and produce long memory. They are nevertheless different RP objects and should not be diagnosed in the same way.

Let $K_\epsilon$ be the Kolmogorov generator of a small-noise hypoelliptic diffusion with smooth invariant measure $\mu_\epsilon$. Suppose that, over the parameter range of interest, the deterministic skeleton has attracting sets
\be
\mathcal A_1^\epsilon,\ldots,\mathcal A_M^\epsilon .
\ee
We shall not identify the metastable sets used below with deterministic basins in a literal topological sense. In nonlinear or chaotic systems, deterministic basin boundaries may be nonsmooth, folded, or fractal \cite{Grebogi1983,Kantz1985,LucariniBodai2017,BodaiLucarini2020}. Such objects are not appropriate boundaries for a local spectral problem. Instead, we use noise-regularized metastable neighborhoods.

Choose smooth core neighborhoods
\be
\mathcal U_a^\epsilon\supset \mathcal A_a^\epsilon,
\qquad
a=1,\ldots,M,
\label{eq:metastable_core_sets}
\ee
with pairwise disjoint closures. These are small smooth target sets around the attracting sets, not basin boundaries. Let
\be
\tau_a^{\epsilon,{\rm hit}} = \inf\{t\ge0:X_t^\epsilon\in \mathcal U_a^\epsilon\}
\label{eq:hitting_time_core}
\ee
be the first hitting time of the $a$th core. The committor to the $a$th core is
\be
q_a^\epsilon(x) = \mathbb P_x
\left(
\tau_a^{\epsilon,{\rm hit}}
<
\min_{b\neq a}\tau_b^{\epsilon,{\rm hit}}
\right).
\label{eq:committor_definition_metastable}
\ee
Equivalently, on the complement of the cores,
\be
K_\epsilon q_a^\epsilon=0,
\qquad
q_a^\epsilon|_{\partial\mathcal U_b^\epsilon}=\delta_{ab},
\label{eq:committor_boundary_value_problem}
\ee
with $q_a^\epsilon=1$ inside $\mathcal U_a^\epsilon$ and $q_a^\epsilon=0$ on the competing cores. Thus $q_a^\epsilon(x)$ is the probability that the noisy trajectory reaches the $a$th metastable core before any competing core. Under the hypoelliptic regularization assumptions used throughout the paper, $q_a^\epsilon$ is smooth in the accessible interior away from the target boundaries, modulo the usual boundary regularity assumptions.

We then define the metastable neighborhood $\mathcal W_a^\epsilon$ as a regular superlevel set of the committor,
\be
\mathcal W_a^\epsilon = \left\{
x:q_a^\epsilon(x)>1-\eta
\right\},
\qquad
0<\eta\ll1,
\label{eq:committor_metastable_neighborhood}
\ee
or by a smooth cell obtained from the same committor family. Since $q_a^\epsilon=1$ on $\mathcal U_a^\epsilon$, one has
\be
\mathcal A_a^\epsilon
\subset
\mathcal U_a^\epsilon
\subset
\mathcal W_a^\epsilon .
\label{eq:core_contained_in_committor_region}
\ee
For regular values of the threshold $1-\eta$, $\partial\mathcal W_a^\epsilon$ is a smooth hypersurface. This is the boundary used below. It is a noise-regularized interface adapted to the metastable decomposition, not the possibly fractal deterministic basin boundary.

With this convention, the global invariant measure is smooth, but for small noise it is organized around the regions $\mathcal W_a^\epsilon$, with rare transitions between them.

 By Freidlin--Wentzell theory \cite{Freidlin1984}, the invariant density of Eq.~\eqref{Eq_unperturbed_SDE} behaves, as $\sigma\to0$, like $\rho(x)\asymp Z(x)\exp[-V(x)/\sigma^2]$, where $V$ is the global quasipotential, i.e.~the rate function of the small-noise large deviations, and $Z$ is a subexponential prefactor; the normalization of $V$ is the one used in Eq.~\eqref{eq:Kramers_quasipotential_general} below. The quasipotential has a local minimum on each attractor and a saddle-like structure at the edge states lying on the basin boundaries, and it is through these edge states that rare transitions between the regions $\mathcal W_a^\epsilon$ overwhelmingly occur \cite{Graham1991,Zhou2012,Zhou2016,LucariniBodai2019,LucariniBodai2020,Margazoglou2021}.

The RP spectrum of the full generator then contains, in general, both intrawell relaxation blocks and interwell transition blocks. The purpose of the partitioning operation is to separate these two mechanisms.

\subsection{Local Critical Blocks from Killed Dynamics}
Fix a metastable region $\mathcal W_a^\epsilon$ associated with the attracting set $\mathcal A_a^\epsilon$. To study local critical slowing down, one wants the relaxation spectrum inside $\mathcal W_a^\epsilon$, not the escape spectrum out of $\mathcal W_a^\epsilon$. This distinction is essential. The dominant nonzero RP block of the full generator $K_\epsilon$ may be an interwell transition block, especially in a small-noise metastable regime. It then controls residence-time statistics and global decorrelation, but not necessarily local recovery inside a well. The local critical block is obtained only after factoring out exits from $\mathcal W_a^\epsilon$.
 To do so we use the notion of killed process associated with an absorbing boundary which is a standard construction in the theory of Markov processes and probabilistic potential theory
\citep{BlumenthalGetoor1968,Sharpe1988,ColletMartinezSanMartin2013}.

The natural construction is the killed process. Let
\be
\tau_a^\epsilon = \inf\{t\ge0:X_t^\epsilon\notin\mathcal W_a^\epsilon\}
\label{eq:exit_time_local_region}
\ee
be the first exit time from $\mathcal W_a^\epsilon$. The killed semigroup is
\be
P_t^{\epsilon,a,D}f(x) = \mathbb E_x
\left[
f(X_t^\epsilon)
\mathds{1}_{\{t<\tau_a^\epsilon\}}
\right],
\qquad
x\in\mathcal W_a^\epsilon,
\label{eq:killed_semigroup_local}
\ee
where the superscript $D$ indicates the absorbing, or Dirichlet, boundary condition. Its generator is the restriction of $K_\epsilon$ to $\mathcal W_a^\epsilon$ with killing at $\partial\mathcal W_a^\epsilon$; we denote it by $K_{\epsilon,a}^{D}$. This semigroup is sub-Markovian:
\be
P_t^{\epsilon,a,D}\mathds{1}(x) = \mathbb P_x(t<\tau_a^\epsilon)
\le 1.
\label{eq:killed_semigroup_submarkov}
\ee
Thus the killed process follows the true dynamics until exit, but it is not conservative. It does not possess an invariant probability measure on $\mathcal W_a^\epsilon$ in the usual sense, because probability mass is continuously lost through the boundary.

Under the standard spectral assumptions for a metastable killed diffusion, the leading Dirichlet eigenvalue is simple and real. We write it as
\bea
&K_{\epsilon,a}^{D}h_{\epsilon,a} = \lambda_{0}^{D,a}(\epsilon)h_{\epsilon,a},
\qquad
h_{\epsilon,a}>0
\quad\text{in }\mathcal W_a^\epsilon,\\
&h_{\epsilon,a}=0
\quad\text{on }\partial\mathcal W_a^\epsilon,
\label{eq:principal_dirichlet_eigenfunction}
\eea
with $\lambda_{0}^{D,a}(\epsilon)<0$. Its modulus is the exponential escape rate from the region.
 In the weak-noise limit, $h_{\epsilon,a}$ is asymptotically proportional to the probability of relaxing back to the core $\mathcal U_a^\epsilon$ before exiting $\mathcal W_a^\epsilon$. It is therefore nearly constant in the bulk of the well, like the committor $q_a^\epsilon$, and differs from it only inside the $O(\sigma)$ boundary layer where $h_{\epsilon,a}$ must vanish while $q_a^\epsilon$ equals $1-\eta$; the two functions cannot be identified globally.

More precisely, if $\nu_{\epsilon,a}$ is the corresponding left eigenmeasure,
\be
\nu_{\epsilon,a}P_t^{\epsilon,a,D} = e^{\lambda_{0}^{D,a}(\epsilon)t}\nu_{\epsilon,a},
\label{eq:left_principal_eigenmeasure}
\ee
then $\nu_{\epsilon,a}$ is a quasi-stationary distribution: for every measurable $E\subset\mathcal W_a^\epsilon$,
\be
\nu_{\epsilon,a}(E) = \mathbb P_{\nu_{\epsilon,a}}
\left(
X_t^\epsilon\in E
\mid
t<\tau_a^\epsilon
\right),
\label{eq:quasi_stationary_distribution_def}
\ee
and the survival probability decays exponentially,
\be
\mathbb P_{\nu_{\epsilon,a}}
(t<\tau_a^\epsilon) = e^{\lambda_{0}^{D,a}(\epsilon)t}.
\label{eq:qsd_survival_decay}
\ee
Thus the principal killed eigenvalue describes survival, or escape from $\mathcal W_a^\epsilon$,
not relaxation inside $\mathcal W_a^\epsilon$.  Thus $\kappa_a^\epsilon=-\lambda_0^{D,a}(\epsilon)$ is the escape rate, and under the quasi-stationary ensemble $\nu_{\epsilon,a}$ its inverse is the mean exit time \cite{LePeutrec2024,Banerjee2026,bovier2004metastability,meleard2012quasi}.

To remove this survival decay one passes to the process conditioned never to exit, the $Q$-process. It is defined as the long-time conditioning limit
\be
\mathbb P_x^{Q}
\left(
X_t^\epsilon\in E
\right) = \lim_{T\to\infty}
\mathbb P_x
\left(
X_t^\epsilon\in E, t<\tau_a^\epsilon
\mid
T<\tau_a^\epsilon
\right),
\label{eq:Q_process_conditioning_limit}
\ee
whenever the limit exists. In terms of the principal Dirichlet eigenfunction,  the conditioned semigroup (governing the conditioned dynamics) is   obtained through the classical Doob $h$-transform
\citep{Doob1957conditional,Doob1984,Pinsky1995,ColletMartinezSanMartin2013}:
\be
P_t^{\epsilon,a,Q}f(x) = e^{-\lambda_{0}^{D,a}(\epsilon)t}
h_{\epsilon,a}(x)^{-1}
P_t^{\epsilon,a,D}
\left(
h_{\epsilon,a}f
\right)(x).
\label{eq:Q_process_doob_transform}
\ee
This formula shows explicitly how the exponential survival factor is removed. Indeed,
\be
P_t^{\epsilon,a,Q}\mathds{1}(x) = e^{-\lambda_{0}^{D,a}(\epsilon)t}
h_{\epsilon,a}(x)^{-1}
P_t^{\epsilon,a,D}h_{\epsilon,a}(x) = 1.
\label{eq:Q_process_conservative}
\ee
Thus the $Q$-process is conservative on $\mathcal W_a^\epsilon$. Its generator is
\be
K_{\epsilon,a}^{Q}f = h_{\epsilon,a}^{-1}
K_{\epsilon,a}^{D}
\left(
h_{\epsilon,a}f
\right) - \lambda_{0}^{D,a}(\epsilon)f .
\label{eq:Q_process_generator}
\ee
For a diffusion generator $K_\epsilon=b_\epsilon\cdot\nabla+\frac{\sigma^2}{2}a_\epsilon:\nabla^2$, this becomes, formally in the interior,
\be
K_{\epsilon,a}^{Q}f = K_\epsilon f
+
\sigma^2a_\epsilon\nabla\log h_{\epsilon,a}\cdot\nabla f .
\label{eq:Q_process_diffusion_generator}
\ee
The extra drift $\sigma^2a_\epsilon\nabla\log h_{\epsilon,a}$ repels the conditioned process from the absorbing boundary and encodes the conditioning on long survival.

The invariant probability measure of the $Q$-process is obtained from the left and right principal Dirichlet eigenvectors. If $\nu_{\epsilon,a}$ is normalized so that $\nu_{\epsilon,a}(h_{\epsilon,a})<\infty$, then
\be
\d\mu_{\epsilon,a}^{Q} = \frac{
h_{\epsilon,a}\d\nu_{\epsilon,a}
}{
\nu_{\epsilon,a}(h_{\epsilon,a})
}
\label{eq:Q_process_invariant_measure}
\ee
is invariant under $P_t^{\epsilon,a,Q}$. Hence the $Q$-process provides a genuine conservative local dynamics and a local invariant measure on $\mathcal W_a^\epsilon$. Far from the boundaries, the invariant probability measure of the $Q$-process is approximately proportional to the global invariant measure, because $h^\epsilon_a$ is approximately constant, whilst it vanishes at the boundaries of $\mathcal{W}_a^\epsilon$ and outside $\mathcal{W}_a^\epsilon$. This measure can be practically constructed by running a large number of simulations initialised within $\mathcal{W}^\epsilon_a$ and, each time an ensemble member escapes through the basin boundaries, another ensemble member is generated by cloning one of the (randomly chosen) remaining ones.\ This is the Fleming--Viot particle scheme, whose empirical measure converges to the quasi-stationary distribution $\nu_{\epsilon,a}$ and, after $h_{\epsilon,a}$-reweighting, to $\mu_{\epsilon,a}^{Q}$ \citep{meleard2012quasi,ColletMartinezSanMartin2013}; it is the ensemble counterpart of the single-trajectory Doob simulation of Sec.~\ref{subsec:Q_process_Girsanov_Doob}.

The relation between the killed spectrum and the conditioned spectrum is explicit. If
\be
K_{\epsilon,a}^{D}\phi_j^{D} = \lambda_j^{D,a}(\epsilon)\phi_j^{D},
\label{eq:killed_eigenvalue_problem}
\ee
then
\be
K_{\epsilon,a}^{Q}
\left(
\frac{\phi_j^{D}}{h_{\epsilon,a}}
\right) = \left(
\lambda_j^{D,a}(\epsilon) - \lambda_0^{D,a}(\epsilon)
\right)
\frac{\phi_j^{D}}{h_{\epsilon,a}}.
\label{eq:Q_process_spectral_shift}
\ee
The principal killed eigenvalue is therefore shifted to zero and becomes the invariant mode of the $Q$-process. The nonzero spectrum of $K_{\epsilon,a}^{Q}$ describes relaxation conditional on remaining in $\mathcal W_a^\epsilon$. In particular, the local relaxation gap is not the escape rate $-\lambda_0^{D,a}(\epsilon)$, but the gap above the principal killed eigenvalue , where $\lambda_1^{D,a}(\epsilon)$ denotes the nonprincipal Dirichlet eigenvalue of largest real part:
\be
\Delta_{\rm loc}^{(a)}(\epsilon) = -\Re
\left(
\lambda_1^{D,a}(\epsilon) - \lambda_0^{D,a}(\epsilon)
\right),
\label{eq:local_gap_killed_Q_process}
\ee
or, equivalently, the spectral gap of the conservative conditioned generator $K_{\epsilon,a}^{Q}$ after removing constants.

This is the precise sense in which the local spectral problem differs from the full dynamics. The killed process follows the true dynamics until exit, but it is nonconservative and its leading eigenvalue measures survival. The $Q$-process is an auxiliary conservative dynamics obtained by conditioning on non-exit for arbitrarily long times; it removes the survival decay and isolates intrawell relaxation. The resulting local critical block is therefore not the dominant nonzero RP block of the full generator $K_\epsilon$. It is the dominant nonzero RP block of the conditioned local generator
\be
K_{\epsilon,a}^{\rm loc}
\equiv
K_{\epsilon,a}^{Q},
\qquad
P_t^{\epsilon,a,{\rm loc}}
\equiv
P_t^{\epsilon,a,Q},
\qquad
\mu_{\epsilon,a}
\equiv
\mu_{\epsilon,a}^{Q}.
\label{eq:local_generator_Q_choice}
\ee
A reflected no-flux process could be used as a computational local closure, but it changes the trajectories at the boundary; the killed/$Q$-process route is the canonical construction when the goal is to condition the original dynamics on long residence in $\mathcal W_a^\epsilon$.

The local spectral problem is now posed in $L^2_{\mu_{\epsilon,a}}(\mathcal W_a^\epsilon)$. A \emph{local critical block} is a leading nonzero RP block of $K_{\epsilon,a}^{\rm loc}$,
\bea
&\left(
\lambda_{\rm loc}^{(a)}(\epsilon),
\Pi_{\rm loc}^{(a),\epsilon},
N_{\rm loc}^{(a),\epsilon}
\right),\\
&N_{\rm loc}^{(a),\epsilon} = \left(
K_{\epsilon,a}^{\rm loc} - \lambda_{\rm loc}^{(a)}(\epsilon)I
\right)
\Pi_{\rm loc}^{(a),\epsilon}.
\label{eq:local_critical_RP_block}
\eea
Its nonzero eigenfunctions are localized in $\mathcal W_a^\epsilon$ and are centered with respect to $\mu_{\epsilon,a}$. Hence this block describes redistribution of probability inside the metastable region, not transfer of probability mass between different regions.

The block becomes critical when the local RP spectral gap closes. For the local conservative process, the eigenvalue $0$ corresponds to constant eigenfunctions and to the local invariant measure $\mu_{\epsilon,a}$. The local critical block is the dominant nonzero block after this invariant component has been removed. Thus, for every precritical value of $\epsilon$,
\be
\Re\lambda_{\rm loc}^{(a)}(\epsilon)<0.
\ee
Criticality means that the local relaxation gap
\be
\Delta_{\rm loc}^{(a)}(\epsilon) = -\Re\lambda_{\rm loc}^{(a)}(\epsilon)
\label{eq:local_gap_definition_precise}
\ee
tends to zero. Equivalently, $\Re\lambda_{\rm loc}^{(a)}(\epsilon)\uparrow0$ should be read as RP-gap closure for the conditioned local Markov generator, not as an eigenvalue-crossing criterion. No universal exponent for this gap is assumed here.

The residue-conditioned EWS criterion has the same structure locally. For an observable $f$, define
\be
f_{\epsilon,a}^c = f-\mu_{\epsilon,a}(f)\mathds{1},
\qquad
B_{\epsilon,a}^{\rm loc} = \partial_\epsilon K_{\epsilon,a}^{\rm loc}.
\label{eq:local_centering_and_perturbation}
\ee
If the local critical block is semisimple, it is visible in the response of the local statistic only if
\be
\Rstyle_{{\rm loc},0}^{(a),f}(\epsilon) = \mu_{\epsilon,a}\left(
B_{\epsilon,a}^{\rm loc}
\Pi_{\rm loc}^{(a),\epsilon}
f_{\epsilon,a}^{c}
\right)
\neq0.
\label{eq:local_semisimple_residue_condition}
\ee
For a nonsemisimple block, the local response-residue ladder is given, for $0\le k\le m_{\rm loc}^{(a)}-1$, as:
\be
\Rstyle_{{\rm loc},k}^{(a),f}(\epsilon) = \mu_{\epsilon,a}\left(
B_{\epsilon,a}^{\rm loc}
\left(N_{\rm loc}^{(a),\epsilon}\right)^k
\Pi_{\rm loc}^{(a),\epsilon}
f_{\epsilon,a}^{c}
\right).
\label{eq:local_response_residue_ladder}
\ee
At least one of these local response residues must be nonzero for the closing local RP gap to be visible in the response of the chosen statistics.
 In practice this means that an observable meant to anticipate tipping should resolve the attempted escapes from $\mathcal{W}^\epsilon_a$; observables that track the position of the edge state relevant for the escape have indeed been found to anticipate transitions markedly better than generic ones \cite{Lohmann2025}.

These response residues should be distinguished from the coefficients controlling local autocorrelations. The local conditioned autocorrelation is
\be
C_{{\rm loc},a}^{f,\epsilon}(t) = \mu_{\epsilon,a}\left(
f_{\epsilon,a}^c
P_t^{\epsilon,a,{\rm loc}}
f_{\epsilon,a}^c
\right).
\label{eq:local_conditioned_correlation}
\ee
Using the local RP expansion of $P_t^{\epsilon,a,{\rm loc}}$, one obtains
\be
C_{{\rm loc},a}^{f,\epsilon}(t) = \sum_j
e^{\lambda_j^{(a)}(\epsilon)t}
\sum_{k=0}^{m_j^{(a)}-1}
\frac{t^k}{k!}
\mathcal A_{j,k}^{(a),f}(\epsilon)
+
\text{residual},
\label{eq:local_conditioned_correlation_RP_expansion}
\ee
where, for the local critical block in particular,
\be
\mathcal A_{{\rm loc},k}^{(a),f}(\epsilon) = \mu_{\epsilon,a}\left(
f_{\epsilon,a}^c
\left(N_{\rm loc}^{(a),\epsilon}\right)^k
\Pi_{\rm loc}^{(a),\epsilon}
f_{\epsilon,a}^c
\right).
\label{eq:local_correlation_residue_defined}
\ee
 The associated \emph{local conditioned power spectral density} is the Fourier transform of this conditioned correlation,
\be
S_{{\rm loc},a}^{f,\epsilon}(\omega)
=
\int_{-\infty}^{+\infty}
e^{-i\omega t}\,
C_{{\rm loc},a}^{f,\epsilon}(|t|)\d t ,
\label{eq:local_conditioned_PSD}
\ee
which, for a real semisimple local critical block, reduces to the Lorentzian
\be
S_{{\rm loc},a}^{f,\epsilon}(\omega)
\simeq
\mathcal A_{{\rm loc},0}^{(a),f}(\epsilon)\,
\frac{2\Delta_{\rm loc}^{(a)}(\epsilon)}
{\left(\Delta_{\rm loc}^{(a)}(\epsilon)\right)^2+\omega^2}.
\label{eq:local_conditioned_PSD_lorentzian}
\ee
We stress that $S_{{\rm loc},a}^{f,\epsilon}$ is \emph{not} the power spectrum of the raw signal: it is the spectrum of the $Q$-process, i.e.~of trajectories conditioned on not having escaped. Escape events, which contribute the interwell part of the raw spectrum, have been removed by construction. This is what makes Eq.~\eqref{eq:local_conditioned_PSD_lorentzian} a clean diagnostic of local recovery, and it is also why it cannot be read off a raw time series without the conditioning step of Sec.~\ref{sec:Girsanov_Doob}.

Thus $\mathcal A_{{\rm loc},k}^{(a),f}$ and $\Rstyle_{{\rm loc},k}^{(a),f}$ involve the same local RP block but different scalar contractions. The correlation residue uses the functional
\be
g\mapsto
\mu_{\epsilon,a}(f_{\epsilon,a}^c g),
\ee
whereas the response residue uses
\be
g\mapsto
\mu_{\epsilon,a}(B_{\epsilon,a}^{\rm loc}g).
\ee
A local critical block may therefore be visible in autocorrelation but weakly visible in parameter response, or conversely.

This is the precise sense in which partitioning reveals local critical slowing down. The full generator $K_\epsilon$ answers a global question: which RP block controls relaxation of the invariant dynamics on the whole state space? In a metastable regime, that answer may be an interwell transition block. The local generator $K_{\epsilon,a}^{\rm loc}$ answers a different question: conditional on long residence in the metastable region associated with $\mathcal A_a^\epsilon$, which RP block controls recovery inside that region? For local tipping of $\mathcal A_a^\epsilon$, the second question is the relevant one.

Thus the local RP early-warning statement is structural: construct a smooth noise-regularized metastable region $\mathcal W_a^\epsilon$, kill the original dynamics at $\partial\mathcal W_a^\epsilon$, pass to the associated $Q$-process to factor out survival, identify the local critical block of $K_{\epsilon,a}^{\rm loc}$, track the gap $\Delta_{\rm loc}^{(a)}(\epsilon)$, and test through which residue it is visible. Autocorrelation warnings require nonzero local correlation residues $\mathcal A_{{\rm loc},k}^{(a),f}$. Susceptibility warnings require nonzero local response residues $\Rstyle_{{\rm loc},k}^{(a),f}(\epsilon)$.

The killed process and the associated absorbing semigroup are classical objects in the theory of Markov processes and probabilistic potential theory
\citep{BlumenthalGetoor1968,Sharpe1988}. In the present metastable setting, they are used in the standard quasi-stationary framework: the killed semigroup encodes survival before exit, its principal Dirichlet eigenvalue gives the exponential escape clock, and the corresponding Doob $h$-transform yields the $Q$-process, i.e. the process conditioned on never being absorbed
\citep{Doob1957conditional,Doob1984,Pinsky1995,ColletMartinezSanMartin2013,ChampagnatVillemonais2016}.

\subsection{Transition Blocks}
\label{subsec:transition_blocks}
Transition blocks are global RP blocks of the full generator $K_\epsilon$. They do not describe relaxation inside a single metastable region. They describe slow exchange of probability mass between metastable regions. This is why they must be distinguished from the local critical blocks defined above. The local construction used the killed process and then passed to the $Q$-process in order to remove survival decay and isolate intrawell relaxation. The transition construction uses the same killed local problems in the complementary way: it keeps the principal escape clocks and combines them with committor-based destination probabilities to build an effective jump process on metastable labels.

Assume that the smooth metastable neighborhoods $\mathcal W_1^\epsilon,\ldots,\mathcal W_M^\epsilon$ are pairwise disjoint and carry most of the invariant measure,
\begin{equation}
\mu_\epsilon\left( \bigcup_{a=1}^M\mathcal W_a^\epsilon \right) \simeq 1.
\label{eq:metastable_mass_condition_precise}
\end{equation}
Let
\begin{equation}
\pi_a^\epsilon = \mu_\epsilon(\mathcal W_a^\epsilon),
\label{eq:well_weights_transition}
\end{equation}
and define the projection onto metastable labels by
\begin{equation}
(\mathcal P_\epsilon f)_a = \frac{1}{\pi_a^\epsilon} \int_{\mathcal W_a^\epsilon} f \, \mathrm{d}\mu_\epsilon, \qquad \mathcal P_\epsilon:L^2_{\mu_\epsilon}\to\mathbb R^M .
\label{eq:coarse_projection_to_vector}
\end{equation}
Conversely, the lifting map $\mathcal I_\epsilon:\mathbb R^M\to L^2_{\mu_\epsilon}$ sends a vector $h=(h_1,\ldots,h_M)$ to the piecewise constant observable
\begin{equation}
(\mathcal I_\epsilon h)(x) = h_a, \qquad x\in\mathcal W_a^\epsilon .
\label{eq:coarse_injection_to_observables}
\end{equation}
The conditional expectation onto metastable labels is therefore
\begin{equation}
\Cstyle_\epsilon = \mathcal I_\epsilon\mathcal P_\epsilon ,
\label{eq:coarse_projection_factorization}
\end{equation}
that is,
\begin{equation}
\Cstyle_\epsilon f = \sum_{a=1}^M \left( \frac{1}{\pi_a^\epsilon} \int_{\mathcal W_a^\epsilon}f \, \mathrm{d}\mu_\epsilon \right) \mathds{1}_{\mathcal W_a^\epsilon}.
\label{eq:coarse_well_projection}
\end{equation}
It replaces an observable by its conditional mean on each metastable set.
A global RP block $(\lambda_j(\epsilon),\Pi_j^\epsilon,N_j^\epsilon)$ is an \emph{interwell transition block} when its generalized eigenspace is almost contained in the range of $\Cstyle_\epsilon$, namely in the space of observables that are approximately constant on each metastable set:
\begin{equation}
\left\| (I-\Cstyle_\epsilon)\Pi_j^\epsilon \right\|_{L^2_{\mu_\epsilon}\to L^2_{\mu_\epsilon}} \ll \left\| \Pi_j^\epsilon \right\|_{L^2_{\mu_\epsilon}\to L^2_{\mu_\epsilon}}.
\label{eq:transition_block_projection_criterion}
\end{equation}
For a simple eigenvalue, this reduces to the statement that the corresponding right Kolmogorov eigenfunction is nearly piecewise constant on the wells:
\begin{equation}
\left\| (I-\Cstyle_\epsilon)\psi_j^\epsilon \right\|_{L^2_{\mu_\epsilon}} \ll \left\| \psi_j^\epsilon \right\|_{L^2_{\mu_\epsilon}}.
\label{eq:simple_transition_eigenfunction_criterion}
\end{equation}
Such a block does not represent local flattening of the drift inside a well. It represents slow exchange of probability mass between wells. To make this statement operational, we now construct the induced dynamics on metastable labels. The construction uses the killed local problems in each $\mathcal W_a^\epsilon$ to extract two quantities: the rate at which a locally equilibrated trajectory exits the region, and the probability that this exit is ultimately committed to each competing metastable core.

For each metastable region, let
\begin{equation}
\tau_a^\epsilon = \inf\{t\ge0:X_t^\epsilon\notin\mathcal W_a^\epsilon\}
\label{eq:transition_exit_time}
\end{equation}
and let $K_{\epsilon,a}^{D}$ be the generator of the killed process in $\mathcal W_a^\epsilon$. Denote its principal eigenpair by
\begin{align}
K_{\epsilon,a}^{D}h_{\epsilon,a} &= \lambda_0^{D,a}(\epsilon)h_{\epsilon,a}, \qquad h_{\epsilon,a}>0 \quad\text{in }\mathcal W_a^\epsilon, \nonumber \\
h_{\epsilon,a} &= 0 \quad\text{on }\partial\mathcal W_a^\epsilon, \qquad \lambda_0^{D,a}(\epsilon)<0.
\label{eq:transition_principal_killed_eigenpair}
\end{align}
The principal killed eigenvalue gives the escape clock
\begin{equation}
\kappa_a^\epsilon = -\lambda_0^{D,a}(\epsilon).
\label{eq:local_escape_rate_from_killed}
\end{equation}
Let $\nu_{\epsilon,a}$ be the corresponding quasi-stationary distribution,
\begin{equation}
\nu_{\epsilon,a}P_t^{\epsilon,a,D} = e^{-\kappa_a^\epsilon t}\nu_{\epsilon,a}.
\label{eq:transition_qsd_survival}
\end{equation}
Then, when the process is initialized according to $\nu_{\epsilon,a}$,
\begin{equation}
\mathbb P_{\nu_{\epsilon,a}}(t<\tau_a^\epsilon) = e^{-\kappa_a^\epsilon t}.
\label{eq:qsd_escape_exponential}
\end{equation}
Thus $\kappa_a^\epsilon$ is the rate at which a locally equilibrated trajectory leaves the $a$th metastable region.
To assign exits to destination wells, let $\omega_{\epsilon,a}$ be the exit distribution from $\mathcal W_a^\epsilon$ under the quasi-stationary ensemble:
\begin{equation}
\omega_{\epsilon,a}(E) = \mathbb P_{\nu_{\epsilon,a}} \left( X_{\tau_a^\epsilon}^\epsilon\in E \right), \qquad E\subset\partial\mathcal W_a^\epsilon .
\label{eq:qsd_exit_distribution}
\end{equation}
Let $q_b^\epsilon$ be the committor to the $b$th core. The probability that an exit from $\mathcal W_a^\epsilon$ is ultimately committed to the $b$th metastable core is
\begin{equation}
p_{ab}^\epsilon = \int_{\partial\mathcal W_a^\epsilon} q_b^\epsilon(y) \, \mathrm{d}\omega_{\epsilon,a}(y), \qquad b\neq a.
\label{eq:exit_destination_probability_committor}
\end{equation}
This distinction is essential: leaving the high-committor neighborhood $\mathcal W_a^\epsilon$ does not necessarily mean that the trajectory has completed a transition to another well. The committor in \eqref{eq:exit_destination_probability_committor} assigns each exit point to its eventual destination probability. The effective transition rates are therefore
\begin{equation}
q_{ab}(\epsilon) = \kappa_a^\epsilon p_{ab}^\epsilon, \qquad b\neq a,
\label{eq:transition_rate_from_killed_committor}
\end{equation}
with
\begin{equation}
q_{aa}(\epsilon) = -\sum_{b\neq a}q_{ab}(\epsilon).
\label{eq:transition_generator_diagonal}
\end{equation}
The finite-state backward generator $\mathsf{Q}_\epsilon=(q_{ab}(\epsilon))$ acts on label observables $h=(h_1,\ldots,h_M)$ by
\begin{equation}
(\mathsf{Q}_\epsilon h)_a = \sum_{b\neq a} q_{ab}(\epsilon)(h_b-h_a).
\label{eq:finite_state_backward_generator_precise}
\end{equation}

The finite-state generator $\mathsf{Q}_\epsilon$ should therefore not be understood as an arbitrary low-dimensional fit of the full Markov semigroup. It is the generator induced on metastable labels by two operations: first, rapid local equilibration inside each $\mathcal W_a^\epsilon$; second, committor-based assignment of each rare exit event to its eventual destination metastable core. The killed local problem supplies the exit clock $\kappa_a^\epsilon=-\lambda_0^{D,a}(\epsilon)$, while the committor-weighted exit distribution supplies the destination probabilities $p_{ab}^\epsilon$. Their product
\begin{equation}
q_{ab}(\epsilon) = \kappa_a^\epsilon p_{ab}^\epsilon
\label{eq:q_ab_clock_destination_short}
\end{equation}
is the effective transition rate from label $a$ to label $b$.
Equivalently, if $h=(h_1,\ldots,h_M)$ is an observable depending only on the metastable label, then $(\mathsf{Q}_\epsilon h)_a$ is the infinitesimal expected change of $h$ for a trajectory that has locally equilibrated in $\mathcal W_a^\epsilon$:
\begin{equation}
(\mathsf{Q}_\epsilon h)_a = \lim_{t\downarrow0} \frac{ \mathbb E_{\nu_{\epsilon,a}} \left[ h_{\ell(X_t)} \right] - h_a }{t},
\label{eq:Q_generator_label_limit}
\end{equation}
where $\ell(X_t)$ denotes the metastable label assigned after exit and committor-based commitment.

Thus $\mathsf{Q}_\epsilon$ is the reduced generator induced by killed escape rates and destination probabilities, not an ad hoc Markov matrix. It is the backward generator of the coarse jump process produced by rare exits from quasi-stationary local ensembles.

The reduction is valid in the metastable regime where conditioned intrawell relaxation is fast compared with successful interwell transitions. In the killed notation, the local relaxation spectrum is obtained by subtracting the principal killed eigenvalue $\lambda_0^{D,a}(\epsilon)$ from the rest of the Dirichlet spectrum. Thus, if $\lambda_1^{D,a}(\epsilon)$ denotes the nonprincipal killed eigenvalue with largest real part after $\lambda_0^{D,a}(\epsilon)$, the conditioned local gap is
\begin{equation}
\Delta_{\mathrm{loc}}^{(a)}(\epsilon) = -\Re\left( \lambda_1^{D,a}(\epsilon) - \lambda_0^{D,a}(\epsilon) \right).
\label{eq:local_Q_gap_recalled_for_transition}
\end{equation}
Equivalently, in the presence of several competing nonprincipal eigenvalues, one may write
\begin{equation}
\Delta_{\mathrm{loc}}^{(a)}(\epsilon) = -\max_{j\ge1} \Re\left( \lambda_j^{D,a}(\epsilon) - \lambda_0^{D,a}(\epsilon) \right).
\label{eq:local_Q_gap_general}
\end{equation}
This is the spectral gap of the conditioned local generator $K_{\epsilon,a}^{Q}$ after its invariant eigenvalue at zero has been removed.
By contrast, the total successful transition rate out of the $a$th metastable label is
\begin{equation}
\kappa_{\mathrm{tr},a}^\epsilon = \sum_{b\neq a}q_{ab}(\epsilon) = \kappa_a^\epsilon \sum_{b\neq a}p_{ab}^\epsilon .
\label{eq:successful_transition_rate}
\end{equation}
The finite-state reduction is appropriate when
\begin{equation}
\max_a\kappa_{\mathrm{tr},a}^\epsilon \ll \min_a\Delta_{\mathrm{loc}}^{(a)}(\epsilon),
\label{eq:metastable_spectral_separation_killed}
\end{equation}
together with the assumption that the time spent in transition channels between metastable neighborhoods is short compared with residence times. In words, trajectories first relax to their quasi-stationary local ensemble, then wait a long time before a successful transition occurs. This separation is what allows the continuous dynamics to be replaced, on metastable timescales and on label observables, by the finite-state generator $\mathsf{Q}_\epsilon$.

On the corresponding metastable time window, the projected semigroup is approximated by the finite-state semigroup:
\begin{equation}
\mathcal P_\epsilon P_t^\epsilon\mathcal I_\epsilon \simeq e^{t\mathsf{Q}_\epsilon}.
\label{eq:coarse_semigroup_finite_state_approx_precise}
\end{equation}
Equivalently, on observables that are nearly constant on the metastable sets,
\begin{equation}
\Cstyle_\epsilon P_t^\epsilon \Cstyle_\epsilon \simeq \mathcal I_\epsilon e^{t\mathsf{Q}_\epsilon}\mathcal P_\epsilon .
\label{eq:coarse_semigroup_finite_state_approx}
\end{equation}
This is the precise sense in which the high-dimensional Markov semigroup is reduced to a finite-state transition process. The reduction is a semigroup approximation on the metastable subspace, not a pointwise replacement of the full generator.
Here $\mathsf{Q}_\epsilon$ is a continuous-time Markov generator , it is not a Markov transition matrix. Its off-diagonal entries are nonnegative transition rates, its rows sum to zero, and its eigenvalues satisfy
\begin{equation}
\Re\zeta_p^\epsilon\le0, \qquad \zeta_0^\epsilon=0.
\label{eq:Q_generator_spectral_location}
\end{equation}
Thus the eigenvalues of $\mathsf{Q}_\epsilon$ live in the left half-plane, like Kolmogorov-generator eigenvalues. It is the finite-time transition matrix $e^{t\mathsf{Q}_\epsilon}$ whose eigenvalues $e^{t\zeta_p^\epsilon}$ lie in the unit disk.
Let
\begin{equation}
\mathsf{Q}_\epsilon \vq_p^\epsilon = \zeta_p^\epsilon \vq_p^\epsilon, \qquad p=0,\ldots,M-1,
\label{eq:Q_eigenvector_transition}
\end{equation}
with $\zeta_0^\epsilon=0$ corresponding to stationarity on the metastable labels. Lifting a nonzero eigenvector gives the piecewise constant observable
\begin{equation}
\psi_{\mathrm{tr},p}^{\epsilon,0} = \mathcal I_\epsilon \vq_p^\epsilon, \qquad p=1,\ldots,M-1.
\label{eq:lifted_transition_eigenfunction}
\end{equation}
Then \eqref{eq:coarse_semigroup_finite_state_approx} implies the projected relation
\begin{equation}
\Cstyle_\epsilon P_t^\epsilon \psi_{\mathrm{tr},p}^{\epsilon,0} \simeq e^{\zeta_p^\epsilon t} \psi_{\mathrm{tr},p}^{\epsilon,0},
\label{eq:lifted_transition_mode_projected_semigroup}
\end{equation}
on the metastable time window. More explicitly, the error is measured after projection onto the metastable-label subspace:
\begin{equation}
\left\| \Cstyle_\epsilon P_t^\epsilon \psi_{\mathrm{tr},p}^{\epsilon,0} - e^{\zeta_p^\epsilon t} \psi_{\mathrm{tr},p}^{\epsilon,0} \right\|_{L^2_{\mu_\epsilon}} \ll \left\| \psi_{\mathrm{tr},p}^{\epsilon,0} \right\|_{L^2_{\mu_\epsilon}}.
\label{eq:projected_transition_mode_error}
\end{equation}
This replaces the vague statement that the equality holds ``modulo intrawell components.'' The intrawell components are precisely those removed by $I-\Cstyle_\epsilon$.
The corresponding global RP transition eigenfunction, when it exists as an isolated RP mode, is not exactly the lifted vector $\psi_{\mathrm{tr},p}^{\epsilon,0}$ but a small deformation of it,
\begin{equation}
\psi_{\mathrm{tr},p}^{\epsilon} = \psi_{\mathrm{tr},p}^{\epsilon,0} + \chi_{\mathrm{in},p}^{\epsilon}, \;\; \left\| \chi_{\mathrm{in},p}^{\epsilon} \right\|_{L^2_{\mu_\epsilon}} \ll \left\| \psi_{\mathrm{tr},p}^{\epsilon,0} \right\|_{L^2_{\mu_\epsilon}}.
\label{eq:transition_mode_inwell_correction}
\end{equation}
Equivalently,
\begin{equation}
\left\| (I-\Cstyle_\epsilon) \psi_{\mathrm{tr},p}^{\epsilon} \right\|_{L^2_{\mu_\epsilon}} \ll \left\| \psi_{\mathrm{tr},p}^{\epsilon} \right\|_{L^2_{\mu_\epsilon}}.
\label{eq:transition_eigenfunction_piecewise_constant}
\end{equation}
This is the defining signature of an interwell transition block.
Therefore the eigenvalues $\zeta_p^\epsilon$ of the continuous-time reduced generator approximate the global RP resonances associated with nearly piecewise constant eigenfunctions:
\begin{equation}
\lambda_{\mathrm{tr}}^{(p)}(\epsilon) \simeq \zeta_p^\epsilon, \qquad p=1,\ldots,M-1.
\label{eq:transition_RP_finite_state}
\end{equation}
Equivalently, at the level of finite-time transfer operators,
\begin{equation}
e^{t\lambda_{\mathrm{tr}}^{(p)}(\epsilon)} \simeq e^{t\zeta_p^\epsilon}.
\label{eq:transition_RP_semigroup_eigenvalues}
\end{equation}
If one works instead with a discrete lag-$\tau$ transition matrix
\begin{equation}
T_{\tau,\epsilon} \simeq \mathcal P_\epsilon P_\tau^\epsilon\mathcal I_\epsilon \simeq e^{\tau\mathsf{Q}_\epsilon},
\label{eq:discrete_lag_transition_matrix}
\end{equation}
and if $\rho_p^\epsilon(\tau)$ denotes a nontrivial eigenvalue of $T_{\tau,\epsilon}$, then the corresponding generator-level resonance is recovered as
\begin{equation}
\lambda_{\mathrm{tr}}^{(p)}(\epsilon) \simeq \frac{1}{\tau} \log \rho_p^\epsilon(\tau),
\label{eq:discrete_to_generator_transition_resonance}
\end{equation}
with the branch chosen according to continuity in $\epsilon$ and the relevant spectral window.
The approximation \eqref{eq:transition_RP_finite_state} is therefore not a claim that the full generator $K_\epsilon$ has been replaced by $\mathsf{Q}_\epsilon$. It is a spectral reduction on the metastable subspace
\begin{equation}
\mathcal M_\epsilon = \operatorname{Ran}\Cstyle_\epsilon = \left\{ f:\ f|_{\mathcal W_a^\epsilon} \text{ is constant for each }a \right\}.
\label{eq:metastable_coarse_subspace}
\end{equation}
Intrawell RP blocks fail this criterion because their eigenfunctions fluctuate inside the wells and are largely removed by $\Cstyle_\epsilon$. Transition blocks satisfy it because their eigenfunctions primarily encode changes of probability mass between metastable labels.

In the small-noise regime, the entries of the reduced generator inherit Freidlin--Wentzell/Kramers asymptotics. With noise amplitude $\sigma$, the successful transition rate from the $a$th metastable region to the $b$th one has the exponential form
\begin{equation}
q_{ab}(\epsilon) \asymp \Gamma_{ab}(\epsilon) \exp\left( -\frac{\Delta\mathcal V_{ab}(\epsilon)}{\sigma^2} \right), \qquad b\neq a.
\label{eq:Kramers_quasipotential_general}
\end{equation}
Here $\Delta\mathcal V_{ab}(\epsilon)$ is the Freidlin--Wentzell quasipotential barrier for a successful transition from the metastable state associated with $\mathcal A_a^\epsilon$ to the one associated with $\mathcal A_b^\epsilon$,
 i.e.~the increment of the quasipotential $V$ of Sec.~\ref{Sec_metastability_partition} between $\mathcal A_a^\epsilon$ and the edge state through which the $a\to b$ transition proceeds \cite{Margazoglou2021}.  More explicitly, if $\mathcal S_\epsilon[\phi]$ denotes the Freidlin--Wentzell action of a path $\phi$, then $\Delta\mathcal V_{ab}(\epsilon)$ is the minimal action among paths that start in the $a$th attracting set and reach the $b$th metastable core before being recommitted to $a$ or to another competing core:
\begin{widetext}
\be
\Delta\mathcal V_{ab}(\epsilon) = \inf \left\{ \mathcal S_\epsilon[\phi]: \phi(0)\in\mathcal A_a^\epsilon, \ \phi(T)\in\mathcal U_b^\epsilon \text{ for some }T>0, \ \phi \text{ realizes a successful } a\to b \text{ transition} \right\}.
\label{eq:successful_transition_quasipotential}
\ee
\end{widetext}
This definition is consistent with the killed/committor construction above. The factor $\kappa_a^\epsilon$ gives the local escape clock, while $p_{ab}^\epsilon$ retains only the fraction of exits that are ultimately committed to the $b$th metastable core. Their product $q_{ab}(\epsilon)=\kappa_a^\epsilon p_{ab}^\epsilon$ is therefore the rate of successful $a\to b$ transitions, and \eqref{eq:Kramers_quasipotential_general} describes its leading small-noise scale.
The factor $\Gamma_{ab}(\epsilon)$ is the subexponential transition prefactor.

 In gradient systems with nondegenerate wells and saddles it reduces to the Eyring--Kramers prefactor, determined by the unstable eigenvalue at the saddle and the Hessian determinants at the well and saddle. In nongradient systems, the minimizing fluctuation path need not be the time reversal of deterministic relaxation, and the relevant transition object is a minimum-action path  \cite{Graham1991,LucariniBodai2020,LucariniBodai2019,Margazoglou2021}.

 In anisotropic or degenerate-noise settings, the action weights fluctuations according to the noise covariance, so the most likely transition channel may be strongly constrained by the directions in which noise is injected. In systems with several competing channels, $q_{ab}$ may receive several exponential contributions,
\begin{equation}
q_{ab}(\epsilon) \asymp \sum_m \Gamma_{ab}^{(m)}(\epsilon) \exp\left( -\frac{\Delta\mathcal V_{ab}^{(m)}(\epsilon)}{\sigma^2} \right),
\label{eq:multiple_transition_channels}
\end{equation}
with the smallest barrier dominating unless two or more channels are nearly degenerate. Thus $\Gamma_{ab}$ should not be read as a universal scalar prefactor; it summarizes the subexponential contribution of the dominant transition geometry, the noise covariance, and possible channel multiplicity.
 Rigorous derivations of the Eyring--Kramers law and of its prefactor, including the nonreversible case, are given in \cite{BouchetReygner2016,BouchetReygner2022}; see also \cite{Banerjee2026} and references therein.

The interpretation is therefore sharper than the statement that ``small gaps produce slow correlations.'' Local critical slowing down and metastable transition risk are controlled by two different spectral reductions of the same dynamics. The local theory kills the process at $\partial\mathcal W_a^\epsilon$ and then passes to the $Q$-process, thereby removing escape and studying the conditioned intrawell gap $\Delta_{\mathrm{loc}}^{(a)}(\epsilon)$. The transition theory keeps the escape clocks $\kappa_a^\epsilon$ and the committor-weighted destination probabilities $p_{ab}^\epsilon$, combines them into the finite-state generator $\mathsf{Q}_\epsilon$, and identifies transition RP resonances through its nonzero generator eigenvalues $\zeta_p^\epsilon$.

A transition warning may therefore appear as a decreasing quasipotential barrier $\Delta\mathcal V_{ab}(\epsilon)$, an increasing successful-transition rate $q_{ab}(\epsilon)$, or a changing coarse generator eigenvalue $\zeta_p^\epsilon$. It need not appear as local critical slowing down inside any single metastable region. Conversely, a closing conditioned intrawell gap $\Delta_{\mathrm{loc}}^{(a)}(\epsilon)$ signals loss of local recovery near $\mathcal A_a^\epsilon$, even if the dominant global transition resonance is controlled by rare interwell exchange. This is the RP distinction between local critical blocks and transition blocks.

\subsection{A Metastable Fold Scenario}
\label{subsec:metastable_fold_scenario}
We close this section by showing how the preceding construction diagnoses a tipping scenario that generalizes the classical saddle-node picture. Let $\epsilon$ be the control parameter and write the local parameter perturbation in the sense of Eq.~\eqref{eq:summary_Keps_expansion}, with $B_\epsilon=\partial_\epsilon K_\epsilon$.

Suppose that, for $\epsilon<\epsilon_c$, the deterministic skeleton has at least two attracting invariant objects. One of them, denoted $\mathcal A_0^\epsilon$, is the low state and contains the origin; another, denoted $\mathcal A_+^\epsilon$, lies away from the origin. These attracting objects need not be steady states. They may be periodic, chaotic, or otherwise extended invariant sets. The intended geometry is an S-shaped continuation: the low attracting branch bends as $\epsilon\uparrow\epsilon_c$, loses local stability at a critical parameter, and the noisy system may subsequently jump to the remote attracting state. A concrete instance in which the bending branch consists of periodic orbits rather than equilibria arises in delay models of the El Ni\~no--Southern Oscillation: a saddle-node bifurcation of periodic orbits coexists there with a subcritical Hopf and a homoclinic bifurcation, and stochastic solutions driven slowly through the critical delay display transition paths between the coexisting invariant sets that organize the interannual variability \cite{chekroun2024effective}. The construction below applies verbatim to such a periodic-orbit fold: the low state $\mathcal A_0^\epsilon$ is the limit cycle, its committor-defined neighborhood is a tubular region around it, and the two warning channels are its conditioned in-cycle recovery and its committor-weighted escape rate.

Thus, the RP formulation does not reduce this situation to a deterministic equilibrium eigenvalue. For each precritical $\epsilon$, one first constructs smooth metastable cores and committor-defined neighborhoods exactly as in Eqs.~\eqref{eq:metastable_core_sets}--\eqref{eq:committor_metastable_neighborhood}. In the present notation, we simply relabel the low-state region by $\mathcal W_0^\epsilon$ and the remote region by $\mathcal W_+^\epsilon$. The condition
\be
\mathcal A_0^\epsilon\subset\mathcal U_0^\epsilon\subset\mathcal W_0^\epsilon
\label{eq:fold_low_region_inclusion}
\ee
is the low-state version of Eq.~\eqref{eq:core_contained_in_committor_region}. Thus $\mathcal W_0^\epsilon$ is not a deterministic basin. It is a noise-regularized neighborhood of $\mathcal A_0^\epsilon$, defined by the probability of returning to the low core before reaching competing cores.

The first diagnostic is local recovery inside $\mathcal W_0^\epsilon$. One kills the process at $\partial\mathcal W_0^\epsilon$, as in Eqs.~\eqref{eq:exit_time_local_region}--\eqref{eq:killed_semigroup_submarkov}, and then passes to the associated $Q$-process. Thus the local generator, local semigroup, and local invariant measure are those defined in Eq.~\eqref{eq:local_generator_Q_choice}, with $a=0$. Equivalently, in the diffusion case, the conditioned generator has the interior form given in Eq.~\eqref{eq:Q_process_diffusion_generator}. This step is essential: it removes the survival decay of the killed process and isolates the recovery spectrum of trajectories conditioned on long residence near the low state.
The local critical block for the metastable fold is therefore the leading nonzero RP block of $K_{\epsilon,0}^{\mathrm{loc}}$ in the sense of Eq.~\eqref{eq:local_critical_RP_block}. The approach to the bend is detected by closure of the conditioned local gap,
\be
\Delta_{\mathrm{loc}}^{(0)}(\epsilon)\longrightarrow0,
\qquad
\epsilon\uparrow\epsilon_c,
\label{eq:fold_local_gap_closure}
\ee
where $\Delta_{\mathrm{loc}}^{(0)}$ is defined equivalently by Eq.~\eqref{eq:local_gap_killed_Q_process} in terms of killed eigenvalues or by Eq.~\eqref{eq:local_gap_definition_precise} in terms of the local conditioned RP resonance. This is the generalized critical-slowing-down statement. It says that, after escape has been factored out, the conditioned low-state dynamics loses its recovery rate. It does not assert that a deterministic steady-state eigenvalue crosses zero.
The perturbation direction relevant for this local diagnosis is also local. It is $B_{\epsilon,0}^{\mathrm{loc}}=\partial_\epsilon K_{\epsilon,0}^{\mathrm{loc}}$, as introduced in Eq.~\eqref{eq:local_centering_and_perturbation}. This operator contains the direct change of the physical generator and, through the $Q$-process construction, the change of the conditioning field that keeps trajectories inside $\mathcal W_0^\epsilon$.

 If the region $\mathcal W_0^\epsilon$ is moved with $\epsilon$, one must also account for the corresponding boundary or coordinate variation. In practice, this can be avoided by working in fixed local coordinates, or by using a fixed data-defined metastable neighborhood over the precritical parameter window.
Let $f$ be an observable designed to monitor the low state, for instance an order parameter that is small near $\mathcal A_0^\epsilon$ and large near $\mathcal A_+^\epsilon$. The residue-conditioned warning is then read directly from the local response ladder in Eq.~\eqref{eq:local_response_residue_ladder}, with $a=0$.

 If the local block is semisimple, visibility reduces to the nonvanishing condition in Eq.~\eqref{eq:local_semisimple_residue_condition}. If the block is nonsemisimple, the active nilpotent level is selected by the highest nonzero local residue in Eq.~\eqref{eq:local_response_residue_ladder}. Thus the bending of the low branch is visible in the local mean $\mu_{\epsilon,0}(f)$ only if the observable and the perturbation direction activate the closing local block.

The autocorrelation version is different but parallel. The local conditioned correlation is the object defined in Eq.~\eqref{eq:local_conditioned_correlation}, and its RP expansion is given in Eq.~\eqref{eq:local_conditioned_correlation_RP_expansion}. For the closing low-state block, the relevant amplitudes are the local correlation residues in Eq.~\eqref{eq:local_correlation_residue_defined}. Hence a growing local autocorrelation time requires nonzero $\mathcal A_{\mathrm{loc},k}^{(0),f}$, while a large parameter susceptibility requires nonzero $\Rstyle_{\mathrm{loc},k}^{(0),f}$. The two are contractions of the same local critical block, but they need not vanish or grow together.

The second diagnostic is the probability of an actual jump from the low state to the remote state. This is not measured by the conditioned gap $\Delta_{\mathrm{loc}}^{(0)}$. It is measured by the killed escape clock and by the committor-weighted destination probability. In the notation of the transition-block construction, the successful low-to-remote transition rate is the $0\to+$ entry of Eq.~\eqref{eq:transition_rate_from_killed_committor},
\be
q_{0+}(\epsilon) = \kappa_0^\epsilon p_{0+}^\epsilon .
\label{eq:fold_successful_transition_rate}
\ee
Here $\kappa_0^\epsilon$ is the principal killed escape rate, as in Eq.~\eqref{eq:local_escape_rate_from_killed}, and $p_{0+}^\epsilon$ is the committor-weighted exit-destination probability, as in Eq.~\eqref{eq:exit_destination_probability_committor}. In words, $\kappa_0^\epsilon$ measures how often locally equilibrated trajectories leave $\mathcal W_0^\epsilon$, while $p_{0+}^\epsilon$ measures what fraction of those exits are ultimately committed to the remote core.
Thus a metastable fold contains two distinct warning channels. The first is local: the conditioned gap $\Delta_{\mathrm{loc}}^{(0)}$ closes and produces local critical slowing down if the relevant local residues are nonzero. The second is global: the successful transition rate $q_{0+}$ increases when exits become more likely or more strongly committed to the remote state. On a fixed time horizon $T$ and for an ensemble locally equilibrated near the low state, the leading metastable jump probability is
\be
\mathbb P
\left(
0\to+ \text{ before }T
\right)
\simeq
1-\exp\left[-Tq_{0+}(\epsilon)\right],
\label{eq:fold_fixed_horizon_jump_probability}
\ee
and under a slow ramp $\epsilon=\epsilon(t)$ this becomes the corresponding hazard integral
\be
\mathbb P
\left(
0\to+ \text{ before }T
\right)
\simeq
1-\exp\left[
-\int_0^T q_{0+}(\epsilon(s)) \, \mathrm{d}s
\right].
\label{eq:fold_ramped_jump_probability}
\ee
These formulas are not replacements for the RP theory; they are its metastable consequence once the reduced transition generator has been constructed.
In the small-noise regime, the rate $q_{0+}$ inherits the quasipotential scaling of Eq.~\eqref{eq:Kramers_quasipotential_general}. The corresponding barrier is the special case $\Delta\mathcal V_{0+}(\epsilon)$ of Eq.~\eqref{eq:successful_transition_quasipotential}. It is the minimum action of a path that leaves the low attracting set and reaches the remote core before being recommitted to the low state or to another competitor. As the bend is approached, this barrier may decrease even while the local conditioned gap is also closing. These effects are related in a tipping scenario, but they are not the same RP object.

The conclusion is that an S-shaped metastable tipping event should not be summarized by a single scalar warning. The local branch bends when the conditioned low-state dynamics develops a closing local RP block. This is diagnosed through $\Delta_{\mathrm{loc}}^{(0)}$ and through the local residues in Eqs.~\eqref{eq:local_response_residue_ladder} and \eqref{eq:local_correlation_residue_defined}. The actual jump risk is diagnosed through $q_{0+}$, equivalently through the appropriate entry of the finite-state transition generator, or through the quasipotential barrier $\Delta\mathcal V_{0+}$. Thus the RP framework separates three quantities that classical EWS often conflates: local loss of recovery, statistical visibility through the measured observable, and successful escape to the remote state.

\section{Conditioned Dynamics, Rare Escape, and the Dual Girsanov--Doob View}
\label{sec:Girsanov_Doob}

\subsection{The $Q$-process as a Girsanov/Doob Transform}
\label{subsec:Q_process_Girsanov_Doob}

The local $Q$-process has a concrete probabilistic interpretation. It is the infinite-horizon survival-conditioning analogue of the Girsanov controls used for optimal importance sampling of Koopman expectations \cite{sikorski2024learning}. This connection is useful because it explains not only what the $Q$-process means, but also how it can be simulated.

Consider a diffusion
\be
\d X_t = b(X_t)\d t + \sigma D(X_t)\d W_t, \quad a(x)=D(x)D(x)^T .
\label{eq:girsanov_original_diffusion}
\ee
A controlled diffusion with the same noise covariance is
\be
\d X_t^u = \left[ b(X_t^u) + \sigma D(X_t^u)u(X_t^u,t) \right]\d t + \sigma D(X_t^u)\d W_t .
\label{eq:girsanov_controlled_diffusion}
\ee
For admissible $u$, the controlled path measure is absolutely continuous with respect to the original path measure on finite time intervals. Girsanov's theorem gives the Radon--Nikodym weight \cite[Theorem 10.14, p. 290]{daprato_zabczyk_2014}. Thus, by simulating \eqref{eq:girsanov_controlled_diffusion} and reweighting, one obtains unbiased estimators of expectations under the original dynamics.
The optimal-control viewpoint identifies special choices of $u$ that reduce the variance of such estimators. For a finite-time Koopman expectation
\be
(P_T h)(x) = \mathbb E_x[h(X_T)],
\label{eq:finite_time_koopman_expectation}
\ee
the zero-variance control has the logarithmic-gradient form
\be
u^*(x,t) = \sigma D(x)^T\nabla \log \left( P_{T-t}h \right)(x).
\label{eq:girsanov_koopman_control}
\ee
This is the finite-horizon formula used in optimal importance sampling of Koopman evaluations. If $h$ is an eigenfunction of the semigroup, then $P_{T-t}h$ is a scalar multiple of $h$, and the scalar disappears after applying $\nabla\log$. The optimal control then becomes time independent.

The killed local problem is the survival-conditioned analogue of this construction. Let $P_t^{\epsilon,a,D}$ be the killed semigroup in $\mathcal W_a^\epsilon$, and let $h_{\epsilon,a}$ be the positive principal Dirichlet eigenfunction defined in Eq.~\eqref{eq:transition_principal_killed_eigenpair}. Then
\be
P_t^{\epsilon,a,D}h_{\epsilon,a} = e^{\lambda_0^{D,a}(\epsilon)t} h_{\epsilon,a}.
\label{eq:killed_eigenfunction_semigroup_relation}
\ee
Therefore the finite-time Koopman control associated with the killed observable $h_{\epsilon,a}$ is independent of time:
\be
u_Q(x) = \sigma D_\epsilon(x)^T\nabla\log h_{\epsilon,a}(x).
\label{eq:Q_process_girsanov_control}
\ee
The controlled drift becomes
\be
b_\epsilon(x)+\sigma D_\epsilon(x)u_Q(x) = b_\epsilon(x) + \sigma^2a_\epsilon(x)\nabla\log h_{\epsilon,a}(x).
\label{eq:Q_process_controlled_drift}
\ee
Thus the corresponding controlled generator is exactly
\be
K_{\epsilon,a}^{Q}f = K_\epsilon f + \sigma^2a_\epsilon\nabla\log h_{\epsilon,a}\cdot\nabla f,
\label{eq:Q_process_generator_from_Girsanov}
\ee
which is the $Q$-process generator in Eq.~\eqref{eq:Q_process_diffusion_generator}. In other words, the $Q$-process is the Doob transform selected by the same logarithmic-gradient mechanism that produces optimal Girsanov controls.
The relation can also be seen from survival probabilities. For the killed process,
\be
P_T^{\epsilon,a,D}\mathds{1}(x) = \mathbb P_x(T<\tau_a^\epsilon).
\label{eq:killed_survival_probability}
\ee
The finite-horizon optimal control for estimating this survival probability is
\be
u_T^*(x,t) = \sigma D_\epsilon(x)^T \nabla \log \left( P_{T-t}^{\epsilon,a,D}\mathds{1} \right)(x).
\label{eq:finite_horizon_survival_control}
\end{equation}
As $T-t\to\infty$, the killed semigroup is dominated by its principal Dirichlet eigenpair:
\be
P_{T-t}^{\epsilon,a,D}\mathds{1}(x) \sim c_a(\epsilon) e^{\lambda_0^{D,a}(\epsilon)(T-t)} h_{\epsilon,a}(x).
\label{eq:killed_survival_asymptotics}
\ee
Taking $\nabla\log$ removes both the exponential factor and the scalar prefactor. Hence
\be
u_T^*(x,t) \longrightarrow \sigma D_\epsilon(x)^T\nabla\log h_{\epsilon,a}(x) = u_Q(x).
\label{eq:finite_horizon_control_to_Q_control}
\ee
This shows that the $Q$-process drift is the long-horizon limit of the optimal finite-time control for survival inside $\mathcal W_a^\epsilon$.
There are therefore two different but complementary ways to use the Girsanov construction.

First, one may use it as an \emph{unbiased rare-event estimator}. One simulates the controlled process \eqref{eq:girsanov_controlled_diffusion}, records the Girsanov weight $G_{0,T}^u$, and estimates killed expectations by
\be
P_T^{\epsilon,a,D}\varphi(x) = \mathbb E_x^u \left[ \varphi(X_T^u) \mathds{1}_{\{T<\tau_a^u\}} G_{0,T}^u \right].
\label{eq:killed_expectation_girsanov_estimator}
\ee
With $\varphi=\mathds{1}$, this gives the survival probability; with general $\varphi$, it gives killed Koopman expectations; with exit observables, it gives exit distributions and committor-weighted destination probabilities. This is the importance-sampling use: the control changes which paths are frequently observed, while the Girsanov weight restores the original probability law.

Second, one may use the same logarithmic-gradient drift as an \emph{unweighted simulator of the conditioned dynamics}. If the control is the exact $Q$-control \eqref{eq:Q_process_girsanov_control}, then the process
\bea
\d X_t^{Q} = [b_\epsilon(X_t^{Q}) + \sigma^2a_\epsilon(X_t^{Q}) \nabla\log &h_{\epsilon,a}(X_t^{Q}) ]\d t \\
&+ \sigma D_\epsilon(X_t^{Q})\d W_t,
\label{eq:Q_process_simulation_SDE}
\eea
is the $Q$-process itself. Long trajectories of \eqref{eq:Q_process_simulation_SDE} sample the invariant measure $\mu_{\epsilon,a}^{Q}$ from Eq.~\eqref{eq:Q_process_invariant_measure}. They can therefore be used to estimate local conditioned means, local conditioned correlations, and the local RP spectrum of $K_{\epsilon,a}^{\mathrm{loc}}$ without contamination by escape events.

This distinction is important. If one keeps the Girsanov weights, the controlled simulation estimates observables of the original killed process. If one discards the weights and uses the exact Doob drift, the simulation estimates observables of the conditioned $Q$-process. The first route is appropriate for escape probabilities, killed semigroups, and transition rates. The second route is appropriate for local recovery, conditioned correlations, and local critical blocks.

In practice, the exact function $h_{\epsilon,a}$ is rarely known in high dimension. The Girsanov/ISOKANN strategy suggests a practical approximation route \cite{sikorski2024learning}. One learns a positive surrogate $\widehat h_{\epsilon,a}$ for the relevant killed or Koopman eigenfunction, or for a positive $\chi$-function spanning the dominant local subspace, and uses
\be
\widehat u_Q(x) = \sigma D_\epsilon(x)^T\nabla\log \widehat h_{\epsilon,a}(x)
\label{eq:approximate_Q_control}
\ee
as a pseudo-optimal control. With Girsanov weights retained, estimates remain unbiased for the original killed process, provided the control is admissible and the weights are computed correctly. Without weights, the controlled process is an approximation of the $Q$-process; its quality is controlled by how well $\widehat h_{\epsilon,a}$ approximates the principal Dirichlet eigenfunction.

This gives a concrete simulation workflow for the local RP theory. For a chosen metastable neighborhood $\mathcal W_a^\epsilon$, one may: construct or learn a positive approximation of $h_{\epsilon,a}$; simulate the controlled drift \eqref{eq:Q_process_simulation_SDE}; estimate $\mu_{\epsilon,a}^{Q}$ and the local conditioned autocorrelation $C_{\mathrm{loc},a}^{f,\epsilon}(t)$; extract the leading local RP gap $\Delta_{\mathrm{loc}}^{(a)}(\epsilon)$ from the decay of this conditioned correlation; and, separately, use weighted controlled trajectories to estimate survival probabilities, exit distributions, and committor-weighted transition rates. The Girsanov viewpoint therefore supplies the computational bridge between the abstract killed/$Q$-process construction and numerical early-warning diagnostics.

\subsection{A unified RP theory of B- and N-tipping}
\label{Sec_RP_B_tipping_N_tipping}

The distinction between local critical slowing down and escape risk can now be stated in the standard language of tipping theory. In the usual classification, \emph{B-tipping} refers to bifurcation-induced tipping: a slowly varying control parameter drives the frozen system toward a loss of stability of the attracting state. The classical early-warning picture is then local. Recovery from perturbations becomes slower because a stability eigenvalue approaches the imaginary axis, and the associated critical slowing down may be detected through autocorrelation, variance, susceptibility, or low-frequency power.

In the present RP formulation, this mechanism is not tied to a fixed point. The attracting object may be an equilibrium, a periodic orbit, a chaotic invariant set, or a noise-regularized metastable region. What matters is the loss of local recovery inside the metastable set. The corresponding RP object is the leading nonzero spectral block of the conditioned local generator
\be
K_{\epsilon,a}^{\rm loc} \equiv K_{\epsilon,a}^{Q},
\ee
obtained by killing the original process at the boundary of the committor-defined metastable neighborhood $\mathcal W_a^\epsilon$ and then passing to the Doob $Q$-process. Thus the RP signature of B-tipping is the closure of the local conditioned gap
\be
\Delta_{\rm loc}^{(a)}(\epsilon)
=
-\Re \lambda_{\rm loc}^{(a)}(\epsilon)
\longrightarrow 0,
\ee
provided the corresponding local RP block is visible through nonzero correlation or response residues. In this sense, the local conditioned RP block is the stochastic counterpart of the classical soft mode.

\emph{N-tipping}, by contrast, refers to noise-induced tipping: the system escapes from a metastable region before the local attracting set has lost stability. The local recovery spectrum may remain uniformly stable, so there need not be any closing of $\Delta_{\rm loc}^{(a)}(\epsilon)$. The warning mechanism is instead global and probabilistic. It is controlled by rare excursions toward the committor channel, by the quasipotential barrier separating metastable regions, and by the escape rates and destination probabilities entering the reduced metastable generator
\be
\mathsf Q_\epsilon=(q_{ab}(\epsilon)),
\qquad
q_{ab}(\epsilon)=\kappa_a^\epsilon p_{ab}^\epsilon,
\quad b\neq a.
\label{eq:BN_reduced_generator}
\ee
The associated RP blocks are interwell transition blocks of the full generator $K_\epsilon$: their eigenfunctions are approximately constant on metastable regions and describe slow exchange of probability mass between them, not local recovery inside one region.

The RP distinction is therefore sharp. B-tipping is diagnosed by a local conditioned critical block of $K_{\epsilon,a}^{\rm loc}$; N-tipping is diagnosed by transition blocks of the full generator, or equivalently by the escape clocks, committors, and quasipotential barriers summarized in $\mathsf Q_\epsilon$. Both mechanisms may produce long memory, enhanced low-frequency variability, or large excursions in observed time series, but they do so for different spectral reasons. A pre-switch excursion is therefore not, by itself, evidence of local loss of stability. It is an escape-risk signal unless it can be traced to the closing of a local conditioned RP gap and to nonzero local residues.

\subsection{A Rare Pre-Switch Excursion Is Not a Local Tipping Criterion}
\label{subsec:rare_excursion_not_tipping}

The Girsanov/Doob construction above also clarifies a common diagnostic ambiguity. There are two distinct uses of conditioned or controlled trajectories. With Girsanov weights retained, controlled simulations estimate observables of the original killed process: survival probabilities, exit distributions, committor-weighted destination probabilities, and transition rates. With the exact Doob drift and no weights, the simulation samples the conditioned $Q$-process and therefore estimates local conditioned statistics: local correlations, local recovery rates, and local RP blocks. Confusing these two uses leads directly to confusing transition risk with local critical slowing down.

The purpose of an early-warning signal is to assess proximity to tipping before the observed trajectory has switched. In that regime, a single finite trajectory has not, by itself, sampled the full interwell transition process. One should therefore not infer the global transition block from one pre-switch realization alone. What may be visible instead is pathwise exploration of a committor channel, approach to an edge-state neighborhood, or a finite-time excursion that is statistically associated with increased escape probability. Such an event is an escape-risk signal; it is not automatically evidence that the local attracting set has lost recovery.

In the terminology of the unified RP picture, this is the distinction between $N$-tipping and $B$-tipping. A local $B$-tipping warning is tied to the conditioned local generator $K_{\epsilon,a}^{\rm loc}$, obtained from the killed process and its Doob $Q$-process. Its spectral signature is the closure of the local conditioned gap $\Delta_{\rm loc}^{(a)}(\epsilon)$, defined in Eq.~\eqref{eq:local_gap_definition_precise},
together with nonzero local response residues
(Eqns.~\eqref{eq:local_semisimple_residue_condition}--\eqref{eq:local_response_residue_ladder})
or nonzero local correlation residues (Eq.~\eqref{eq:local_correlation_residue_defined}). In that case, the observed slowing down is genuinely an in-well recovery phenomenon.

By contrast, an $N$-tipping or escape-risk warning may occur while the local conditioned gap remains bounded away from zero. Its relevant objects are not the local critical block, but the committor geometry, the killed escape clocks $\kappa_a^\epsilon$, the destination probabilities $p_{ab}^\epsilon$, the transition rates $q_{ab}(\epsilon)$ in Eq.~\eqref{eq:transition_rate_from_killed_committor}, and the quasipotential barriers $\Delta\mathcal V_{ab}(\epsilon)$. These are ensemble or model-based quantities. They can be estimated from many trajectories, from rare-event sampling, or from weighted Girsanov simulations, but they should not be read off naively from a single pre-switch record.

This distinction also clarifies the role of low-frequency signals. A large low-frequency component in a finite record does not say, by itself, whether the visible block is a local recovery block or an interwell transition block. It may reflect local critical slowing down, but it may also reflect residence-time persistence, intermittent visits to the committor channel, or finite-time contamination by escape-risk dynamics. The RP question is therefore not simply whether correlations are large or spectra are red. It is which spectral object is responsible for the signal, and through which residue or transition observable it is visible.

The diagnostic prescription is consequently mechanism-dependent. For local loss of recovery, one should construct the committor-defined neighborhood $\mathcal W_a^\epsilon$, pass to the $Q$-process, estimate the local conditioned RP gap, and test the corresponding local residues. For transition risk, one should estimate survival probabilities, committors, exit distributions, transition rates, and quasipotential barriers, using weighted controlled trajectories when rare-event sampling is required. A rare pre-switch excursion is therefore a meaningful warning only after its mechanism has been identified: it is a local tipping signal if it is tied to the local conditioned critical block, and an escape-risk signal if it is tied to committor-channel exploration or to the metastable transition generator $\mathsf Q_\epsilon$.
\section{A 1D Metastable Fold Example: Tipping as a Competition between two Different Processes}
\label{sec:1D_fold_example}

Consider the scalar diffusion
\begin{equation}
\mathrm{d}X_t = \left( \epsilon+X_t-X_t^3 \right) \mathrm{d}t + \sigma \, \mathrm{d}W_t .
\label{eq:scalar_fold_SDE}
\end{equation}
This is a gradient diffusion,
\be
\mathrm{d}X_t = -U_\epsilon'(X_t)  \mathrm{d}t + \sigma  \mathrm{d}W_t, \label{eq:scalar_fold_potential}
\ee
with $ U_\epsilon(x) = \frac{x^4}{4} - \frac{x^2}{2} - \epsilon x.$

The deterministic equilibria satisfy $x^3-x = \epsilon.$
For
\begin{equation}
|\epsilon| < \epsilon_c, \qquad \epsilon_c = \frac{2}{3\sqrt{3}},
\label{eq:fold_critical_parameter}
\end{equation}
there are two stable equilibria separated by one saddle. With the sign convention in Eq.~\eqref{eq:scalar_fold_SDE}, as $\epsilon\uparrow\epsilon_c$ the left stable equilibrium and the intermediate saddle coalesce at
\begin{equation}
x_c = -\frac{1}{\sqrt{3}}.
\label{eq:fold_collision_point}
\end{equation}
After this collision, only the right stable state remains. This is the one-dimensional analogue of the metastable fold discussed above , with the low state $\mathcal A_0^\epsilon$ of Sec.~\ref{subsec:metastable_fold_scenario} realized by the left equilibrium; we accordingly relabel the index $0\to-$ throughout this section, so that $\mathcal W_0^\epsilon\to\mathcal W_-^\epsilon$ and $q_{0+}\to q_\pm$.
Let $x_-^\epsilon$ denote the left stable equilibrium,  $x_+^\epsilon$ the right one, and $x_s^\epsilon$ the  intermediate saddle. A concrete local metastable region for the left state is, for example,
\begin{equation}
\mathcal W_-^\epsilon = (-\infty,x_s^\epsilon),
\label{eq:left_well_region}
\end{equation}
or a nearby committor-defined subregion as in Eq.~\eqref{eq:committor_metastable_neighborhood}. The killed generator is
\begin{equation}
K_{\epsilon,-}^{D}f = \left( \epsilon+x-x^3 \right)f' + \frac{\sigma^2}{2}f'', \qquad f|_{\partial\mathcal W_-^\epsilon}=0.
\label{eq:fold_killed_generator}
\end{equation}
Let $h_{\epsilon,-}$ be the positive principal Dirichlet eigenfunction. The conditioned local generator is therefore
\begin{equation}
K_{\epsilon,-}^{\mathrm{loc}}f = \left[ \epsilon+x-x^3 + \sigma^2\partial_x\log h_{\epsilon,-}(x) \right]f' + \frac{\sigma^2}{2}f'' .
\label{eq:fold_Q_generator_explicit}
\end{equation}
The additional drift
\begin{equation}
\sigma^2\partial_x\log h_{\epsilon,-}
\label{eq:fold_conditioning_drift}
\end{equation}
pushes trajectories away from the absorbing boundary and realizes the process conditioned on long residence in the left well.
Because the original diffusion is reversible with stationary density
\begin{equation}
\rho_\epsilon(x) = Z_\epsilon^{-1} \exp\left( -\frac{2U_\epsilon(x)}{\sigma^2} \right),
\label{eq:fold_stationary_density}
\end{equation}
the invariant density of the conditioned local process is, up to normalization,
\begin{equation}
\mathrm{d}\mu_{\epsilon,-}^{Q}(x) \propto h_{\epsilon,-}(x)^2 \rho_\epsilon(x) \, \mathrm{d}x .
\label{eq:fold_Q_invariant_measure}
\end{equation}
This is the one-dimensional reversible specialization of Eq.~\eqref{eq:Q_process_invariant_measure}.
 Far from $x_s^\epsilon$, where $h_{\epsilon,-}$ is nearly constant, this density is proportional to the global invariant density, while it vanishes for $x\geq x_s^\epsilon$.
The local critical block is the leading nonzero RP block of $K_{\epsilon,-}^{\mathrm{loc}}$.  The conditioning drift \eqref{eq:fold_conditioning_drift} is exponentially small in the bulk of the well, where $h_{\epsilon,-}$ is nearly constant, and becomes $O(1)$ only in a boundary layer of width $O(\sigma)$ around $x_s^\epsilon$; since the conditioned invariant measure \eqref{eq:fold_Q_invariant_measure} puts exponentially little mass there, the leading conditioned relaxation rate is that of the unconditioned dynamics linearized at the well bottom. In the small-noise regime, and away from the narrow stochastic scaling layer of the fold, its leading relaxation rate is well approximated by the linearization at the left stable equilibrium:
\begin{equation}
\lambda_{\mathrm{loc}}^{(-)}(\epsilon) \approx b_\epsilon'(x_-^\epsilon) = 1-3(x_-^\epsilon)^2 < 0.
\label{eq:fold_local_linear_rate}
\end{equation}
As $\epsilon\uparrow\epsilon_c$, set
\begin{equation}
\delta = \epsilon_c-\epsilon.
\end{equation}
Then
\begin{equation}
x_-^\epsilon = x_c - \sqrt{\frac{\delta}{\sqrt{3}}} + o(\sqrt{\delta}), \quad x_s^\epsilon = x_c + \sqrt{\frac{\delta}{\sqrt{3}}} + o(\sqrt{\delta}),
\label{eq:fold_root_asymptotics}
\end{equation}
and therefore
\begin{equation}
\Delta_{\mathrm{loc}}^{(-)}(\epsilon) = -\lambda_{\mathrm{loc}}^{(-)}(\epsilon) \approx 2 \cdot 3^{1/4}\sqrt{\epsilon_c-\epsilon}.
\label{eq:fold_local_gap_scaling}
\end{equation}
This is the classical square-root critical-slowing-down law, but now interpreted as a local conditioned RP gap. The warning is observable only through the local residues of Eqs.~\eqref{eq:local_response_residue_ladder} and \eqref{eq:local_correlation_residue_defined}.
For example, if $f(x)=x$ and the local critical block is semisimple, then local autocorrelation has leading form
\begin{equation}
C_{\mathrm{loc},-}^{f,\epsilon}(t) \sim \mathcal A_{\mathrm{loc},0}^{(-),f}(\epsilon) e^{\lambda_{\mathrm{loc}}^{(-)}(\epsilon)t},
\label{eq:fold_local_correlation_simple}
\end{equation}
where $\mathcal A_{\mathrm{loc},0}^{(-),f}$ is the local correlation residue in Eq.~\eqref{eq:local_correlation_residue_defined}. The local mean susceptibility is controlled instead by the response residue
\begin{equation}
\Rstyle_{\mathrm{loc},0}^{(-),f}(\epsilon) = \mu_{\epsilon,-} \left( B_{\epsilon,-}^{\mathrm{loc}} \Pi_{\mathrm{loc}}^{(-),\epsilon} f_{\epsilon,-}^c \right),
\label{eq:fold_local_response_semisimple}
\end{equation}
as in Eq.~\eqref{eq:local_semisimple_residue_condition}.

Since
\begin{equation}
K_\epsilon f = (\epsilon+x-x^3)f' + \frac{\sigma^2}{2}f'',
\end{equation}
the unrestricted perturbation is simply
\begin{equation}
B_\epsilon f = f'.
\label{eq:fold_unrestricted_B}
\end{equation}
The local perturbation $B_{\epsilon,-}^{\mathrm{loc}}$ also contains the derivative of the conditioning drift in Eq.~\eqref{eq:fold_Q_generator_explicit}, because the $Q$-process itself depends on $\epsilon$ through $h_{\epsilon,-}$.
The escape component is different. The killed principal eigenvalue gives the escape clock
\begin{equation}
\kappa_-^\epsilon = -\lambda_0^{D,-}(\epsilon),
\label{eq:fold_escape_clock}
\end{equation}
and the successful transition rate to the right well is
\begin{equation}
q_{\pm}(\epsilon) = \kappa_-^\epsilon p_{\pm}^\epsilon ,
\label{eq:fold_successful_rate}
\end{equation}
where $p_{\pm}^\epsilon$ is the committor-weighted probability that an exit from the left region reaches the right core before returning to the left core.

It is instructive to evaluate $p_{\pm}^\epsilon$ explicitly, because it is here that the committor weighting of Eq.~\eqref{eq:exit_destination_probability_committor} earns its keep. Absorbing exactly at the saddle does \emph{not} mean that every exit is a successful transition. For the reversible dynamics \eqref{eq:scalar_fold_potential} the committor to the right core is
\be
q_+^\epsilon(x)
=
\frac{\int_{x_-^\epsilon}^{x}e^{2U_\epsilon(y)/\sigma^2}\d y}
{\int_{x_-^\epsilon}^{x_+^\epsilon}e^{2U_\epsilon(y)/\sigma^2}\d y},
\label{eq:fold_committor_explicit}
\ee
and both integrals are dominated, by Laplace's method, by the same symmetric neighborhood of the maximum of $U_\epsilon$ at $x_s^\epsilon$. Since $x_s^\epsilon$ splits that neighborhood into two equal halves,
\bea
p_\pm^\epsilon= q_+^\epsilon(x_s^\epsilon)
&=
\frac12+O\!\left(\frac{\sigma}{\left|U_\epsilon''(x_s^\epsilon)\right|^{3/2}}\right),\\
\text{so}\quad
q_\pm(\epsilon)&\simeq\tfrac12\,\kappa_-^\epsilon .
\label{eq:fold_committor_half}
\eea
The correction, which originates from the cubic term of $U_\epsilon$ at the saddle, is of relative size $\sigma(\epsilon_c-\epsilon)^{-3/4}$ by Eq.~\eqref{eq:fold_curvature_scaling}; it is negligible exactly on the Kramers window \eqref{eq:fold_Kramers_validity} and becomes $O(1)$ precisely inside the stochastic fold layer $\epsilon_c-\epsilon\sim\sigma^{4/3}$, where the notion of a well-defined transition state ceases to make sense.
This factor $\tfrac12$ is not a detail: it is exactly the classical factor $2$ between the mean first-passage time to the barrier top and the Kramers interwell transition time. Indeed, the principal Dirichlet eigenvalue on $\mathcal W_-^\epsilon=(-\infty,x_s^\epsilon)$ obeys \cite{matkowski1981}:
\bea
\kappa_-^\epsilon
&\simeq
\frac{\sqrt{U_\epsilon''(x_-^\epsilon)\left|U_\epsilon''(x_s^\epsilon)\right|}}{\pi}
\,e^{-\Delta\mathcal V_\pm(\epsilon)/\sigma^2}\\
&= 2\,\Gamma_\pm(\epsilon)\,e^{-\Delta\mathcal V_\pm(\epsilon)/\sigma^2},
\label{eq:fold_killed_eigenvalue_kramers}
\eea
with $\Gamma_\pm$ the Eyring--Kramers prefactor recalled in Eq.~\eqref{eq:fold_EK_prefactor} below, so that $q_\pm=\kappa_-^\epsilon p_\pm^\epsilon$ reproduces the Eyring--Kramers rate \eqref{eq:fold_kramers_rate} exactly. The escape clock alone would overestimate the successful-transition rate by a factor two.

A second, more structural remark follows. The individual factors $\kappa_a^\epsilon$ and $p_{ab}^\epsilon$ both depend on where the regularized boundary $\partial\mathcal W_a^\epsilon$ is placed: pushing the boundary deeper into the well increases the escape clock and decreases the destination probability. Their product $q_{ab}=\kappa_a^\epsilon p_{ab}^\epsilon$ is, to leading exponential order and to leading prefactor order, \emph{independent} of that choice, because both factors are controlled by the same Laplace neighborhood of the transition state. This is the precise sense in which the reduced generator $\mathsf Q_\epsilon$ of Sec.~\ref{subsec:transition_blocks} is an intrinsic object and not an artifact of the committor threshold $\eta$ in Eq.~\eqref{eq:committor_metastable_neighborhood}.

In the small-noise limit, the successful transition rate inherits the Eyring--Kramers/Freidlin--Wentzell scale
\begin{equation}
q_{\pm}(\epsilon) \asymp \Gamma_{\pm}(\epsilon) \exp\left[ -\frac{\Delta\mathcal V_{\pm}(\epsilon)}{\sigma^2} \right],
\label{eq:fold_kramers_rate}
\end{equation}
where the quasipotential barrier is
\begin{equation}
\Delta\mathcal V_{\pm}(\epsilon) = 2\left[ U_\epsilon(x_s^\epsilon) - U_\epsilon(x_-^\epsilon) \right].
\label{eq:fold_quasipotential_barrier}
\end{equation}
The factor $2$ appears because the noise convention in Eq.~\eqref{eq:scalar_fold_SDE} is $\sigma \, \mathrm{d}W_t$, so that the stationary density is proportional to $\exp[-2U_\epsilon(x)/\sigma^2]$.
The prefactor $\Gamma_{\pm}(\epsilon)$ is not needed to identify the leading exponential barrier, but it is needed for quantitative transition rates and finite-time jump probabilities. In the present one-dimensional gradient case it is explicit. The Eyring--Kramers prefactor is
\begin{equation}
\Gamma_{\pm}(\epsilon) = \frac{1}{2\pi} \sqrt{ U_\epsilon''(x_-^\epsilon) \left| U_\epsilon''(x_s^\epsilon) \right| }
\label{eq:fold_EK_prefactor}
\end{equation}
up to lower-order corrections in the small-noise limit. Since
\begin{equation}
U_\epsilon''(x) = 3x^2-1,
\label{eq:fold_potential_second_derivative}
\end{equation}
and since Eq.~\eqref{eq:fold_root_asymptotics} gives
\begin{align*}
x_-^\epsilon &= x_c - \sqrt{\frac{\delta}{\sqrt{3}}} + o(\sqrt{\delta}), \\
x_s^\epsilon &= x_c + \sqrt{\frac{\delta}{\sqrt{3}}} + o(\sqrt{\delta}), \qquad \delta=\epsilon_c-\epsilon,
\end{align*}
one obtains
\begin{equation}
U_\epsilon''(x_-^\epsilon) \sim 2 \cdot 3^{1/4}\delta^{1/2}, \qquad \left| U_\epsilon''(x_s^\epsilon) \right| \sim 2 \cdot 3^{1/4}\delta^{1/2}.
\label{eq:fold_curvature_scaling}
\end{equation}
Therefore
\begin{equation}
\Gamma_{\pm}(\epsilon) \sim \frac{3^{1/4}}{\pi} (\epsilon_c-\epsilon)^{1/2}.
\label{eq:fold_prefactor_scaling}
\end{equation}
This prefactor decreases as the saddle and well coalesce, but the dominant acceleration of escape comes from the collapse of the exponential barrier.
We now compute the barrier. Let $y=x-x_c$, with $x_c=-1/\sqrt{3}$, and set $\delta=\epsilon_c-\epsilon$. Near the fold, the potential has the normal-form expansion
\begin{equation}
U_\epsilon(x_c+y) = \text{constant} + \delta y - \frac{\sqrt{3}}{3}y^3 + \frac{1}{4} y^4 + \cdots .
\label{eq:fold_potential_normal_form}
\end{equation}
Using Eq.~\eqref{eq:fold_root_asymptotics}, write
\bea
\alpha_\delta = \sqrt{\frac{\delta}{\sqrt{3}}}, &\quad x_-^\epsilon = x_c - \alpha_\delta + o(\sqrt{\delta}), \\
&\quad x_s^\epsilon = x_c + \alpha_\delta + o(\sqrt{\delta}).
\label{eq:fold_alpha_delta}
\eea
The quartic contribution is even and cancels to leading order in the difference $U_\epsilon(x_s^\epsilon)-U_\epsilon(x_-^\epsilon)$. Hence
\begin{align*}
U_\epsilon(x_s^\epsilon)-U_\epsilon(x_-^\epsilon) &= 2\delta\alpha_\delta - \frac{2\sqrt{3}}{3}\alpha_\delta^3 + o(\delta^{3/2}) \\
&= \frac{4}{3^{5/4}} \delta^{3/2} + o(\delta^{3/2}).
\end{align*}
Multiplying by the factor $2$ in Eq.~\eqref{eq:fold_quasipotential_barrier} gives
\begin{equation}
\Delta\mathcal V_{\pm}(\epsilon) \sim \frac{8}{3^{5/4}} (\epsilon_c-\epsilon)^{3/2}.
\label{eq:fold_barrier_scaling}
\end{equation}
Combining Eqs.~\eqref{eq:fold_prefactor_scaling} and \eqref{eq:fold_barrier_scaling}, the transition rate has the more explicit asymptotic form
\begin{equation}
q_{\pm}(\epsilon) \asymp \frac{3^{1/4}}{\pi} (\epsilon_c-\epsilon)^{1/2} \exp\left[ -\frac{8}{3^{5/4}} \frac{(\epsilon_c-\epsilon)^{3/2}}{\sigma^2} \right],
\label{eq:fold_transition_rate_explicit}
\end{equation}
as long as the Kramers regime remains valid, namely while
\begin{equation}
\frac{(\epsilon_c-\epsilon)^{3/2}}{\sigma^2}\gg1.
\label{eq:fold_Kramers_validity}
\end{equation}
Inside the stochastic fold layer $(\epsilon_c-\epsilon)\sim\sigma^{4/3}$, the barrier is of the same order as the noise, and the classical Kramers expansion is no longer the correct asymptotic description. In that layer, one should return to the killed eigenvalue problem itself, or to controlled rare-event sampling, rather than rely on the prefactor formula.
Thus the same approach to the fold produces two distinct scalings:
\begin{equation}
\Delta_{\mathrm{loc}}^{(-)}(\epsilon) \sim 2 \cdot 3^{1/4} (\epsilon_c-\epsilon)^{1/2}, \qquad \Delta\mathcal V_{\pm}(\epsilon) \sim \frac{8}{3^{5/4}} (\epsilon_c-\epsilon)^{3/2}.
\label{eq:fold_two_scalings}
\end{equation}
The first quantity is the local conditioned RP gap, see Eq.~\eqref{eq:local_Q_gap_general}. It controls the decay rate of perturbations inside the left well, after escape has been removed by the $Q$-process. If the local critical block is residue-visible for the observable $f$, then the local conditioned spectrum has the leading low-frequency form
\begin{equation}
S_{\mathrm{loc},-}^{f,\epsilon}(\omega) \sim \mathcal A_{\mathrm{loc},0}^{(-),f}(\epsilon) \frac{ 2\Delta_{\mathrm{loc}}^{(-)}(\epsilon) }{ \left(\Delta_{\mathrm{loc}}^{(-)}(\epsilon)\right)^2+\omega^2 },
\label{eq:fold_local_reddening_lorentzian}
\end{equation}
in the semisimple real-resonance case , which is the one-dimensional instance of the local conditioned power spectral density introduced in Eqs.~\eqref{eq:local_conditioned_PSD}--\eqref{eq:local_conditioned_PSD_lorentzian}.

Hence, as $\epsilon\uparrow\epsilon_c$, the local spectral peak narrows toward $\omega=0$, the local correlation time grows like
\begin{equation}
\tau_{\mathrm{loc},-}(\epsilon) \sim \frac{1}{\Delta_{\mathrm{loc}}^{(-)}(\epsilon)} \sim \frac{1}{ 2 \cdot 3^{1/4} } (\epsilon_c-\epsilon)^{-1/2},
\label{eq:fold_local_correlation_time}
\end{equation}
and the zero-frequency local power grows, provided the residue $\mathcal A_{\mathrm{loc},0}^{(-),f}$ does not vanish.  For $f(x)=x$ this residue is itself the conditioned in-well variance, and in the harmonic approximation of the left well
\be
\mathcal A_{\mathrm{loc},0}^{(-),f}(\epsilon)
\simeq
\frac{\sigma^2}{2\Delta_{\mathrm{loc}}^{(-)}(\epsilon)}
\propto
(\epsilon_c-\epsilon)^{-1/2},
\label{eq:fold_local_variance_scaling}
\ee
so that the zero-frequency local power diverges one order faster than the correlation time,
\be
S_{\mathrm{loc},-}^{f,\epsilon}(0)
\simeq
\frac{\sigma^2}{\left(\Delta_{\mathrm{loc}}^{(-)}(\epsilon)\right)^{2}}
\propto
(\epsilon_c-\epsilon)^{-1}.
\label{eq:fold_zero_frequency_power_scaling}
\ee
Variance, correlation time and zero-frequency power therefore carry three \emph{different} exponents ($-1/2$, $-1/2$ and $-1$) of the same closing local block --- a concrete instance of the statement of Sec.~\ref{subsec:EWS_vs_CSD} that classical EWS metrics are distinct contractions of one RP object rather than interchangeable symptoms. All three scalings hold only in the Kramers window \eqref{eq:fold_Kramers_validity}; inside the stochastic fold layer $\epsilon_c-\epsilon\lesssim\sigma^{4/3}$ the local gap saturates at $\Delta_{\mathrm{loc}}^{(-)}=O(\sigma^{2/3})$ and the divergences are cut off. This is genuine local critical slowing down, and its frequency-domain manifestation is genuine local spectral reddening.  In the taxonomy of Sec.~\ref{subsec:intro_tipping}, this is the B-tipping channel.

The second quantity, $\Delta\mathcal V_{\pm}$, controls successful escape. It enters the transition rate $q_{\pm}$, not the conditioned local recovery rate. As the barrier collapses like $(\epsilon_c-\epsilon)^{3/2}$, the escape rate increases approximately according to Eq.~\eqref{eq:fold_transition_rate_explicit}. This means that the risk of an actual jump to the remote well can become large even before the deterministic fold is reached. On a finite observation window $T$, the relevant transition-risk diagnostic is therefore the hazard
\begin{equation}
1-\exp[-Tq_{\pm}(\epsilon)],
\label{eq:fold_transition_hazard}
\end{equation}
or, under ramping, the corresponding integrated hazard in Eq.~\eqref{eq:fold_ramped_jump_probability}.  In the same taxonomy, this is the N-tipping channel.

This distinction has a direct spectral consequence. The local critical block produces a resonance moving toward the origin:
\begin{equation}
\lambda_{\mathrm{loc}}^{(-)}(\epsilon) \simeq -\Delta_{\mathrm{loc}}^{(-)}(\epsilon) \uparrow 0.
\label{eq:fold_local_resonance_to_zero}
\end{equation}
This is the RP source of local reddening. By contrast, the transition process is governed by a coarse transition rate. In the one-way escape regime from the disappearing well, its characteristic generator scale is of order
\begin{equation}
\lambda_{\mathrm{tr}}(\epsilon) \sim -q_{\pm}(\epsilon).
\label{eq:fold_transition_resonance_scale}
\end{equation}
Far from the fold, $q_{\pm}$ is exponentially small, so transition observables can show very long residence-time correlations. But as the barrier decreases, $q_{\pm}$ increases. The transition peak then broadens rather than narrows. Thus increasing transition risk need not appear as spectral reddening; it may appear as a rising escape hazard and a decreasing quasipotential barrier.

For the scalar fold, the two warnings therefore have different meanings. A local observable that resolves fluctuations inside the left well should show local critical slowing down through Eq.~\eqref{eq:fold_local_correlation_time}, provided its local correlation residue is nonzero. A basin-label or threshold-crossing observable should instead be sensitive to the transition mechanism through $q_{\pm}$ and $\Delta\mathcal V_{\pm}$.

 This is the unified picture of B- and N-tipping announced in the introduction, now made quantitative. A global lag-one autocorrelation or low-frequency power estimate can mix these two effects. It may be large because the local RP gap is closing, or because rare switching produces long residence times, or because both mechanisms coexist. The RP framework makes the distinction operational: identify whether the visible block is the local conditioned block of $K_{\epsilon,-}^{\mathrm{loc}}$ or the transition block encoded by the coarse jump rate $q_{\pm}$, and then test whether the corresponding residue is nonzero.

\section{Discussion}
\label{sec:discussion}
\subsection{What has been established}
The starting point of this paper is a simple algebraic observation with substantial consequences. When the first-order sensitivity of an invariant statistic $\mu_\epsilon(f)$ is resolved on the RP blocks of the Kolmogorov generator, it takes the form
\bes
\pd{}{\epsilon}\mu_\epsilon(f)
=
\sum_{j}\sum_{k=0}^{m_j-1}
\frac{\Rstyle_{j,k}^{f}(B_\epsilon)}{(-\lambda_j)^{k+1}}
+\ \text{residual},
\ees
so that the quantity usually monitored --- the RP gap $|\Re\lambda_j|$ --- appears only in the denominators. The numerators $\Rstyle_{j,k}^{f}(B_\epsilon)=\mu_\epsilon(B_\epsilon N_j^k\Pi_j f_\epsilon^c)$ are contractions of the block against \emph{both} the measured observable and the perturbation direction, and nothing prevents them from vanishing. A closing gap is thus a necessary condition critical-slowing-down-induced divergent susceptibility, never a sufficient one. Theorem~\ref{thm:RP_sensitivity_residue_selection} makes this precise, including the selection of the active nilpotent level by the highest surviving residue.

The worked counterexample of Sec.~\ref{sec:nonnormal} shows that this is not a fussy caveat. There, a two-dimensional linear diffusion has a spectrum that is constant by construction, and yet variance, autocorrelation, integrated autocorrelation and low-frequency power all diverge. The classical approach cannot fail more completely, and it fails for exactly the reason the formula above predicts: the denominators never move and the numerators grow without bound. The same decomposition then supplies the two repairs --- residue selection, which produces an observable that is exactly blind to the spurious signal, and the index $\mathcal N$ of Eq.~\eqref{eq:nonnormality_index}, whose reversible bound (Proposition~\ref{prop:nonnormality_bound}) turns ``the residues are cancelling'' into a falsifiable inequality on measurable quantities.

Three further consequences deserve emphasis. First, the classical EWS approach is not a set of interchangeable symptoms: variance is a zero-resolvent quantity, integrated autocorrelation is a one-resolvent quantity weighted by \emph{correlation} residues, parameter susceptibility is a one-resolvent quantity weighted by \emph{response} residues, and frequency-resolved sensitivity is a two-resolvent quantity weighted by \emph{pairwise} residues. They can, and in the scalar fold of Sec.~\ref{sec:1D_fold_example} demonstrably do, scale with different exponents of the same closing block; in Sec.~\ref{sec:nonnormal} they are separated more brutally still, since the variance carries no denominator at all and is therefore the one indicator that residue selection cannot rescue. Second, algebraic multiplicity matters: a nonsemisimple critical block can amplify by $|\lambda_j|^{-m_j}$, but only if the top nilpotent residue is nonzero. Third, in a metastable regime the dominant nonzero RP block of the full generator is generically an interwell transition block, so that ``the'' spectral gap of the observed dynamics is not the object that classical CSD theory has in mind.

The resolution proposed here is to split the problem with the killed process.
 This is in the spirit of recent work on quasi-stationary and quasi-ergodic measures for metastable systems, in which Lamb and collaborators adapt classical notions of dynamical systems theory, such as Lyapunov exponents and bifurcations, to trajectories conditioned to remain in a prescribed domain \cite{Engel2019a,Engel2019b,Castro2024a,Castro2024b}. Killing at a committor-defined, noise-regularized boundary and passing to the Doob $Q$-process removes survival decay and produces a genuine conservative in-well dynamics whose leading nonzero block is the local critical block; this is B-tipping. Keeping the same killed problems but retaining the escape clocks and weighting them by committor destination probabilities produces the reduced generator $\mathsf Q_\epsilon$ on metastable labels; this is N-tipping. Both reductions are built from the same ingredients, which is why they are so easily conflated, and both are needed, which is why a single scalar indicator cannot serve.

\subsection{Limitations and caveats}
Several restrictions should be stated plainly. (i) The spectral framework is developed in $L^2_{\mu_\epsilon}$ under hypoellipticity and dissipativity, so that the invariant density is smooth. For deterministic chaotic systems the appropriate RP spectrum lives on anisotropic Banach spaces \cite{baladi2000positive,dyatlov2019mathematical}, and the residue-selection statements would have to be reformulated there. (ii) The finite-state reduction requires the metastability separation \eqref{eq:metastable_spectral_separation_killed}, namely that a trajectory relaxes to its quasi-stationary in-well ensemble long before it escapes. This is exactly what fails at the end of an approach to a fold. In the scalar example of Sec.~\ref{sec:1D_fold_example} the conditioned local gap decreases only algebraically, $\Delta_{\rm loc}^{(-)}\sim(\epsilon_c-\epsilon)^{1/2}$, whereas the escape rate increases exponentially, $q_\pm\sim\exp[-\mathrm{const}\,(\epsilon_c-\epsilon)^{3/2}/\sigma^2]$; the two therefore cross at the edge of the Kramers window, $\epsilon_c-\epsilon\sim\sigma^{4/3}$. Physically, the well has become so shallow that the residence time is no longer long compared with the in-well relaxation time: the system leaves before it has had time to sample its own metastable state. Beyond that point ``in-well statistics'' ceases to be a well-defined notion, the quasi-stationary distribution loses its meaning, and B- and N-tipping stop being separable --- not because the diagnostics are imperfect, but because the two mechanisms have genuinely merged into a single overdamped slide out of the well. This is a real physical crossover rather than a defect of the framework, but it does delimit its purpose: the RP separation is a \emph{precursor} tool, sharp while the metastable state still exists, and silent about the endgame itself. Locating the crossover --- equivalently, estimating when $\kappa_{\mathrm{tr},a}^\epsilon$ overtakes $\Delta_{\rm loc}^{(a)}(\epsilon)$ --- is therefore arguably the practically relevant question, since it marks the moment after which no in-well early warning can be constructed. (iii) The metastable neighborhoods $\mathcal W_a^\epsilon$ depend on the committor threshold $\eta$; we have argued in Sec.~\ref{sec:1D_fold_example} that the product $\kappa_a^\epsilon p_{ab}^\epsilon$ is insensitive to this choice, but the individual factors are not, and the conditioned spectrum inherits a mild dependence through the boundary layer. When $\epsilon$ is ramped, $\mathcal W_a^\epsilon$ itself moves, and $B_{\epsilon,a}^{\rm loc}$ then contains a boundary-variation term that we have only indicated. (iv) Everything here is first-order response theory around a frozen parameter value; genuine rate-induced tipping \cite{ashwin2012tipping,alkhayuon2018rate}, in which the transition is caused by the \emph{speed} of the forcing rather than by its endpoint, falls outside the present formulation and requires a nonautonomous (pullback) extension \cite{csg11}.

\subsection{Practical prescriptions}
The theory is deliberately operational, and it can be summarized as a checklist. To diagnose loss of \emph{local} recovery of an attracting set: construct a committor-based neighborhood; kill the dynamics on its boundary; pass to the $Q$-process, either by learning $h_{\epsilon,a}$ and simulating the Doob drift \eqref{eq:Q_process_simulation_SDE}, or by running an unweighted Fleming--Viot ensemble; estimate the conditioned correlation $C_{{\rm loc},a}^{f,\epsilon}$ and hence $\Delta_{\rm loc}^{(a)}(\epsilon)$; and finally test whether the observable at hand has nonvanishing local residues. To diagnose \emph{escape risk}: estimate committors, killed escape clocks, destination probabilities, transition rates and quasipotential barriers, using weighted Girsanov trajectories or splitting algorithms \cite{cerou2007adaptive,bouchet2019rare,ragone2018computation}. A rising low-frequency signal that has not been assigned to one of these two mechanisms is not yet an early-warning signal; it is an observation awaiting interpretation.

This also suggests how to design observables rather than merely test them. The residue $\Rstyle_{{\rm loc},k}^{(a),f}$ is a bilinear pairing, so maximizing it over a family of admissible observables is a well-posed optimization problem whose solution is, essentially, the projection of the perturbation functional onto the critical generalized eigenspace. The empirical finding that edge-state-aligned observables anticipate transitions far better than generic ones \cite{Lohmann2025} is precisely what this variational reading predicts, and it also explains why the same system can look ``safe'' in one coordinate and ``critical'' in another.

\subsection{Open directions}
 Five extensions seem within reach. First, the non-normality index should be taken beyond the linear Gaussian setting in which we have validated it. Nothing in Proposition~\ref{prop:nonnormality_bound} is linear: $\mathcal N(f)=\Delta_{\rm RP}{\rm Var}_\mu(f)/\mathcal E(f,f)$ is defined for any diffusion with a smooth invariant density, its three ingredients are all estimable from data, and its reversible bound is just the Poincar\'e inequality with the optimal constant. What remains open is its behaviour in a genuinely metastable regime, where the leading RP gap is a transition rate rather than a relaxation rate, and where we would expect $\mathcal N$ to be controlled by the ratio of residence time to in-well relaxation time --- that is, to measure metastability rather than non-normality. Disentangling the two contributions to $\mathcal N$, presumably by computing it for the conditioned generator $K_{\epsilon,a}^{\rm loc}$ rather than for $K_\epsilon$, would extend the certificate to the metastable setting where it is most needed. Second, the frequency-resolved two-resolvent formula \eqref{eq:freq_sensitivity_blocks} predicts \emph{pairwise} block interference in spectral EWS, a signature that has, to our knowledge, never been looked for in data. Third, the whole construction should be tested on a physically meaningful metastable model where both channels are active and independently computable, and where, unlike in Sec.~\ref{sec:nonnormal}, the answer is not known in closed form in advance; the Stommel--Cessi box model of the thermohaline circulation, whose tipping transitions have been analyzed through optimal parameterizing manifolds \cite{chekroun2023optimal,chekroun2023transitions}, is a natural candidate, since its quasipotential and its conditioned in-well spectrum are both accessible. Fourth, the nonautonomous extension needed for ramped forcing, in which the frozen RP blocks are replaced by the spectral objects of a pullback or Fokker--Planck two-parameter evolution family, would close the gap between the hazard integral \eqref{eq:fold_ramped_jump_probability} and the genuinely time-dependent theory, possibly leading to an improved understanding of R- and P-tipping.

 Fifth, and least developed, is the computational question of how any of this is to be done in high dimension. The three objects the framework actually requires --- the principal Dirichlet eigenfunction $h_{\epsilon,a}$ that generates the Doob drift, the committor $q_a^\epsilon$ that defines the metastable neighborhoods, and the leading RP blocks with their spectral projectors --- are all instances of the same computational problem: extracting a dominant invariant subspace of a Markov semigroup from trajectory data. The classical route is a Galerkin projection onto a dictionary of observables, as in transfer-operator and extended dynamic mode decomposition methods \cite{dellnitz1999approximation,williams2015data,klus2018data,budivsic2012applied,Chekroun_al_RP2}, and its well-known bottleneck is the dictionary itself. Bollt made precisely this the object of study rather than a preprocessing step, treating dictionary choice geometrically through the cardinality and conditioning of the representation and through the notion of a primary eigenfunction \cite{bollt2021geometric,bollt2013applied}, and, with coauthors, replacing hand-designed dictionaries by learned ones \cite{li2017extended}. A complementary strategy dispenses with the dictionary altogether and parameterizes the dominant invariant subspace directly, evolving it with short trajectory bursts, as in the ISOKANN construction \cite{rabben2020isokann,sikorski2024learning} --- which is also, as Sec.~\ref{subsec:Q_process_Girsanov_Doob} noted, what supplies the Girsanov control in practice.

 Kernel-based variants of extended dynamic mode decomposition offer a third route: they dispense with an explicit dictionary altogether, the feature map being induced implicitly by the kernel and adapting to the geometry of the data, which is a decisive practical advantage in high dimension \cite{DeGennaro2019,Klus2018,Klus2019,Nateghi2024,Colbrook2024Multi,Zagli2026}; rigorous approximation-error bounds are moreover available for them \cite{Philipp2025,Kohne2025}, a guarantee that few data-driven methods can offer.

The residue-conditioned viewpoint raises the accuracy bar on all of these methods in a specific and, we think, underappreciated way. A dictionary may reproduce the leading eigenvalues accurately while representing the associated spectral projectors badly, and the entire content of Secs.~\ref{sec:L2mu_setting}--\ref{sec:nonnormal} is that the warning lives in the projectors: in $\Pi_jf^c$, in its pairing against the perturbation, and in the residues that result. Section~\ref{sec:nonnormal} makes the point unusually sharp, since there the eigenvalues are exactly independent of the control parameter, so an approximation that is eigenvalue-accurate and projector-inaccurate is blind to the whole phenomenon by construction; and the quantity that grows, ${\rm cond}(\Pi_\pm)\sim\kappa$, is exactly the conditioning that governs how badly a Galerkin truncation can misrepresent a non-normal spectral decomposition. Dictionaries and learned subspaces should therefore be selected and validated against residue accuracy rather than eigenvalue accuracy --- a well-posed objective that, to our knowledge, no current method optimizes. A second and more speculative thread concerns interpretability. Residue selection tells us \emph{which} observable to measure, which is only useful if the answer can be read; a neural parameterization returns a black box, whereas architectures that yield symbolic expressions, such as Kolmogorov--Arnold networks \cite{liu2024kan}, would return the residue-selected observable as a formula in the physical variables. Combining a residue-aware objective with such an architecture would deliver what the theory asks for and current practice does not: not merely a warning, but the named observable through which it is visible.

The broader message is a change of question. The RP framework does not ask whether correlations are large or whether spectra are red. It asks which block produces the signal, and through which residue it is visible. Only once that question has been answered does an observed slowing down  can become evidence about tipping.

\section*{Data Availability Statement}
The data that support the findings of this study are available from the corresponding author upon reasonable request. 

\begin{acknowledgments}
This work has been supported by the European Research Council (ERC) under the European Union's Horizon 2020 research and innovation program (grant agreement No.~810370), by the Knell Family Institute for Artificial Intelligence, Weizmann Institute of Science, and by the Institute for Environmental Sustainability (IES) at the Weizmann Institute of Science. This work has also been partially supported by the Office of Naval Research (ONR) Multidisciplinary University Research Initiative (MURI) Grant N00014-20-1-2023, and by the National Science Foundation Grant DMS-2407484. VL wishes to thank J. Lamb for useful exchanges on quasi-stationary and quasi-ergodic measures and D. Sornette for useful exchanges on non-normal dynamics, and MDC wishes to thank  Manuel Santos Guti\'errez for inspiring discussions on non-normality.
VL acknowledges the partial support provided by the Horizon Europe Projects Past2Future (Grant No. 101184070) and ClimTIP (Grant No. 100018693), by the ARIA SCOP-PR01-P003 - Advancing Tipping Point Early Warning AdvanTip project, and by the European Space Agency Project PREDICT (Contract 4000146344/24/I-LR). VL acknowledges the support received as Simons Fellow of the I. Newton Institute (FCPW01 programme).
\end{acknowledgments}

\appendix
{\small

\section{Direct covariance perturbation}
\label{subsec:direct_covariance_perturbation}
It is useful to distinguish perturbations of the diffusion amplitude from direct first-order perturbations of the covariance tensor. Let
\be
a_\eps
= a+\eps\Sigma,
\label{eq:direct_covariance_perturbation}
\ee
where $\Sigma=\Sigma(x)$ is a symmetric tensor field, chosen so that $a_\eps$ remains an admissible covariance tensor for $|\eps|$ small. We assume also that the perturbed diffusion retains the hypoellipticity and dissipativity properties used above, so that its invariant measure remains smooth:
\be
\d\mu_\eps(x)=\rho_\eps(x)\d x.
\label{eq:cov_pert_density}
\ee
At the level of the Kolmogorov generator, the perturbation is
\be
B_\Sigma\varphi
= \frac{\sigma^2}{2}
\Sigma:\nabla^2\varphi .
\label{eq:BSigma_def}
\ee
Thus Eq.~\eqref{eq:linear_response_poisson} gives
\bea
\left.\pd{}{\eps}\mu_\eps(f)\right|_{\eps=0}
&=
-\mu(B_\Sigma u_f)\\
&=
-\frac{\sigma^2}{2}
\int_{\mathbb R^d}
\Sigma(x):\nabla^2 u_f(x)\rho(x)\d x .
\label{eq:cov_response_poisson}
\eea
Equivalently, using the perturbation functional introduced in Eq.~\eqref{eq:perturbation_functional_L2},
\be
\mathscr{L}_{B_\Sigma}(g)
= \mu(B_\Sigma g)
= \frac{\sigma^2}{2}
\int_{\mathbb R^d}
\Sigma(x):\nabla^2g(x)\rho(x)\d x .
\label{eq:ell_BSigma}
\ee
The RP sensitivity formula \eqref{eq:RP_sensitivity_residue_form} gives
\bea
\left.\pd{}{\eps}\mu_\eps(f)\right|_{\eps=0}
&=
\sum_{j=1}^N
\sum_{k=0}^{m_j-1}
\frac{\Rstyle_{j,k}^{f}(B_\Sigma)}{(-\lambda_j)^{k+1}}\\
&\quad
- \mathscr{L}_{B_\Sigma}
\left(
\mathcal R_N^{(0)}f^c
\right),
\label{eq:cov_response_RP_residue}
\eea
where, by Eq.~\eqref{eq:RP_residues_general_B},
\be
\Rstyle_{j,k}^{f}(B_\Sigma)
= \frac{\sigma^2}{2}
\mu\left(
\Sigma:\nabla^2
\big(
N_j^k\Pi_j f^c
\big)
\right).
\label{eq:covariance_RP_residue}
\ee
Hence, written explicitly,
\bea
\left.\pd{}{\eps}\mu_\eps(f)\right|_{\eps=0}
&=
\frac{\sigma^2}{2}
\sum_{j=1}^N
\sum_{k=0}^{m_j-1}
\frac{1}{(-\lambda_j)^{k+1}}
\mu\left(
\Sigma:\nabla^2
\big(
N_j^k\Pi_j f^c
\big)
\right)\\
&\quad
-\frac{\sigma^2}{2}
\mu\left(
\Sigma:\nabla^2
\big(
\mathcal R_N^{(0)}f^c
\big)
\right).
\label{eq:cov_response_RP}
\eea
Thus a direct covariance perturbation probes the RP blocks through the Hessians of the projected observable components. For the $j$th RP block, the relevant scalar residues are not simply the projections $\Pi_j f^c$, but
\be
\mu\left(
\Sigma:\nabla^2
\big(
N_j^k\Pi_j f^c
\big)
\right),
\qquad
0\le k\le m_j-1.
\label{eq:covariance_residue_levels}
\ee
A small RP denominator therefore produces a large covariance sensitivity only if at least one of these Hessian residues is nonzero.

\section{Differentiating invariant statistics}
\label{sec:differentiating_statistics}
We now perturb the unperturbed diffusion introduced above. Let
\be
\d X_t^\eps
= F_\eps(X_t^\eps)\d t
+
D_\eps(X_t^\eps)\d W_t,
\label{eq:perturbed_SDE_general}
\ee
with
\be
F_0=F,
\qquad
D_0=D.
\label{eq:unperturbed_coefficients}
\ee
The corresponding Kolmogorov generator is denoted by
\be
K_\eps \varphi
= F_\eps\cdot\nabla\varphi
+
\frac{\sigma^2}{2} a_\eps:\nabla^2\varphi,
\qquad
a_\eps=D_\eps D_\eps^T.
\label{eq:Keps_general}
\ee
We assume that, for $|\eps|$ small, the perturbed diffusion satisfies the same hypoellipticity and dissipativity assumptions as the unperturbed one. Thus $K_\eps$ admits a unique invariant probability measure $\mu_\eps$, absolutely continuous with respect to Lebesgue measure
$\d\mu_\eps(x)=\rho_\eps(x)\d x,$
with $\rho_\eps$ smooth. At $\eps=0$ we take $\mu_0=\mu$ with $\mu$ given by Eq.~\eqref{eq:mu_density}.
We view the perturbation at $\eps=0$ on the unperturbed Hilbert space $\mathcal{H}=L^2_\mu(\mathbb R^d).$

Assume that there exists a common core $\mathscr C\subset D(K)\cap D(B)$ such that
\be
K_\eps\varphi
= K\varphi+\eps B\varphi+o(\eps),
\qquad
\varphi\in\mathscr C,
\label{eq:Keps_expansion}
\ee
where
\be
B
= \left.\pd{K_\eps}{\eps}\right|_{\eps=0}
\label{eq:B_def}
\ee
is the first-order generator perturbation. For example, a drift perturbation $F_\eps=F+\eps G$ gives
\be
B\varphi=G\cdot\nabla\varphi,
\label{eq:B_drift_example}
\ee
whereas a direct covariance perturbation $a_\eps=a+\eps\Sigma$ gives
\be
B\varphi
= \frac{\sigma^2}{2}\Sigma:\nabla^2\varphi.
\label{eq:B_cov_example}
\ee
The invariant density satisfies the stationary identity
\be
\int_{\mathbb R^d} K_\eps\varphi(x)\rho_\eps(x)\d x=0,
\qquad
\varphi\in\mathscr C.
\label{eq:stationary_diff_start_density}
\ee
Assume that $\eps\mapsto\rho_\eps$ is differentiable at $\eps=0$ in the weak sense, and write
\be
\dot\rho
= \left.\pd{\rho_\eps}{\eps}\right|_{\eps=0},
\qquad
\dot\mu(\varphi)
= \int_{\mathbb R^d}\varphi(x)\dot\rho(x)\d x.
\label{eq:rho_dot_mu_dot}
\ee
Since each $\mu_\eps$ is a probability measure, $\dot\mu(\mathds{1})=0.$
Differentiating Eq.~\eqref{eq:stationary_diff_start_density} at $\eps=0$ gives
\be
\dot\mu(K\varphi)
+
\mu(B\varphi)
= 0,
\qquad
\varphi\in\mathscr C.
\label{eq:linearized_stationarity}
\ee
Let $f\in L^2_\mu$ and let $u_f$ be the Poisson corrector
\be
Ku_f=f^c,
\qquad
\mu(u_f)=0,
\label{eq:poisson_for_response}
\ee
with $u_f\in D(B)$ and with $u_f$ admissible in \eqref{eq:linearized_stationarity}. Since
$f^c=f-\mu(f)\mathds{1},$
and $\dot\mu(\mathds{1})=0$, we obtain
\bea
\left.\pd{}{\eps}\mu_\eps(f)\right|_{\eps=0}
&=
\dot\mu(f)
= \dot\mu(f^c)
= \dot\mu(Ku_f)\\
&=
-\mu(Bu_f).
\label{eq:linear_response_poisson}
\eea
Equivalently, using the invariant density $\rho$,
\be
\left.\pd{}{\eps}\mu_\eps(f)\right|_{\eps=0}
= -\int_{\mathbb R^d}
(Bu_f)(x)\rho(x)\d x.
\label{eq:linear_response_density}
\ee
Combining \eqref{eq:linear_response_poisson} with the RP block expansion \eqref{eq:Poisson_RP_L2} yields,  after using $-(-1)^k\lambda_j^{-(k+1)}=(-\lambda_j)^{-(k+1)}$,
\bea
\left.\pd{}{\eps}\mu_\eps(f)\right|_{\eps=0}
&=
\sum_{j=1}^N
\sum_{k=0}^{m_j-1}
\frac{1}{(-\lambda_j)^{k+1}}
\mu\left(
B N_j^k\Pi_j f^c
\right)\\
&\quad
- \mu\left(
B\mathcal R_N^{(0)}f^c
\right).
\label{eq:RP_sensitivity_formula}
\eea
This is the basic $L^2_\mu$ RP-resolvent sensitivity formula.

It is useful to isolate the scalar functional through which the perturbation acts. Define
\be
\mathscr{L}_B(g)
= \mu(Bg)
= \int_{\mathbb R^d}
(Bg)(x)\rho(x)\d x,
\qquad
g\in D(B).
\label{eq:perturbation_functional_L2}
\ee
Then the RP residues associated with the $j$th RP block are
\bea
\Rstyle_{j,k}^{f}(B)
&= \mathscr{L}_B\left(N_j^k\Pi_j f^c\right)
= \mu\left(
B N_j^k\Pi_j f^c
\right),\\
&\qquad\qquad\qquad\qquad 0\le k\le m_j-1.
\label{eq:RP_residues_general_B}
\eea
With this notation,
\bea
\left.\pd{}{\eps}\mu_\eps(f)\right|_{\eps=0}
&=
\sum_{j=1}^N
\sum_{k=0}^{m_j-1}
\frac{\Rstyle_{j,k}^{f}(B)}{(-\lambda_j)^{k+1}}\\
&\quad
- \mathscr{L}_B\left(\mathcal R_N^{(0)}f^c\right).
\label{eq:RP_sensitivity_residue_form}
\eea
For each RP block, the dominant singular order is selected by the highest nilpotent level with nonzero residue. More precisely, if
\be
\kappa_j^f(B)
= \max
\left\{
0\le k\le m_j-1:
\Rstyle_{j,k}^{f}(B)\neq0
\right\}
\label{eq:dominant_nilpotent_level}
\ee
is well-defined, then the leading contribution of the $j$th block scales as
\be
\frac{\Rstyle_{j,\kappa_j^f(B)}^{f}(B)}{(-\lambda_j)^{\kappa_j^f(B)+1}} .
\label{eq:dominant_block_contribution}
\ee
If all residues $\Rstyle_{j,k}^{f}(B)$ vanish, then the $j$th RP block does not contribute to the first-order response of $\mu(f)$, even though $f^c$ may have a nonzero projection on $\Pi_jL^2_\mu$.

This distinction is important. The factor $\Pi_j f^c$ determines whether the observable has a component in the generalized eigenspace of $\lambda_j$. The functional $\mathscr{L}_B=\mu\circ B$ determines whether the perturbation couples that component to the invariant statistic. Thus a small RP denominator produces a large response only when at least one scalar residue
\be
\mu\left(
B N_j^k\Pi_j f^c
\right)
\label{eq:nonzero_scalar_residue}
\ee
is nonzero.

A direct $L^2_\mu$ bound follows from \eqref{eq:RP_sensitivity_formula}. Since $\mu$ is a probability measure,
\be
|\mu(Bg)|
\le
|Bg|_{L^2_\mu},
\qquad
g\in D(B).
\label{eq:L2_mu_B_bound}
\ee
Therefore
\bea
\left|
\left.\pd{}{\eps}\mu_\eps(f)\right|_{\eps=0}
\right|
&\le
\sum_{j=1}^N
\sum_{k=0}^{m_j-1}
\frac{
\left|
B N_j^k\Pi_j f^c
\right|_{L^2_\mu}
}{
|\lambda_j|^{k+1}
}\\
&\quad
+
\left|
B\mathcal R_N^{(0)}f^c
\right|_{L^2_\mu}.
\label{eq:RP_sensitivity_bound_L2}
\eea
If, in addition, the residual reduced resolvent is $B$-admissible in the sense that
\begin{widetext}
\be
\left|
B\mathcal R_N^{(0)}
\left(
I-\Pi_0-\sum_{j=1}^N\Pi_j
\right)g
\right|_{L^2_\mu}
\le
C_{N,B}
\left|
\left(
I-\Pi_0-\sum_{j=1}^N\Pi_j
\right)g
\right|_{L^2_\mu},
\label{eq:B_admissible_remainder}
\ee
\end{widetext}
then
\be
\left|
B\mathcal R_N^{(0)}f^c
\right|_{L^2_\mu}
\le
C_{N,B}
\left|
\left(
I-\Pi_0-\sum_{j=1}^N\Pi_j
\right)f^c
\right|_{L^2_\mu}.
\label{eq:remainder_bound_L2_B}
\ee
Thus a global perturbation bound based only on the norm of the reduced inverse $K^{-1}(I-\Pi_0)$ is recovered by estimating all RP blocks and the residual contribution by a single constant. The refinement of \eqref{eq:RP_sensitivity_formula} is that it decomposes this inverse into RP blocks and exposes the observable-dependent and perturbation-dependent residues that actually determine the sensitivity.

}

\bibliographystyle{ieeetr}
\bibliography{Super_Bib_v2MC_VL}
\end{document}